\documentclass[twocolumn,pra,aps,superscriptaddress]{revtex4}

\usepackage{color}
\usepackage{epsfig}
\usepackage{latexsym}
\usepackage{amssymb}
\usepackage{amsmath}
\usepackage{algorithm}
\usepackage{algorithmic}
\usepackage{wrapfig}
\usepackage[apple mac]{inputenc}
\usepackage[english]{babel}
\usepackage{times}
\usepackage{latexsym}
\usepackage{fancyhdr}
\usepackage{verbatim}
\usepackage{tabularx}
\usepackage{epsfig}
\usepackage{amsmath}
\usepackage{amssymb}
\usepackage{graphicx}
\usepackage{wasysym}

\usepackage{yfonts}

\newcommand{\qed}{\hspace*{\fill}$\square$}

\newcommand{\be}{\begin{equation}}
\newcommand{\ee}{\end{equation}}
\begin{document}

\title{The New SI and the Fundamental Constants of Nature}
\author{Miguel A. Martin-Delgado}
\affiliation{Departamento de F\'{\i}sica Te\'orica, Universidad Complutense, 28040 Madrid, Spain.\\
CCS-Center for Computational Simulation, Campus de Montegancedo UPM, 28660 Boadilla del Monte, Madrid, Spain.}

\begin{abstract} 
The launch in 2019 of the new international system of units is an opportunity to highlight the key role that the fundamental laws of physics and chemistry play in our lives and in all the processes of basic research, industry and commerce. The main objective of these notes is to present the new SI in an accessible way for a wide audience. After reviewing the fundamental constants of nature and its universal laws, the new definitions of SI units are presented using, as a unifying principle, the discrete nature of energy, matter and information in these universal laws. The new SI system is here to stay: although the experimental realizations may change due to technological improvements, the definitions will remain unaffected. Quantum metrology is expected to be one of the driving forces to achieve new quantum technologies of the second generation.\\ 
\\
La puesta en marcha en 2019 del nuevo sistema internacional de unidades es una oportunidad
para resaltar el papel fundamental que las leyes fundamentales de la F\'{\i}sica y la Qu\'{\i}mica juegan
en nuestra vida y en todos los procesos de la investigaci\'on fundamental, la industria y el comercio.
El principal objetivo de estas notas es presentar el nuevo SI de forma accesible para una audiencia amplia. Tras repasar las constantes fundamentales de la naturaleza y sus leyes universales, se presentan las nuevas definiciones de las unidades SI utilizando como principio unificador la naturaleza discreta de la energ\'{\i}a, la materia y la informaci\'on en esas leyes universales.
El nuevo sistema SI tiene vocaci\'on de futuro: aunque las realizaciones experimentales cambien por mejoras tecnol\'gicas, las definiciones permanecer\'an inalteradas. La Metrolog\'{\i}a cu\'antica est\'a llamada a ser uno de las fuerzas motrices para conseguir
nuevas tecnolog\'{\i}as cu\'anticas de segunda generaci\'on.
\end{abstract}

\maketitle

\tableofcontents

%%%%%%%%%%%%%%%%%%%%%%%%%%%%%%%%%%%%%%
%%%%%%%%%%%%%%%%%%%%%%%%%%%%%%%%%%%%%%
\section{Introduction}
\label{sec:intro}
%%%%%%%%%%%%%%%%%%%%%%%%%%%%%%%%%%%%%%
%%%%%%%%%%%%%%%%%%%%%%%%%%%%%%%%%%%%%%

\begin{figure}[t]
  \includegraphics[width=0.3\textwidth]{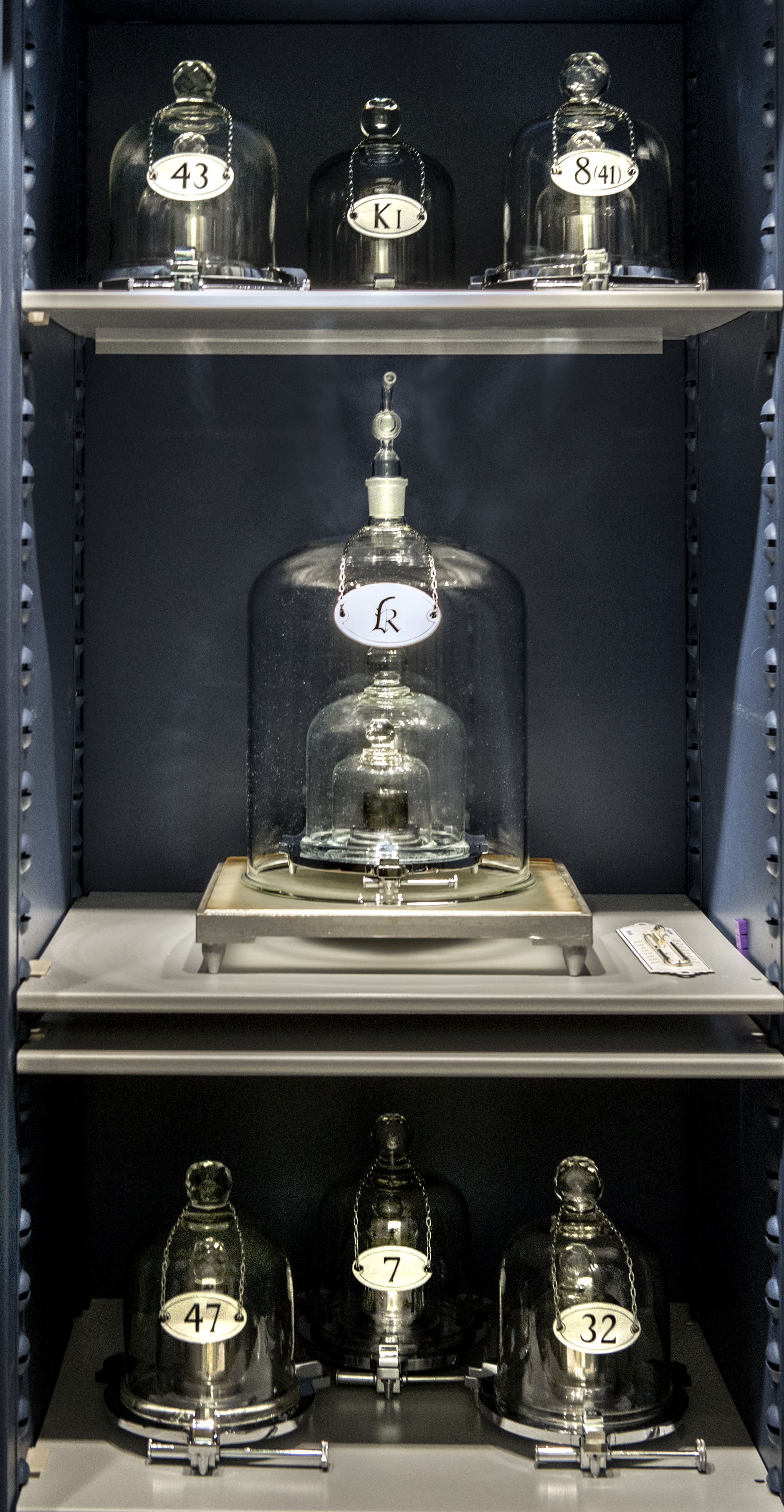}
  \caption{The international prototype of the IPK kilogram, kept at the BIPM near Paris, and its six official copies, {\it t\'emoins}. (Credit: BIPM).}
  \label{fig:IPK}
\end{figure}

On May 20th, 2019, matching with the World Metrology Day, the new international system (SI) of units that was approved at the assembly of the 26th General Conference of Weights and Measures (CGPM) met in Versailles during November 13nd-16th, 2018  \cite{press_kit}. This is a historical achievement. It is the culmination of many efforts during many years of joint work between the national metrology institutes from the member states and the BIPM (Bureau International des Poids et Mesures), providing a wonderful example of international collaboration.

The CGPM approved to review in 2018 four of the base units (the kilogram, the ampere, the kelvin and the mole). In this way, all the basic measurement units are linked to physical constants instead of arbitrary references. This means the retirement of the famous mass pattern, the kilo IPK \cite{press_kit,cem,SI:BIPM}, which was the only standard linked to a remaining material device. Now all the base units are associated with nature's rules to create our measurement rules \cite{press_kit}. What underlies all these redefinitions is the possibility of carrying out measurements at atomic and quantum scales to perform the units at macroscopic scale.

Although removing of artifacts to define base units is useful to better guarantee their stability and universality, however the new system brings with it the enormous challenge of explaining its functioning to society in layman words, and to high schools and universities. Artifacts are tangible (see Fig.\ref{fig:IPK}), while the fundamental laws of nature (physics and chemistry) 
are abstract and harder to grasp by the general public.
In this sense, in \ref{sec:nuevasdefiniciones} a unified treatment of all definitions of SI units is presented using as a common framework the discretization of energy, matter and information that is the fundamental ingredient of the laws of physics and chemistry to which the new SI units are linked.
These notes arise from several introductory lectures to explain the relationship of the fundamental constants with the new unit definitions.

When introducing the constants of nature in section \ref{sec:constantesfundamentales}, a distinguishing feature has been highlighted among the five universal constants associated with fundamental laws of nature. While $h, c$ and $e$ are associated with principles of symmetry, it is not the case of  
Botzmann's $k$ and Avogadro's $N_{\text A}$ constants.

Even though all the new definitions of the base units are presented here, however an exhaustive presentation of all their experimental realizations is avoided for it is far too technical for the purpose of these notes. There is more detailed documentation for that \cite{cem,SI:BIPM,SI:BIPMrealization}. 
The case of the `quantum kilo' deserves a special treatment, as it is so novel and mass something so common in daily life. Thus,  the Kibble balance,  which is the practical realization of the new
kilo, is given a simple description of how it works.

The rest of the article is organized as follows: in section \ref{sec:nuevoSI}, the new methodology of separating unit definitions from their experimental realizations is explained; section  \ref{sec:constantesfundamentales} describes the fundamental constants of nature appearing in the new definition of SI units as preparation for the explicit definition in section \ref{sec:nuevasdefiniciones} of the seven base units. 
In \ref{sec:MetrologiaSI} the role of quantum metrology in the new SI is explained through three examples: quantum clocks, the Kibble balance and the `quantum kilo', and the quantum metrological triangle. In section \ref{sec:anomaliaG} we reflect on the absence of the universal gravitational constant $G$ in the new system of units and its implications. Section \ref{sec:conclusions} is devoted to conclusions.

%\'{\i}

\begin{table}
\begin{center}
    \begin{tabular}{ | c | c |}
    \hline \hline
    Constant & Value  \\ \hline \hline
    $h$ & $6.626 \ 070 \ 15 \times 10^{-34} \text{J s}$  \\ \hline
    $e$ & $1.602 \ 176 \ 634 \times 10^{-19} \text{C}$  \\  \hline
    $k$ & $1.380 \ 649  \times 10^{-23} \text{J} \text{K}^{-1}$  \\  \hline
    $N_{\text{A}}$ &  $6.022 \ 140 \ 76  \times 10^{23} \text{mol}^{-1}$     \\ 
    \hline\hline
    \end{tabular} 
    \label{CODATA2018}
\end{center}
\caption{CODATA 2018 values \cite{CODATA2018,newell2018} for universal constants whose value has been set to define the kilo, ampere, kelvin and mole in the new SI of units enacted by
BIPM since May 20th, 2019.} 
\end{table}

%%%%%%%%%%%%%%%%%%%%%%%%%%%%%%%%%%%%%%
%%%%%%%%%%%%%%%%%%%%%%%%%%%%%%%%%%%%%%
\section{ The New SI of Units}
\label{sec:nuevoSI}
%%%%%%%%%%%%%%%%%%%%%%%%%%%%%%%%%%%%%%
%%%%%%%%%%%%%%%%%%%%%%%%%%%%%%%%%%%%%%

The new International System (SI) of units that is ruling as of
May 20th,  2019 represents a great conceptual and practical revolution since
for the first time all units are linked to natural constants,
many of them universal, and it means a dream of Physics and Chemistry come true.

The foundation of the new SI is based on the following premises \cite{cem,SI:BIPM}:
\begin{enumerate}
\item The separation of unit definitions from their particular experimental realizations.

\item The linking of unit definitions to natural constants.

\item The new unit system is designed to last over time and not be subject
to changes due to the continuous advances in the methods of experimental measurement.
\end{enumerate}
\begin{figure}[h]
  \includegraphics[width=0.32\textwidth]{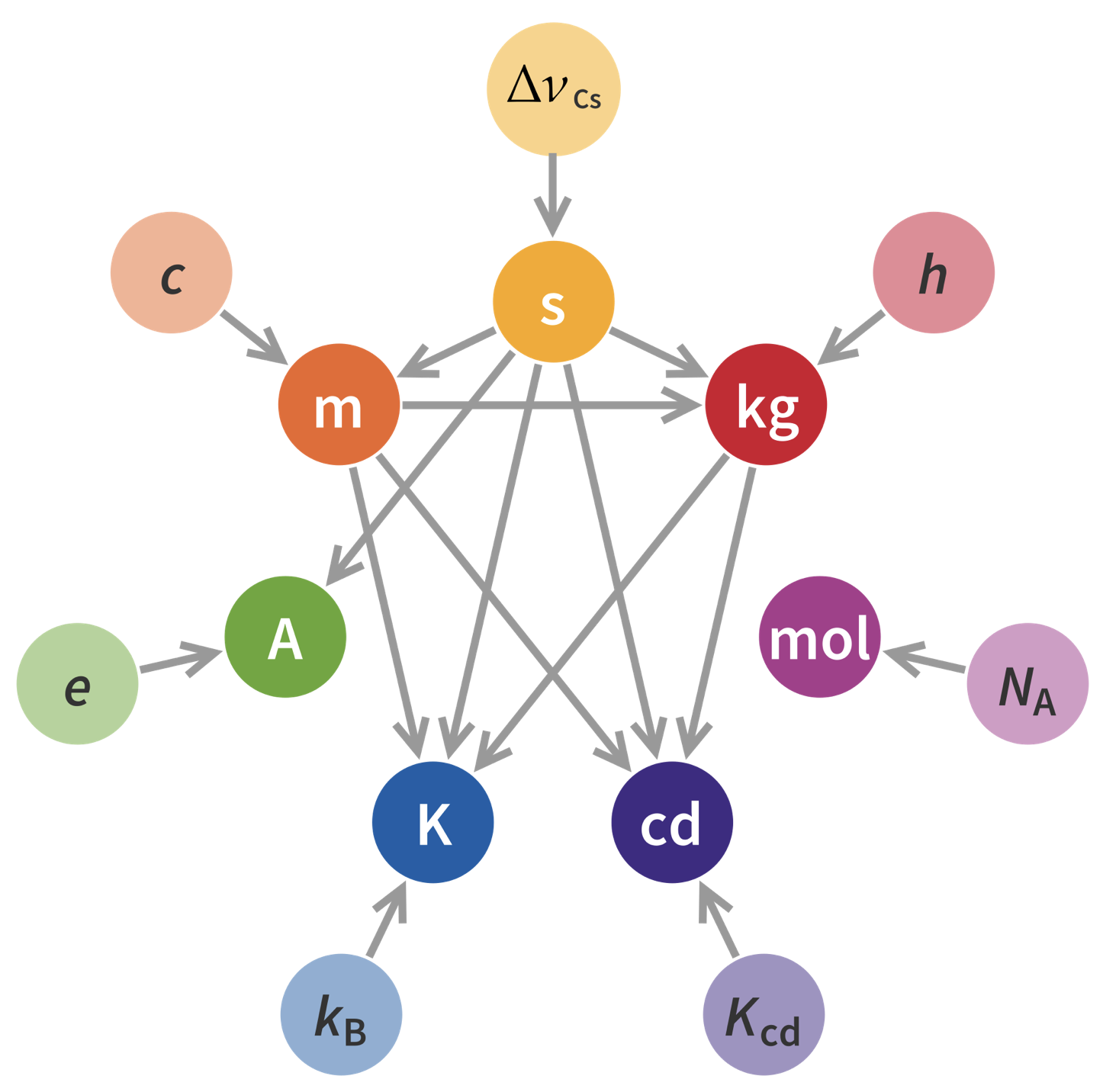}
  \caption{
  Schematic relationship between the base units of the new SI and its associated natural constants.
  In the central part the units and their dependencies among each other appear: the second influences the definition of five units, while the mole appears decoupled. The symbols of the constants used to define the units appear on the outside. See subsection   \ref{sec:nuevasdefiniciones}. (Credit: Emilio Pisanty/Wikipedia).}
   \label{fig:dependencias}
\end{figure}

The great conceptual advance of the new SI consists in separating the practical realization of units
from their definitions. This allows the units to be materialized independently anywhere and
at any time as envisioned by the Committee of Experts of the Decimal Metric System in 1789. This also allows new types of realizations to be added in the future as new technologies get developed, without having to modify the definition of the unit itself. An example of this comes from the new quantum technologies and the development of the quantum clock that will change the experimental realization of a second in the near future 
(see subsection \ref{sec:nuevasdefiniciones}).

In the new SI, the units of mass (kg), electric current (A), temperature (K) and quantity of substance (mol) are redefined by linking them to the four universal constants that appear in table \ref{CODATA2018}, whereas  the units of time (s), length (m) and luminous efficacy (cd)
remain associated with constants of nature as before (see Fig.\ref{fig:dependencias}).

The new dependencies among the units in the new SI are now much more symmetrical than in the previous system as leaps to the eye by taking a look at  Fig.\ref{fig:dependencias}.
The second is still the base unit on which all the others depend, except for the mole that appears decoupled from the rest of the units. The fundamental constants that are set to an exact value appear outside the scheme and their linked units appear inside. These redefinitions have fundamental consequences in certain magnitudes, such as
the electrical $\epsilon_0$ and magnetic $\mu_0$  constants in vacuum that cease to be exact and become experimentally determined in the new SI.
Thus, the magnetic constant is determined by the equation \cite{SI:BIPM}:
\begin{equation}\label{cmagnetica}
\mu_0 = \frac{2 \alpha}{e^2}  \frac{h}{c},
\end{equation}
so that all constants in this equation have a fixed value (see table \ref{CODATA2018})
except for the electromagnetic fine structure constant  $\alpha$ that is experimentally measured.
In turn, this results in a value given by  (CODATA2018)
\begin{align}\label{cmagnetica2}
\mu_0 &=  4\pi[1 + 0.0(6.8) \times 10^{-10}] \times 10^{-7}  \text{N} \text{A}^{-2} \\
 & =1.256 637 062 12 (19) \times 10^{-6} \text{N} \text{A}^{-2}. 
\end{align}
Then, the electric constant in vacuum is obtained from the relationship,
\begin{equation}\label{celectrica}
\epsilon_0 = \frac{1}{\mu_0 c^2},
\end{equation}
yielding the current value (CODATA2018)
\begin{equation}\label{celectrica2}
\epsilon_0 = 8.854 187 8128(13)  \times 10^{-12} \text{F} \text{m}^{-1}.
\end{equation}
However, in our daily life the changes will not cause any trouble because they typically correspond to changes of a part in $10^8 $, or even less. Their effects are very important though in the high precision measurements that are needed in research laboratories and metrology institutes where it is essential to be able to make exact and precise measurements to know whether a new discovery has been really found.

%\'{\i}

%%%%%%%%%%%%%%%%%%%%%%%%%%%%%%%%%%%%%%
%%%%%%%%%%%%%%%%%%%%%%%%%%%%%%%%%%%%%%
\section{The Fundamental Constants of Nature}
%\section{Revisi\'on del SI basada en las Constantes Fundamentales}
\label{sec:constantesfundamentales}
%%%%%%%%%%%%%%%%%%%%%%%%%%%%%%%%%%%%%%
%%%%%%%%%%%%%%%%%%%%%%%%%%%%%%%%%%%%%%

Underlying every universal constant of nature there is one of the fundamental laws
of Physics and Chemistry. Of the seven units of the new SI, five are associated
to universal constants of nature as shown in table \ref{Constantes-Leyes}.

The fundamental constants are like the DNA of our universe. Other universes, if they exist, may have a different set of universal constants. In our universe, depending on the particular physical phenomenon and its scale, we will need some fundamental constants to explain it. With this handful of constants we can describe our physical world from atomic, mesoscopic, microscopic, macroscopic, astronomical to cosmological scales.

In addition to these 5 universal constants, in the new SI there are two extra constants that are used to determine the unit of time and that of luminous efficacy. The first of these is the hyperfine transition frequency of the unperturbed fundamental state of the cesium atom 133, denoted by $\Delta \nu_{\text{Cs}}$.
Although this is a constant of nature, we cannot consider it as fundamental on an equal footing with the other five. If we did, any energy gap in the spectrum of any atom would also be fundamental and we would end up with an infinite number of fundamental constants. In addition, this frequency  is computable in principle using the laws of quantum electrodynamics, while constants such as those in table \ref{CODATA2018} are not calculable from the first principles currently known.
As for the light efficiency $K_{\text{cd}}$, the associated constant is not even universal and is purely conventional. In summary, only 5 of the 7 constants used in the new SI are really fundamental in the sense expressed here.

\begin{table}
\begin{center}
    \begin{tabular}{ | c | l | l |  p{3cm} |}
    \hline \hline
    Symbol & Constant  & Law \\ \hline \hline
    $c$ & speed of light & Theory of  Relativity \\ \hline
    $h$ & Planck  & Quantum Physics \\ \hline
    $k$ & Botzmann  & Thermodynamics \\  \hline
    $e$ & Electron charge & Quantum Electrodynamics \\  \hline
    $N_{\text{A}}$ &  Avogadro   & Atomic Theory \\ 
    \hline\hline
    \end{tabular} 
\end{center}
\caption{The five universal constants of nature and their corresponding laws they are associated with. The laws of Physics and Chemistry allow us to describe natural phenomena once the values of the constants are known.} 
    \label{Constantes-Leyes}
\end{table}

There is another essential aspect that deserves to be highlighted: three of these five universal constants are associated with  symmetry principles of nature.
The speed of light constant $c$  is responsible for the unification of space and time in the theory of Relativity \cite{einstein1905d}, one of the pillars of modern physics.
 What underlies this fundamental law is the Principle of Relativity, which declares all inertial reference frames in relative motion as physically equivalent. It is this symmetry that is responsible for the constancy of the speed of light. If $c$ were not constant, Lorentz's transformations and therefore the Relativity Principle, would be broken.

The Planck constant $h$ is responsible for the physical quantities such as energy, angular momentum, etc. can take on discrete values, called quanta. It is the fundamental constant of Quantum Mechanics, another of the pillars of modern physics. What underlies this fundamental law is the Unitary Principle, enforcing the probability of finding the particles in their quantum state to be preserved throughout their temporal evolution. Even more basic is the linearity of Quantum Mechanics represented by the Superposition Principle of states that is necessary to guarantee unitarity. If $h$ were not constant, the Unitarity Principle would be broken. 
It is the Principle of Superposition (linearity) of Quantum Mechanics that is at the root of all the counterintuitive surprises that quantum physics brings about as R. Feynman \cite{feynmannSIX,doblerendija} teaches us. Precision tests on the possible lack of non-linearity of Quantum Mechanics can be performed using non-linear models that provide theoretical estimates of  linearity validity rates of only $10^{-21}$ error \cite{weinberg1989}, and up to $4 \times 10^{-27}$ with direct measures \cite{wineland1989}. It so happens that to obtain these estimations, the highly exact measurements of the radio-frequency transitions  probed in the frequency standards are used. It turns out that a possible non-linearity would produce a de-tuning of those resonant transitions in the standards.

The charge of the electron $e$ is the value of the elementary (unconfined) source of electric field in Quantum Electrodynamics, the first of the known elementary particle theories and the one that serves as a reference for the rest of fundamental interactions. What underlies this fundamental law is the Principle of Gauge Invariance that describes known elementary interactions. In the case of electromagnetism, the invariance group is the simplest $U(1)$.
It is this symmetry that is responsible for the constancy of the electron charge: if $e$ is not constant, the gauge symmetry breaks down.

In these three examples, the values of the fundamental constants $c, h$ and $e$ are protected by nature's symmetries. An increasingly accurate measurement of them may result in a lack of constancy, and therefore, the violation of one of the fundamental laws of physics. Then metrology is also a source of discovery of new physics through the improvement over time of its measurement methods. A very important example of this fact within the new SI is the so-called quantum metrological triangle that we will see in the subsection \ref{sec:MetrologiaSI}, and also the possible variations in the electromagnetic structure constant $\alpha$ or the ratio of the proton mass to the electron mass (see \ref{sec:MetrologiaSI}).

The Boltzmann constant $k$ is the conversion factor that allows the thermodynamic temperature $T$ of a body to be related to the thermal energy of its microscopic degrees of freedom (constituents). This is the fundamental constant in Statistical Physics, which studies the relationship between macroscopic physics and its microscopic constituents, another of the pillars of physics. The constant $k$ appears in the description of the macroscopic world in the probability $P_i$, or Boltzmann factor, of finding a system in a microscopic state $i$ when it is in thermodynamic equilibrium at the temperature $ T $:
\begin{equation}\label{probabilidad}
P_i = \frac{e^{-\frac{E}{kT}}}{Z},
\end{equation}
where $Z$ is the partition function characteristic of the system.
Boltzmann established the relationship between macroscopic and microscopic worlds in his formula for entropy $S$:
\begin{equation}\label{entropia}
S = k \ln W,
\end{equation}
where $W$ is the number of different microscopic states corresponding to a macroscopic state of the system with given  energy $E$. This is the famous equation that appears in the frontispiece of 
Boltzmann's tomb in Vienna. However, Boltzmann established the relationship \eqref{entropia} as a proportionality law, without explicitly introducing his constant. Historically, Planck was the first to write it in his article where he laid down the black body radiation law, along with his constant $h$ of energy quanta \cite{planck1901}. Planck was the first to give numerical values to these two constants using the experimental values of the universal constants that appear in the Wien law of displacement and the Stefan-Boltzmann's law, which describe essential properties of radiation in thermal equilibrium. These first values turned out to be very close to the current ones \cite{planck1901} (see Table \ref{CODATA2018}):
\begin{equation}\label{probabilidad}
h = 6.55 \times 10^{-27} \ \text{erg s}; \quad k =  1.346 \times 10^{-16}  \text{erg} \ \text{degree}^{-1}.
\end{equation}
The Boltzmann constant has no associated symmetry principle, unlike the other three constants mentioned above.

Avogadro's constant $N_{\text A}$ is a conversion factor that relates the macroscopic amount of a substance to the number of its elementary constituents, whether they are atoms, ions, molecules, etc.
It is a fundamental constant in the Atomic Theory of matter in Physics and Chemistry. The mole is introduced to handle macroscopic quantities of a substance that is made of a huge number of elementary entities. Avogadro's constant $ N_{\text A}$ is the proportionality factor between the mass of one mole of substance (molar mass) and the average mass of one of its molecules, or whatever its elementary constituents. $N_{\text A}$ is also approximately equal to the number of nucleons in a gram of matter. To define the mole, the oxygen atom was initially taken as a reference and then carbon. In the new SI, the mass of one mole of any substance, be it hydrogen, oxygen or carbon, is $N_{\text A}$ times the average mass of each of its constituent particles, which is a physical quantity whose value must be determined experimentally for each substance.

The origin of the word mol comes from the Latin {\it moles} which means mass, and {\it molecula} which means small portion of mass. Avogadro's constant  has also no symmetry principle  associated.

As both the Boltzmann  constant $k$ and Avogadro's $N_ {\text A}$ are conversion factors for macroscopic and microscopic properties, they are also related to one  another:
\begin{equation}\label{gases}
R = N_{\text A} k,
\end{equation}
where $R$ is the constant of the ideal gases that relates pressure, volume and temperature: 
$P V=nRT$, with $n$ the number of moles in the gas. 

The Atomic Hypothesis plays a fundamental role in the description of nature. It states that matter is not a continuum, but is discrete and made of elementary entities called atoms. Feynman considered it the most important idea of all science because it contains a lot of information in a few words, and from which, you can reconstruct many of the properties surrounding us, such as that there are different states of matter depending on the temperature and its phase changes
\cite{feynmannSIX}. At the time of Boltzmann in the second half of XIX century, the existence of atoms and molecules was still under debate and is one of the reasons why the Boltzmann constant was introduced late because then macroscopic energies were used with the gas constant, instead of energies per molecule \eqref{gases} \cite{planck1920}. 
The works on the Brownian motion, theoretical by Einstein and experimental by Perrin, were essential to establish the validity of the Atomic Hypothesis at the beginning of the XX century.

%\'{\i}

\section{Revision of the SI based on the fundamental constants}
\label{sec:revisionSI}
%%%%%%%%%%%%%%%%%%%%%%%%%%%%%%%%%%%%%%
%%%%%%%%%%%%%%%%%%%%%%%%%%%%%%%%%%%%%%

\subsection{The New Definitions}
\label{sec:nuevasdefiniciones}

The explanations of the new SI are greatly facilitated by the new
viewpoint adopted  to separate the definitions of units, which are linked
to the constants of nature, from their concrete experimental realizations. The latter may be 
changing with the technology and the development of new measurement methods in the laboratory (section \ref{sec:nuevoSI}).

The visible universe is made of matter and radiation. Physics is the science devoted to the study of matter and radiation, and their interactions. The new SI uses the discrete nature of matter and radiation to define its units based on natural constants. The discrete character of matter is historically called the atomic hypothesis and the discrete character of radiation, the quantum hypothesis.

Let us start with electromagnetic radiation, one of whose forms is light. Its velocity  $c$  has a property that makes it special for measuring times and distances: it is a universal constant and has the same value for all inertial observers, that is, those who measure physical, i.e. observable, magnitudes.

Since time is the most difficult magnitude to define, then it is defined first as the most basic one.
 For this we use an oscillator that is very stable: the cycles of cesium atoms in an atomic clock. Galileo used pendulums, or even his own pulse, to measure time. The definition of time given by Einstein is famous:
``What is time? Time is what a clock measures". 
It is a very simple and at the same time a very deep definition. In fact, it is a metrological definition of time that fits very well in the new SI: once the time is defined in a generic way in terms of an oscillator or clock, 
the choice of a suitable clock is left for the realization of the unit second (s). According to the rules of the new SI, the definition and realization of the second is as follows:

\noindent {\bf second}:
``The second, symbol s, is defined by setting the fixed numerical value of  the cesium frequency  $\Delta \nu_{\text{Cs}}$, the unperturbed ground-state hyperfine
transition frequency of the caesium 133 atom, in 9 192 631 770, when expressed in the unit of Hz (hertz), equal to 1/s".
\qed

The realization of the second by means of the transition frequency of cesium
\begin{equation}\label{cesio}
\Delta \nu_{\text{Cs}} = 9 192 631 770 \ \text{Hz},
\end{equation}
implies that the second is equal to the duration of 9 192 631 770 periods of the radiation corresponding to the transition between the two hyperfine levels of the unperturbed ground-state of the atom
${}^{133}\text{Cs}$.

This materialization of the time standard is an example of the provisional character of the experimental realizations of the SI units. Standards based on cesium  have been around since the 60s of the XX century. We currently have more precise realizations using quantum clocks and a renewal of the second  using this quantum technology is already planned by the BIPM before 2030 (see \ref{sec:MetrologiaSI}). However, the definition of time will remain unchanged.

The metre is then defined with the time and the speed of light  $c$:

\noindent {\bf metre}:
``The metre, symbol $m$, is defined by taking the fixed numerical value of the speed of light in vacuum 
$c$ as 299 792 458 when expressed in the unit $\text{m} \text{s}^{-1}$, where the second is defined in terms of the cesium frequency $\Delta \nu_{\text{Cs}}$".
\qed

With this definition, a metre is the length of the path traveled by the light in vacuum during a time interval with a duration of 1/299 792 458 of a second. This definition is based on setting the speed of light in vacuum exactly at
\begin{equation}\label{luz}
c = 299 792 458 \ \text{m} \text{s}^{-1}.
\end{equation}
The methods for measuring the speed of light have changed over time, from the initial of Ole R\"{o}mer in 1676, based on the transit of the moon Io of Jupiter measured by a telescope, to modern techniques using laser interferometry.

Next, the natural thing is to define the unit of mass, the kilo.
It turns out that light has also another property that makes it very useful for defining the kilo:
light of a fixed frequency (monochromatic) has a minimum discrete energy called photon
whose energy is proportional to their frequency as discovered by Planck \cite{planck1901}, and then Einstein \cite{einstein1905a}.
That constant of proportionality is Planck's $h$ constant. The units of this constant are the basic three of what was once called the MKS System, precursor of the current SI: metre, kilo and second in the following proportion,
\begin{equation}\label{hunidades}
[h] = \text{kg}\ \text{m}^2 \ \text{s}^{-1}.
\end{equation}
It is important to note that Newton's universal gravitation constant $G$ has also units of the MKS system, although in another proportion:
\begin{equation}\label{Gunidades}
[G] = \text{kg}^{-1}\ \text{m}^3 \ \text{s}^{-2}.
\end{equation}
It turns out that $h$ and $G$ are the only fundamental constants with MKS units. There are other constants associated with fundamental interactions, but do not contain mass but other elementary charges.
However, for $G$ this is not enough to define the unit of mass with the accuracy that is needed in metrology. The problem is that the precision with which $G$ is measured is much worse than that of 
$h$. The `gravitational kilo' is not a good practical metrological unit. This fact is the origin of the `quantum way' for the kilo as we will see. In short, we can use $h$ to define the kilo from the second and the metre that are already defined once the value of $c$ is set.

Were it not for this lack of precision in measuring $G$, the constant $h$ could be decoupled from the kilo and set independently through the purely quantum effects of Hall and Josephson
(see \ref{sec:MetrologiaSI}):
\begin{equation}
h = \frac{4}{K_J^2 R_K} = 6.626068854 \ldots \times 10^{-34} \text{Js}.
\end{equation}
But if this other quantum route were taken, so natural in theory, then we would decouple the kilogram from $h$ and it would be linked to an artifact again: we find ourselves forced to choose  the `quantum kilo'.

Once the path of the `quantum kilo' is chosen, the next question is how to use the Planck  constant $h$ to define it. To do this, we use the prescription from the new SI to use the $h$ units and the second and metre definitions already introduced above. So, the definition of kilo looks like this:

\noindent {\bf kilo}:
``The kilogram, symbol kg, is defined by taking the fixed numerical value of the Planck constant, $h$ as 
$6. 626 070 15 \times 10^{-34}$, when expressed in unit J s, equal to 
$\text{kg} \ \text{m}^2 \ \text {s}^{-1}$, where the metre and the second are defined in terms of $c$ and $\Delta \nu_{\text{Cs}}$".
\qed

\noindent Or in equations,
\begin{equation}\label{kilo}
\begin{matrix}
1 \text{kg} &= \left(\frac{h}{6,626 070 15 \times 10^{-34}}\right)   \text{m}^{-2} \ \text{s} \\
&= 1,475 521 \ldots
\times 10^{40} \frac{h \Delta \nu_{\text{Cs}}}{c^2}
\end{matrix}
\end{equation}
This definition is equivalent to the exact relationship\begin{equation}\label{h}
h = 6. 626 070 15 \times 10^{-34} \ \text{Js}.
\end{equation}

After the definition of the 'quantum kilo', the problem arises as to how to do it experimentally.
The simplest thing at first sight would be to use the fundamental energy relations of Einstein \cite{einstein1905b} and Planck \cite{planck1901}, respectively:
\begin{equation}\label{energia}
E = mc^2 \quad \text{and} \quad E=h \nu.
\end{equation}
The basis of the `quantum kilo' is to have a very precise method to measure $h$ and then use it to define the kilo. But for this, the previous fundamental relationships present a problem. The photon, being a quantum of light energy,  has no mass. If we want the quantum to have a mass, what is better defined is its de Broglie wavelength
\cite{debroglie1924}:
\begin{equation}
E = \frac{h c}{\lambda}.
\end{equation}
However, measuring a wavelength is easy for a plane wave, which is again more typical of monochromatic radiation. To have a real mass $m$, we need a particle with an associated wavelength. This corresponds to a mass localized in space, which is more naturally described with a wave packet.
However, this one  does not have a single wavelength in turn. Thus, using the most basic relations of energy \eqref{energia} is not the most metrologically sensible thing to do.

Hence, the quantum way for the kilo is realized through the Kibble Balance (see \ref{sec:MetrologiaSI}). This leads us to an important question: What kind of mass, inertial or gravitational, appears in the units of $h$, and therefore in the new definition of kilo? In the case of the photon that has no mass, such a distinction does not exist. When we have a particle with mass, then it will depend on the mechanical relationship we use to relate it to $h$, and thus decide wether the kilo we define is inertial or gravitational.
For example, if we use Einstein's relationship, the kilo will be inertial, if we use a balance, then the kilo will be gravitational. Therefore, the Kibble Balance provides us with a definition of a quantum gravitational kilo. Now, the Equivalence Principle tells us that both types of mass are equal and is experimentally proven with an accuracy greater than the measurement of the fundamental constants involved in the SI: an uncertainty of $ (0.3 \pm 1.8) \times 10^ {-13}$ \cite{equivalence}. Thus, we can omit the gravitational term while the precision of the Equivalence Principle  is greater than that of the fundamental constants.

The description of the `quantum kilo' by means of the Kibble Balance belongs to the part of the new SI system corresponding to the practical realization of the kilo unit, not to its definition
(see \ref{sec:MetrologiaSI}). 

To continue defining the other SI units and derive them from the previous ones already defined, we turn to the discrete nature of matter. Thus, we know that  there are atoms (neutral) and electrons (charged). The most elementary charged matter (unconfined) is the electron and with it the ampere is defined using the second already defined:

\noindent {\bf ampere}:
``The ampere, symbol A, is defined by taking the fixed numerical value of the elementary charge, $e$, as:
\begin{equation}
e = 1.602176634 \times 10^{-19} \ \text{C},
\end{equation}
when expressed in the unit coulomb, C, equal to A s, and the second is defined in terms of  $\Delta \nu_{\text{Cs}}$".
\qed

Consequently, an ampere is the electric current corresponding to the flow of 
$1/(1.602 176 634 \times 10^{-19})$ elementary charges per second.
The advantage of the new ampere is that it can be really measured, unlike the old one that had an awful and impracticable definition that actually left it outside the SI system. In addition, it is independent of the kilogram and the uncertainty of the electrical quantities is reduced.

To experimentally realize  the ampere, several techniques have been proposed: a) the most direct is to use the current definition of SI through single electron transport (SET)
(see \ref{sec:MetrologiaSI}) \cite{SI:BIPMrealization,cem} 
although it is still under development to make it competitive; b) using Ohm's law and the Hall and Josephson effects to define volt and ohm (see \ref{sec:MetrologiaSI}) \cite{SI:BIPMrealization,cem}; 
c) the relationship between the electric current and the temporal variation of the bias voltage in a capacitor
\cite{SI:BIPMrealization,cem}.

%\'{\i}

As for atoms, historically, they are the elementary units of substance and with them the mole can be defined as the unit of quantity of a certain substance. Behind this definition is the discrete nature of matter through the atomic hypothesis and the natural constant associated to it is  Avogadro's
\eqref{gases}:

\noindent {\bf mol}:
``The mole, symbol mol, is the unit of the amount of substance. One mole contains exactly $6.02214076 \times 10^{23}$ elementary entities".
\qed

This value comes from setting the numerical value of the Avogadro constant to
\begin{equation}
N_{\text A} = 6.02214076 \times 10^{23} \ \text{mol}^{-1},
\end{equation}
in units $\text{mol}^{-1}$.  
As a consequence, the mole is the amount of substance in a system that contains  $6.022 140 76 \times 10^{23}$ specified elementary entities. The amount of substance in a system is a measure of the number of elementary quantities. An elementary quantity can be an atom, a molecule,
an ion, an electron, any other particle or a specific group of particles.

For the experimental realization of the ampere, several techniques have been proposed \cite{SI:BIPMrealization,cem}: a) the Avogadro project (International Avogadro Coordination), 
b)  gravimetric methods, c) equation of  gas, and d) electrolitic methods.

%\'{\i}
\begin{figure}[t]
  \includegraphics[width=0.25\textwidth]{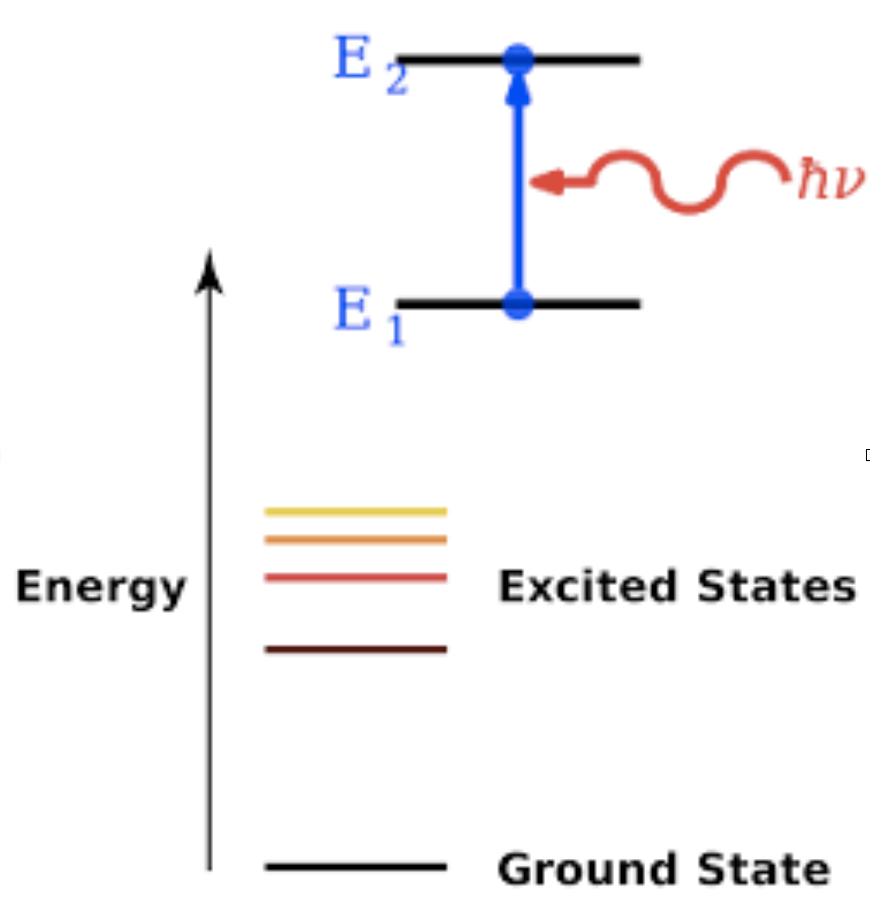}
  \caption{The first quantum revolution of technologies is based on the discrete nature of physical quantities, such as energy states in atoms. Photons of the electromagnetic radiation allow us to manipulate the states of well-defined energy \eqref{energia}. Credit Wikipedia (adapted)..}
  \label{fig:primerarevolucion}
\end{figure}

We still need the fundamental unit to measure temperature, the kelvin. Following the new SI, we must link it to a fundamental constant of nature, in this case the Boltzmann constant $k$. Boltzmann's constant serves as a conversion factor between energy and temperature:
\begin{equation}\label{kT}
E = k T.
\end{equation}
We can also see this constant as a result of nature's discrete character. For example, the atomic hypothesis for ideal noble gases allows to calculate their kinetic energy as:
\begin{equation}\label{nobles}
E_{\text c} = \frac{3}{2} k T.
\end{equation}
The resulting new definition for kelvin as the unit of thermodynamic temperature is:

%\'{\i}

\noindent {\bf kelvin}:
``The kelvin, symbol K, is defined by taking the fixed numerical value of the Boltzman constant, k, as
\begin{equation}\label{Boltzmann}
k = 1.380649  \times 10^{-23} \ \text{J} \text{K}^{-1},
\end{equation}
when it is expressed in units  kg $\text{m}^2$  $\text{s}^{-1}$ $\text{K}^{-1}$,
where the  kilogram, metre and second are defined according to $h, c$ and $\Delta \nu_{\text{Cs}}$".
\qed

\begin{figure}[t]
  \includegraphics[width=0.45\textwidth]{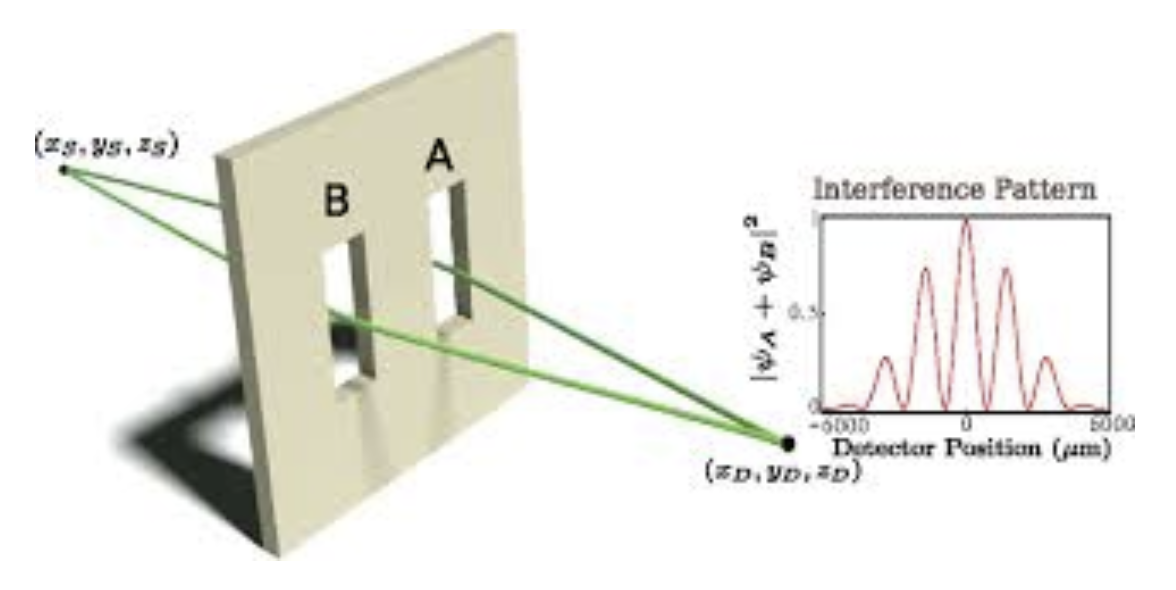}
 \caption{ The second quantum revolution of technologies is based on the principle of superposition of quantum mechanics. The simplest case is exemplified by the two-slit experiment \cite{feynmannSIX,doblerendija} where the properties of a single particle get in superposition. When quantum superposition involves several particles, the resulting phenomenon is quantum entanglement, which is the fundamental resource in quantum information \cite{rmp}. Credit: R. Sawant et al. \cite{segundarevolucion}}
  \label{fig:segundarevolucion}
\end{figure}

This definition implies that the kelvin is equal to a thermodynamic change in temperature resulting in a change in thermal energy kT of $1.380649  \times 10^{-23} \ \text{J}$. As a consequence of the new definition, the triple point of water ceases to have an exact value and now has an uncertainty given by:
\begin{equation}\label{triple}
T_{\text{TPW}} = 273.160 00 \text{K} \pm 0.000 10 \text{K},
\end{equation}
as a result of inheriting the uncertainty that the Boltzmann constant  $k$ had before the new definition.

Following the guiding principle used so far to introduce the new definitions of SI units using the discrete nature of energy \eqref{energia} and matter \eqref{gases}, we can go further on and use the discrete nature of information to introduce the Boltzmann constant. Thus, another way to substantiate the discrete origin of Boltzmann's constant is through the Landauer Principle
\cite{landauer1961} which establishes that the minimum energy dissipated in the form of heat at a temperature $T$ in the simplest elementary system, whether classical (bit) or quantum (qubit), is:
\begin{equation}\label{kT}
E = \ln 2 \ k T.
\end{equation}
The underlying origin of the Landauer Principle is the irreversible nature of information deletion or erasure
\cite{bennett1982, bennett2003} in a system, which leads to a minimum dissipation of energy
\cite{informationengine2014}. 
Its experimental verification has been possible directly in several recent works \cite{expland1,expland2,expland3,expland4}.
In this way, we have a unified vision of all SI units based on the fundamental property that both energy, matter and information possess: their discrete nature in our universe.

Several techniques have been proposed for the experimental realization of the kelvin \cite{SI:BIPMrealization,cem}:
a) by acoustic gas thermometry, b) radiometric spectral band thermometry, c) polarizing gas thermometry and d) Johnson noise thermometry.

%\'{\i}

The seventh base SI unit used to measure the light efficiency of a source deserves a special comment. It is a measure of the goodness of a light source when its visible light is perceived by the human eye. It is clearly conventional and subjective. Average values of human eye behavior are used.
It is quantified by the ratio of the luminous flux to the power, measured in lumens (lm) per watt in the SI. There is no universal law of physics associated with this unit, for the light source does not have to be in thermodynamic equilibrium. The basic concept for the candle is maintained in the new SI:

\begin{figure}[t]
  \includegraphics[width=0.35\textwidth]{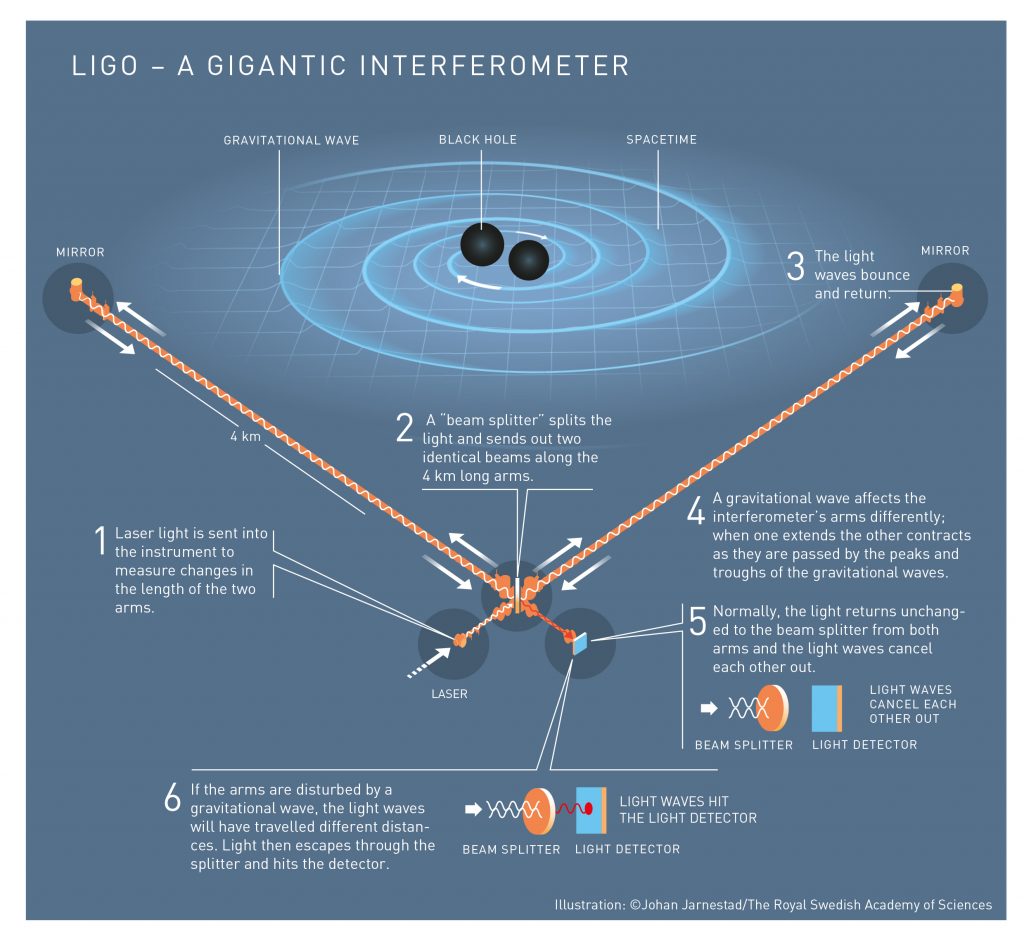}
  \caption{Diagram of a LIGO type gravitational wave detector. The two perpendicular arms of four km each are shown. The gravitational waves caused by the collision of two black holes can be detected in the interference pattern thanks to the extreme sensitivity of the device that allows ressolving distances thousands of times smaller than the atomic nucleus.
  (Credit Johan Jarnestad/The Royal Swedish Academy of Sciences \cite{ligonobel}).}
  \label{fig:LIGO}
\end{figure}

\noindent {\bf candela}:
``The candela, symbol cd, is the SI unit of luminous intensity in a given direction. It is defined by taking the fixed numerical value of the luminous efficacy of monochromatic frequency radiation
$540 \times 10^{12} \text{Hz}$, $K_{\text cd}$, to be 683 when expressed in the  unit
 lm $\text{W}^{-1}$, which is equal to  cd sr  
$\text{W}^{-1}$, or cd sr  $\text{kg}^{-1}$ $\text{m}^{-2}$ $\text{s}^3$, where the kilogram, metre and second
are defined in terms of  $h, c$ and $\Delta \nu_{\text{Cs}}$".
\qed

This definition is based on taking the exact value for the constant
\begin{equation}\label{luminosidad}
K_{\text {cd}} = 683 \ \text{cd}  \ \text{sr} \  \text{kg}^{-1} \ \text{m}^{-2} \ \text{s}^3,
\end{equation}
for monochromatic radiation of frequency $\nu = 540 \times 10^{12 }\text{Hz}$.
As a consequence, a candela is the light intensity, in a given direction, of a source that emits monochromatic radiation of frequency $\nu = 540 \times 10^{12 }\text{Hz}$ and has a radiating intensity in that direction of
 (1/683) W/sr.

For the experimental realization of  the candela, several techniques have been proposed \cite{SI:BIPMrealization,cem}
such as the practical realization of radiometric units, using two types of primary methods: those based on standard detectors such as the electrical replacement radiometer and photodiodes of predictable quantum efficiency, and those based on standard sources such as Planck's radiator and  synchrotron radiation. In practice, a standard lamp with optimized design is more commonly used to emit in a defined direction and at a long distance from the detector.

%\'{\i}

\subsection{Quantum Metrology and the New SI}
\label{sec:MetrologiaSI}

\begin{figure}[t]
  \includegraphics[width=0.45\textwidth]{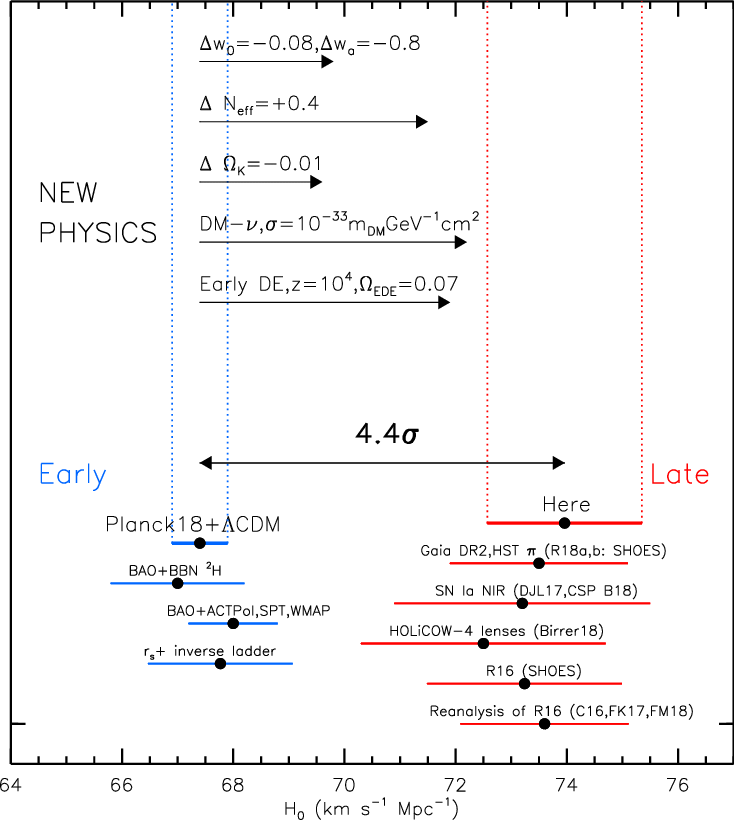}
  \caption{
  Value differences for the Hubble constant $H_0$ measured with the Planck 2018 satellite
in the early universe and extrapolated with the standard cosmological model (blue) \cite{planck2018}, and
local measurements made with the cepheid and supernova distance ladders.
(Credit Riess et al. \cite{H0tension}).}
  \label{fig:Tension}
\end{figure}

The foundation of the laws of quantum mechanics in the first quarter of the twentieth century allowed us to understand nature on an atomic scale. As a result of that better understanding of the atomic world, applications emerged in the form of new quantum technologies. This first period is known as the first quantum revolution and has produced technologies as innovative as the transistor and the laser. Even today's classic computers are a consequence of these first-generation quantum advances. 
With the beginning of the 21st century, we are seeing the emergence of new, second-generation quantum technologies \cite{manifesto} that constitute what is known as the second quantum revolution. Both revolutions are based on exploiting specific aspects of the laws of quantum mechanics. Thus, we can classify these technological revolutions into two groups:

\noindent {\bf First Quantum Revolution}: It is based on the discrete character of the  quantum world's properties: energy quanta (such as photons),  angular momentum quanta, etc. This discrete nature of physical quantities is the first thing that surprises in quantum physics.
(see Fig.\ref{fig:primerarevolucion}).
\qed

\noindent {\bf Second Quantum Revolution}:  It is based on the superposition principle of quantum states. With it,  information can be stored and processed as a result of its quantum entanglement properties
 (see Fig.\ref{fig:segundarevolucion}). 
\qed

In the second quantum revolution, five working areas have been identified that will lead to new technological developments. Enumerating them from least to greatest complexity are: quantum metrology, quantum sensors, quantum cryptography, quantum simulation and quantum computers.

Quantum metrology is considered one of the quantum technologies with most immediate development. It is the part of metrology that deals with how to perform measurements of physical parameters with high resolution and sensitivity using quantum mechanics, especially exploiting their entanglement properties.
A fundamental question in quantum metrology is how accuracy scales when measured in terms of the variance
$\Delta \textfrak{m}$ , used to estimate a physical parameter, with the number of particles used  $N$ or repetitions of the experiment. It turns out that the classical interferometers for light, electrons etc. cannot overcome the so-called shot-noise limit \cite{qmetrology,quantummetrology,quantummetrology2}  given by

\begin{figure}[t]
  \includegraphics[width=0.35\textwidth]{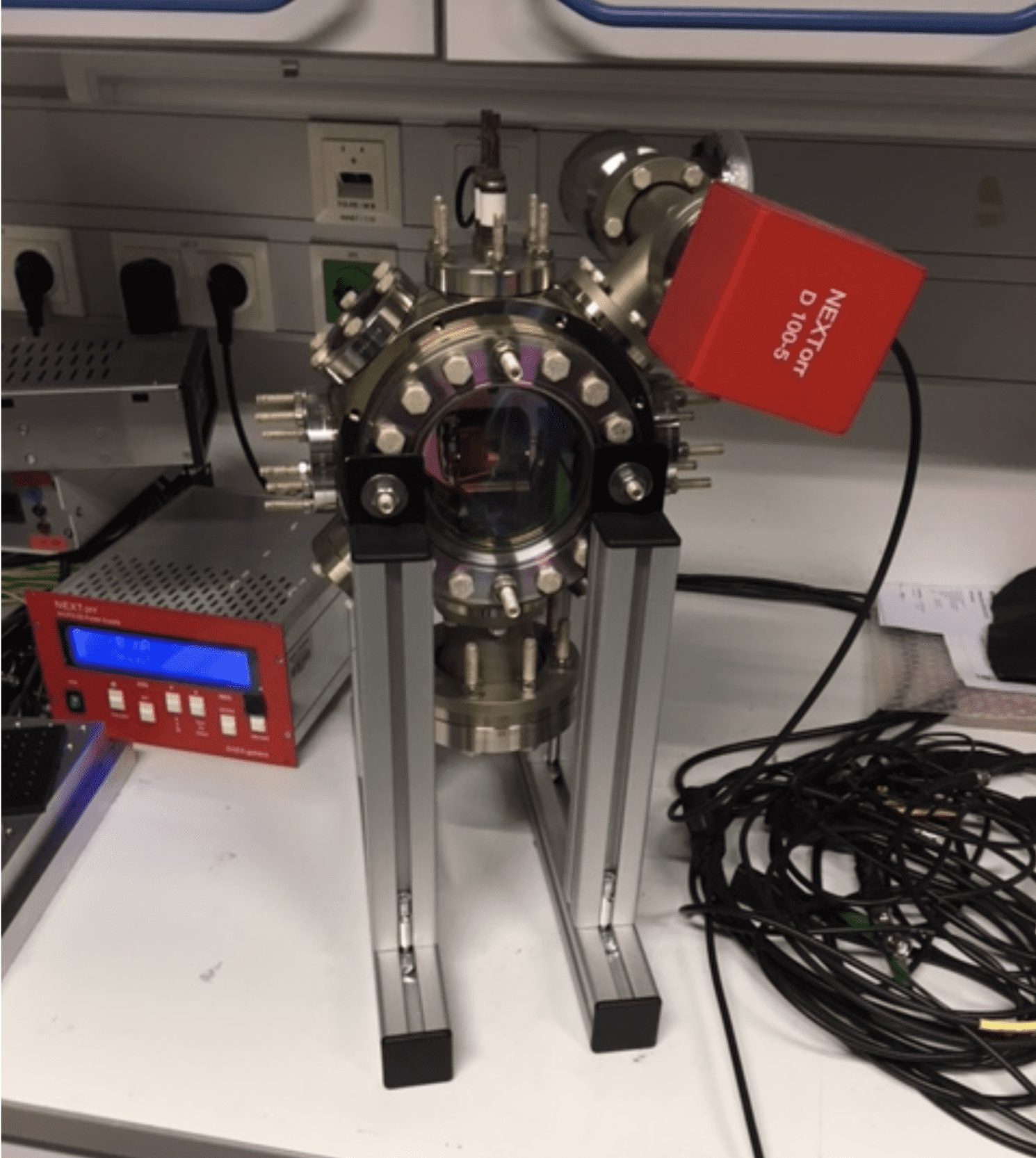}
    \includegraphics[width=0.35\textwidth]{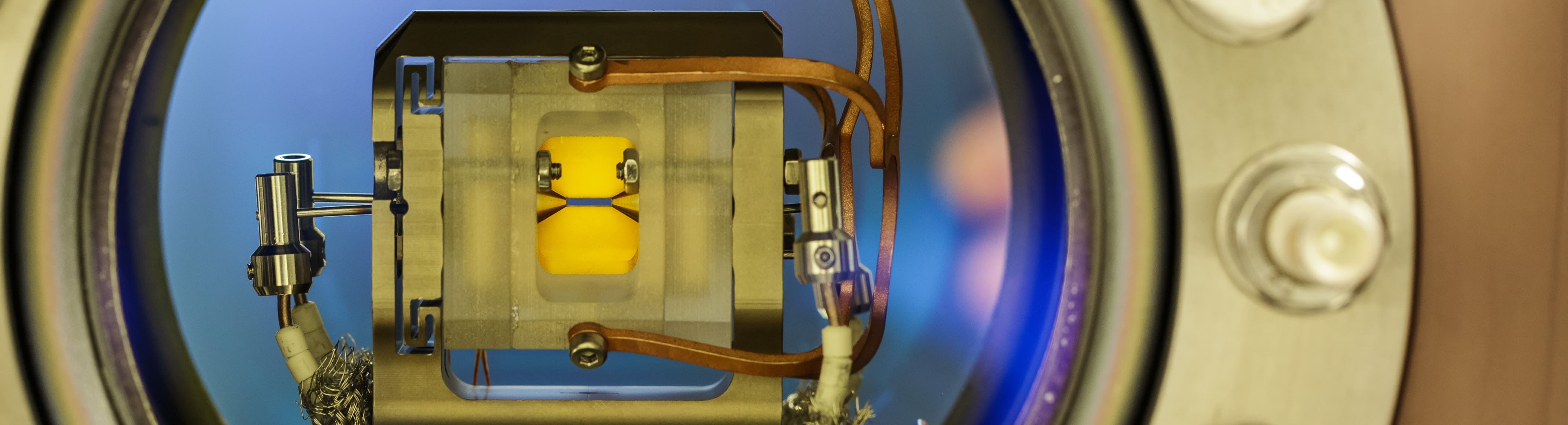}
  \caption{ Above: Overview of a trapped-ions quantum clock developed by Rainer Blatt's group in the  University of Innsbruck laboratory. It is a compact prototype that fits into a small functional space and is marketed by the company Alpine Quantum Technologies \cite{monz2019}. Below: detail of the central part of the clock showing the trap, where ions are confined using electromagnetic fields.
  (Credit: Rainer Blatt Lab).}
  \label{fig:QuantumClock}
\end{figure}

\begin{equation}\label{limiteclasico}
\Delta \textfrak{m} \geq \frac{1}{\sqrt{N}},
\end{equation}
whereas with second-generation quantum metrology it is possible to reach the Heisenberg limit given by
\begin{equation}\label{limitecuantico}
\Delta \textfrak{m} \geq \frac{1}{N}.
\end{equation}
An example of a very important application of quantum metrology to basic research is the detection of gravitational waves by experiments such as LIGO (Laser Interferometer Gravitational-Wave Observatory) 
\cite{ligo2016} (see Fig.\ref{fig:LIGO}) ,
where you need to measure distances between separate masses of the order of a few kilometers with a very high precision (thousandth of the diameter of a proton). These variations in distance occur in the lengths of the interferometer arms when a gravitational wave passes through them. Using `squeezed light', a form of quantum optics \cite{orzag2016}, the sensitivity of interferometers can be improved according to the quantum limit
\eqref{limitecuantico} \cite{squeezed,squeezed_LIGO,squeezed_LIGO2}.

The LIGO-type experiment measures may have another basic application to try to elucidate the controversy that has recently arisen with the Hubble constant $H_0$. In the standard cosmological model denoted by 
$\Lambda \text{CDM}$ ($\Lambda$=dark energy, CDM=cold dark matter) 
$H_0$ measures the speed with which the universe is currently expanding according to Hubble's law. It is a parameter of capital importance in cosmology. There are two sources of measures that give discrepant values. On the one hand, the value provided by the Planck satellite in 2018 by analyzing the microwave radiation background of the early universe and extrapolating the Hubble parameter to its current value with the standard model $\Lambda \text{CDM}$ gives a value of $H_0=67.4 \pm 0.5$ kilometers per second and per megaparsec distance \cite{planck2018}. 
On the other hand, measurements made in the current universe using distance ladders based on cepheids and type 1a supernovae result in $H_0=74.03 \pm 1.42 \ \text{km} \text{s}^{-1}\text{Mpc}^{-1}$ \cite{shoes2019}. This discrepancy implies a statistical confidence of 4.4 sigma (standard deviations), very close to the 5 sigma barrier that is considered as clear evidence that they are different results. If that were the case, it would amount to new physics that the standard model has not taken into account (see Fig.\ref{fig:Tension}).
In addition, to increase the controversy, there is also a measure of the current universe with another distance ladder that yields an intermediate value between the two discrepants, $H_0=69.8 \ \text{km} \text{s}^{-1}\text{Mpc}^{-1}$ \cite{friedman2019}.
All measurements made with the current universe give values above the values obtained with the early universe and the standard model. Then, either there are systematic errors, or it may be the indication of new physics beyond the current cosmological model, such as the existence of dynamic dark energy, to name just one possible example \cite{H0tension}.

LIGO experiments can be very useful in the future to elucidate this controversy, one more, about the Hubble constant. This time, it is not about analyzing black hole collisions as in the original discovery of gravitational waves, but about binary neutron star collisions. It turns out that when two neutron stars merge, the resulting gravitational waves can be used to obtain information about the position of the stars, and hence the galaxies where they are found. By making statistics of at least 50 events of these collisions one could have a direct measure of the Hubble constant with an accuracy not achieved to date and resolve the dispute
\cite{neutronstars1}. 
And recent results improve these expectations. It has already been possible to detect a neutron star fusion event useful to estimate the Hubble constant, whose resulting value is precisely 
$H_0=70 \ \text{km} \text{s}^{-1}\text{Mpc}^{-1}$ with an uncertainty of the order of $\sim 7\%$ \cite{neutronstars2}. It is estimated that with 15 events the error can be reduced to 1\% and begin to resolve the discrepancy.

%\'{\i}

Let us look now at three important examples of applications of quantum metrology to the new SI, both from quantum technologies of first and second revolution.

\subsubsection{Quantum Clocks}

The basic mechanism of a clock consists of a system with very stable periodic oscillations where each period defines the unit of time such that the clock counting those periods, measures time. In the past, natural periodic movements such as that of the Earth around its axis or around the sun have been used, as well as mechanical oscillators such as pendulums or quartz crystal resonators. In the 60s of the last century, atomic oscillations began to be used to measure time using cesium atoms for their greater accuracy and stability than mechanical systems. The current reference to define the time standard is cesium 133 using the resonance frequency corresponding to the energy difference between the two hyperfine levels of its fundamental state 
(see \ref{sec:nuevasdefiniciones}). 
Cesium atomic clocks can typically measure time with an accuracy of one second in 30 million years.
In general, the accuracy and stability of an atomic clock is greater the higher the frequency of the atomic transition and the smaller the width of the electronic transition line.
If we make the frequency of the oscillator bigger we can increase the resolution of the clock by reducing the period we use as a reference.

\begin{figure}[t]
  \includegraphics[width=0.5\textwidth]{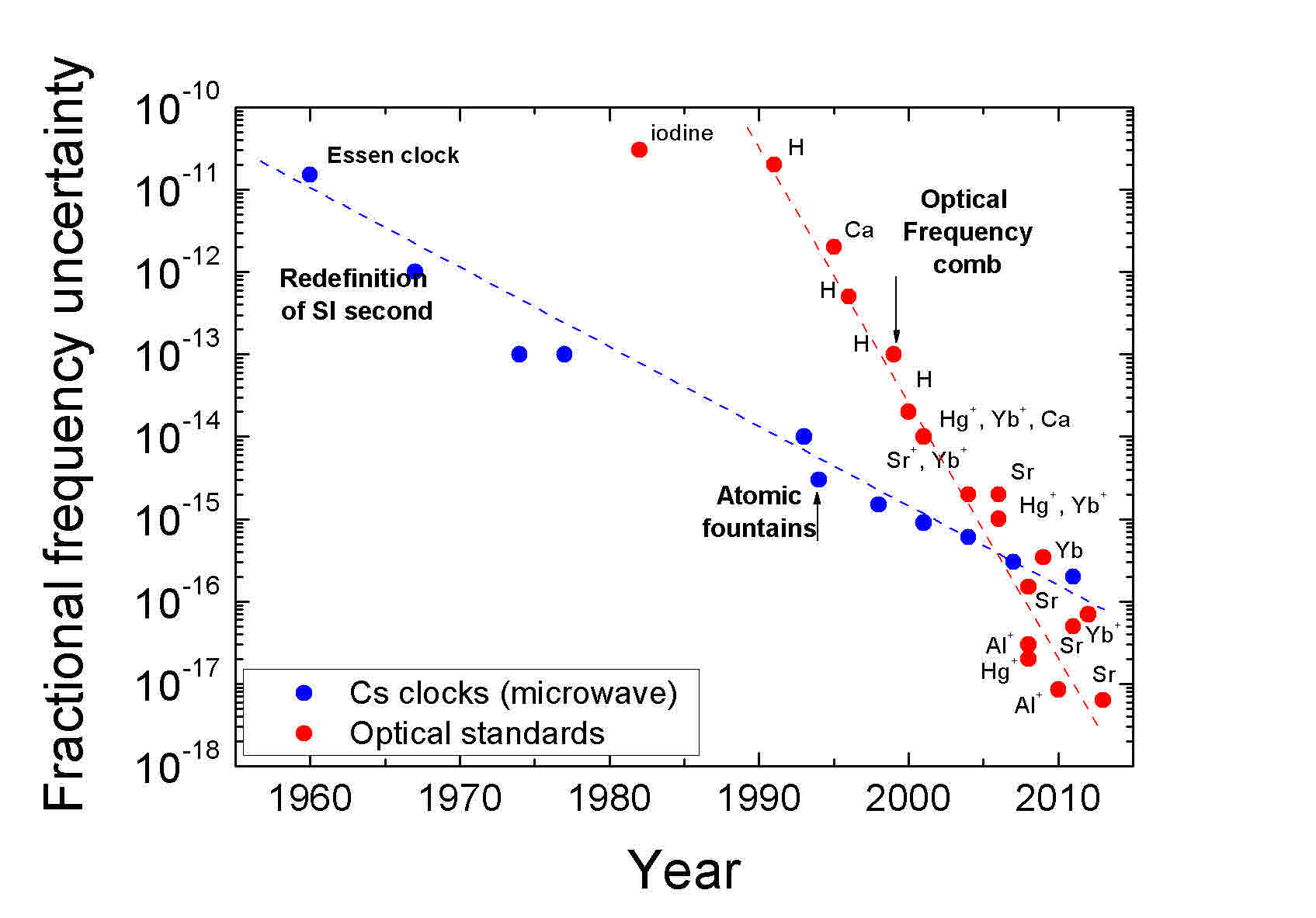}
  \caption{Comparison of the recent temporal evolution of uncertainties in cesium atomic (microwave) clocks and quantum (optical) clocks: trapped ions and neutral atoms in optical networks. 
  There is clearly a change in trend with a gain of accuracy in quantum clocks.
(Credit: Reproduced  with kind permission of  Societ\`a Italiana di Fisica from N. Poli et al. \cite{comparativeclocks}).}
  \label{fig:Comparativa}
\end{figure}
Atomic clocks have allowed us to improve multiple technological developments that we are used to in our daily lives: controlling the frequency of television broadcasting waves, global satellite navigation systems such as GPS, financial transactions, internet, mobile phones etc. .

\begin{figure}[t]
  \includegraphics[width=0.45\textwidth]{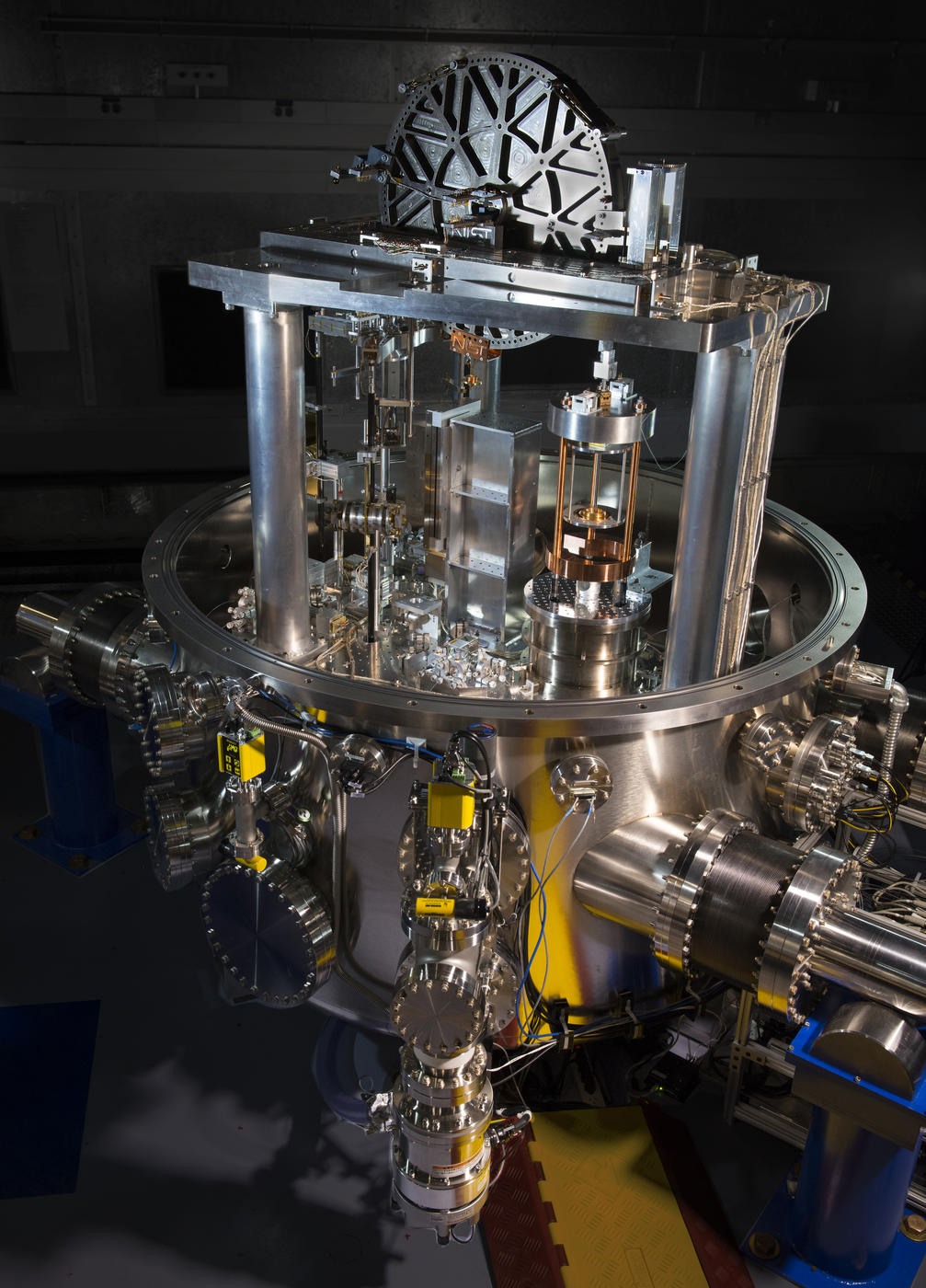}
  \caption{ Kibble NIST-4 balance used to measure the Planck constant $h$ with an uncertainty of 13 parts in one billion in 2017, contributing to the redefinition of the kilogram as a unit of mass in the new SI in 2019.  (Credit NIST).}
  \label{fig:KibbleBalance}
\end{figure}

The new quantum clocks are a type of atomic clocks where the increase in accuracy is due to using atomic transition frequencies in the optical range instead of the microwaves  used by cesium clocks. Optical frequencies of visible light are about five orders of magnitude greater than microwaves. In order to make this leap, it was necessary to use ion traps (see Fig.\ref{fig:QuantumClock}) with quantum logic techniques used in quantum computing \cite{cz1,cz2}, a part of the second revolution quantum technologies \cite{quantumclock1,quantumclock2,quantumclock3}.
Quantum clocks achieve an uncertainty of only one part in $10^{17}$s, which is equivalent to an error of a second in a quantum clock
that would have begun ticking 13.7 billion years ago, what represents the whole age of the universe.
An ion clock is an instance of quantum clock, which is  based on quantum logic and is an example of cooperation between two ions where each provides complementary functionalities. For example, the aluminum ion $\text{Al}^+$ has a transition frequency in the optical range that is useful for the clock reference frequency. However, its atomic level structure makes it a bad candidate to cool it down to the temperatures necessary for stabilization. Instead, this is possible with a beryllium ion $\text{Be}^+$. Using quantum computing protocols, information on the internal state of the spectroscopic ion $\text{Al}^+$, after probing its transition with a laser, can be faithfully transferred to the logical ion $\text{Be}^+$, where this information can be detected with almost 100\% efficiency \cite{blattwineland2003}. 
Each ion species provides a different functionality, the reference frequency or the cooling method, and the quantum entanglement between the states of both ions allows them to function as a quantum clock altogether.

%\'{\i}

Current quantum clocks use either a) one or two trapped ions, or b) ultra-cold atoms confined in electromagnetic fields in the form of optical lattices.

Each of these realizations has pros and cons. Ion clocks have a very high accuracy for they can confine the ion by cooling it down in a trap so that we get very close to the ideal of an isolated system from external disturbances. However, by using only one ion for the absorption signal, less stability is achieved as the ratio of the signal to external noise gets reduced. On the contrary, clocks of atoms in optical lattices can work with a large number of atoms achieving greater stability and a better signal-to-noise ratio.

\begin{figure}[t]
  \includegraphics[width=0.35\textwidth]{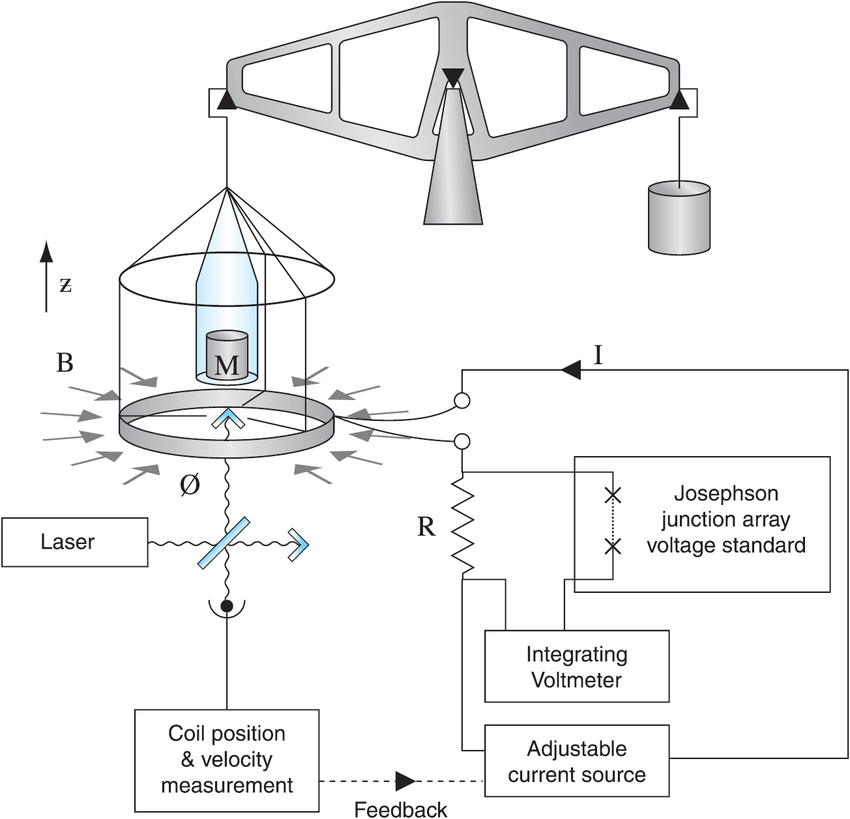}
  \caption{Kibble balance operating in weighing mode. Explanations in the text.  
(Credit Robinson-Schlamminger \cite{kibblebalance}).}
  \label{fig:KibbleModoPesada}
\end{figure}

Different teams working with both options are developing techniques to get better and better performances. The most recent record with trapped ions has an uncertainty of 
$9.4 \times 10^{-19}$s \cite{quantumclockrecord2019}.
Quantum clocks in optical lattices also achieve $10^{-18} \text{s}$ accuracy. There is still time to decide which of the two alternatives will be chosen to realize a new revision of the second, or whether both of them are complementary (see Fig.\ref{fig:Comparativa}).

The improvements in time measurement provided by quantum clocks also have important applications. The technological applications are similar to those mentioned above and it is of great interest to be able to send one of these quantum clocks in space missions and to improve navigation systems.
Another application is the high precision measurement  of the gravitational field for, according to Einstein's general relativity, there is a time dilation due to gravitational effects in addition to the speedup due to velocity. With a quantum clock you can distinguish gravitational fields only 30 cm high \cite{dilataciongravitacional}, and even less.
These measures will allow to better define the heights above sea level, since this is not measured in the same way in different parts of the world and is crucial to know the activity of the oceans. Similarly, these quantum devices can be applied to geodesy, hydrology and telescope network synchronization.

Basic research is one of the first fundamental applications of them. Comparing the operation of several quantum clocks over time we can discover if any of the fundamental constants of physics changes with time, which is essential to find new physics and to define the base units according to the new SI (see \ref{sec:nuevasdefiniciones}). Examples of fundamental constants that can be probed for their temporal dependence are the electromagnetic fine structure constant $\alpha$ \eqref{cmagnetica} and the ratio of the proton to the electron mass $\mu:=m_p/m_e$. In the past there has been controversy over possible temporal variations of $\alpha$ and $\mu$ detected by measures of atomic transitions in distant quasars compared to current measurements in the laboratory \cite{alphavariacion1,alphavariacion2}. 
It so happens that all atomic transitions functionally depend on $\alpha$ and also hyperfine transitions depend significantly on the $\mu$ ratio. It turns out that quantum clocks allow to improve the levels of variation of these fundamental constants.
With these experiments, the ratio of optical frequencies between ions of $\text{Al}^+$ and  $\text{Hg}^+$ 
can be measured  providing a bound to the time variation for $\alpha$ of $-1.6\pm 2.3 \times 10^{-17}$ per year, and with the yterbium ion $\text{Yb}^+$ a bound for  $\mu$ of $0.2\pm 1.1 \times 10^{-16}$ per year,
 which are better by a factor of ten than the astrophysical measurements \cite{alphavariacion3,alphavariacion4}.
 These negative results for the temporal variation of the fundamental constants serve as justification for the new SI unit system and its universality regardless of space and time, at least as long as the experiments continue to confirm such results.

%\'{\i}

\begin{figure}[t]
  \includegraphics[width=0.35\textwidth]{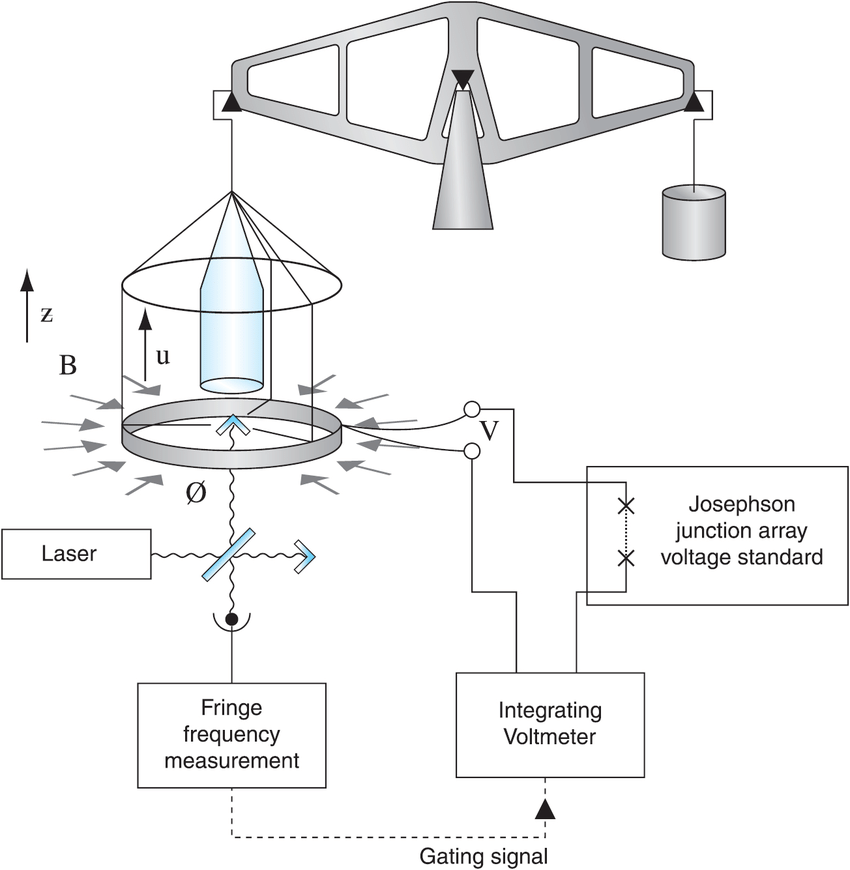}
  \caption{
Kibble balance operating in moving mode. Explanations in the text.  
(Credit: Robinson-Schlamminger \cite{kibblebalance}).}
  \label{fig:KibbleModoVelocidad}
\end{figure}

\subsubsection{ Kibble Balance}

The Kibble balance is the current experimental realization of the unit of mass through the quantum way in the new SI \cite{kibble1,kibble2,kibble3,kibblebalance}. In this way, we are fulfilling the new methodology of separating unit definitions from their practical materialization (see \ref{sec:nuevoSI}). Whereas the definition of the new kilo linked to the Planck constant has already been explained in  \ref{sec:nuevasdefiniciones}, we will now see how to realize it in the laboratory with current technology.

The realization of the `quantum kilo' consists of two distinct parts: a) the Kibble balance and b) the quantum determination of the electric power. The Kibble balance (see Fig.\ref{fig:KibbleBalance}) aims to establish the equivalence of a mechanical power into electrical power. This is then related to Planck's constant through metrology procedures of the first quantum revolution: integer quantum Hall and Josephson effects.

Let us start with the Kibble balance. We present a simplified discussion,  but sufficient to understand its foundations. It looks like an ordinary balance in which it also has two arms, but
while in the ordinary balance two masses are compared, one standard and another unknown, in Kibble's we compare gravitational mechanical forces with electromagnetic forces.
Its functioning consists of two operating modes: i/ Weighing Mode and ii/ Moving Mode.

\begin{figure}
  \includegraphics[width=0.45\textwidth]{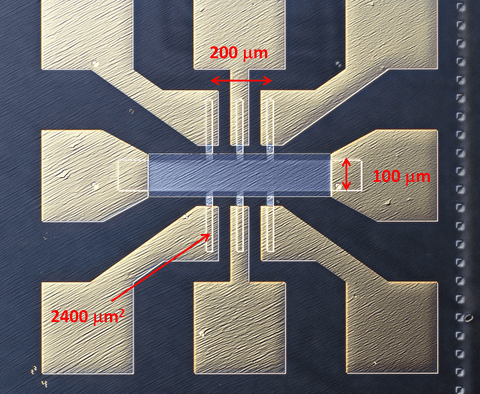}
  \caption{
  An example of an integer quantum Hall effect device used by NIST to measure resistances.
  This Hall bar uses graphene components that are outlined by white lines.
   The source and drain of electrons are at the left and right ends of the bar. The electrical contacts above and below the bar are not shown.   (Credit NIST).}
  \label{fig:Hall_bar}
\end{figure}

\noindent {\bf Weighing Mode}: 
A test mass $m$ is available in one of the arms, which could be for example the IPK standard. 
On the other plate, a circuit of electric coils is mounted where a current $I$ is passed through
(see Fig.\ref{fig:KibbleModoPesada}).
The circuit is suspended in a very strong magnetic field created by magnets with a stationary and permanent field 
$B$. The length of the circuit is $L$. Then, the current induces an electromagnetic field that interacts with the constant magnetic field of the magnet. The resulting vertical electromagnetic force is equal to the weight of the test mass,
\begin{equation}\label{}
B L I = m g. 
\end{equation}
During this operating mode, the intensity of direct electric current is measured very accurately by appropriate instruments (integer quantum Hall effect), and it is proportional to the vertical force. The current is adjusted so that the resulting force equals the weight of the test mass.

\qed

\noindent {\bf Moving Mode}: 
This is a calibration mode that is necessary as the quantity
$B L$ is very difficult to measure accurately. Were it not for this, the weighing mode would suffice. An electric motor is used to move the wire circuit vertically through the external magnetic field at a constant speed $v$
(see Fig.\ref{fig:KibbleModoVelocidad}). 
This movement induces a voltage $V$ in the circuit whose origin is also a Lorentz force and is given by,
\begin{equation}\label{}
B L v= V. 
\end{equation}
During this operating mode, the voltage is measured very accurately by appropriate instruments (Josephson effect), and therefore the magnetic field that is proportional. Laser sensors are also used to monitor the vertical movement of the electrical circuit by interferometry. With this, variations of the order of the laser's semi-wavelength used can be detected. In all, it is ensured that the vertical movement happens at constant speed and the constant magnetic field can be measured.

%\'{\i}
\qed

The result of comparing the weighing mode with the moving mode, eliminating the quantity $BL$, is the equivalence between mechanical and electrical powers:
\begin{equation}\label{potencias}
m g v = I V.
\end{equation}
Although it is usual to call the Kibble balance as a power or watt balance, however, note that the Kibble balance does not measure real, but virtual, powers. This point is of crucial importance in metrology:
were the mechanical power really measured, then the device would be subject to uncontrollable friction losses; otherwise, if the electrical power were measured directly, then it would be subject to heat dissipation. We see that the moving mode is essential and provides adequate calibration.

\begin{figure}
  \includegraphics[width=0.45\textwidth]{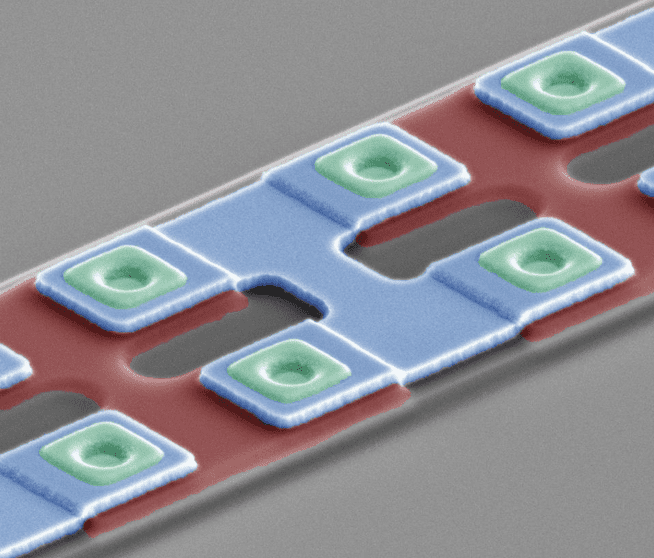}
  \caption{Close-up view of a modern Josephson junction used in NIST. Josephson's junctions are built in green circular wells where the two superconducting layers overlap. See explanation in text. (Credit: M. Malnou/NIST/JILA).}
  \label{fig:Josephson_union}
\end{figure}

It turns out that experimentally it is more accurate to measure resistance than current intensities. Using Ohm's law, we can obtain the mass on the Kibble balance based on resistance and voltage measurements:
\begin{equation}\label{masaKibble}
m  = \frac{V_{\text R} V}{ g v R},
\end{equation}
where $V_{\text R}$ and $V$ are the two necessary voltage measurements.

In the second part of the experimental realization of the `quantum kilo' we need to relate the electrical power in \eqref{potencias} to the Planck  constant $h$. This is done through the measurement of the electric intensity $I$ in the weighing mode and the voltage $V$ in the moving mode of the Kibble balance. For this, the following quantum effects are used.

\begin{figure}[t]
  \includegraphics[width=0.45\textwidth]{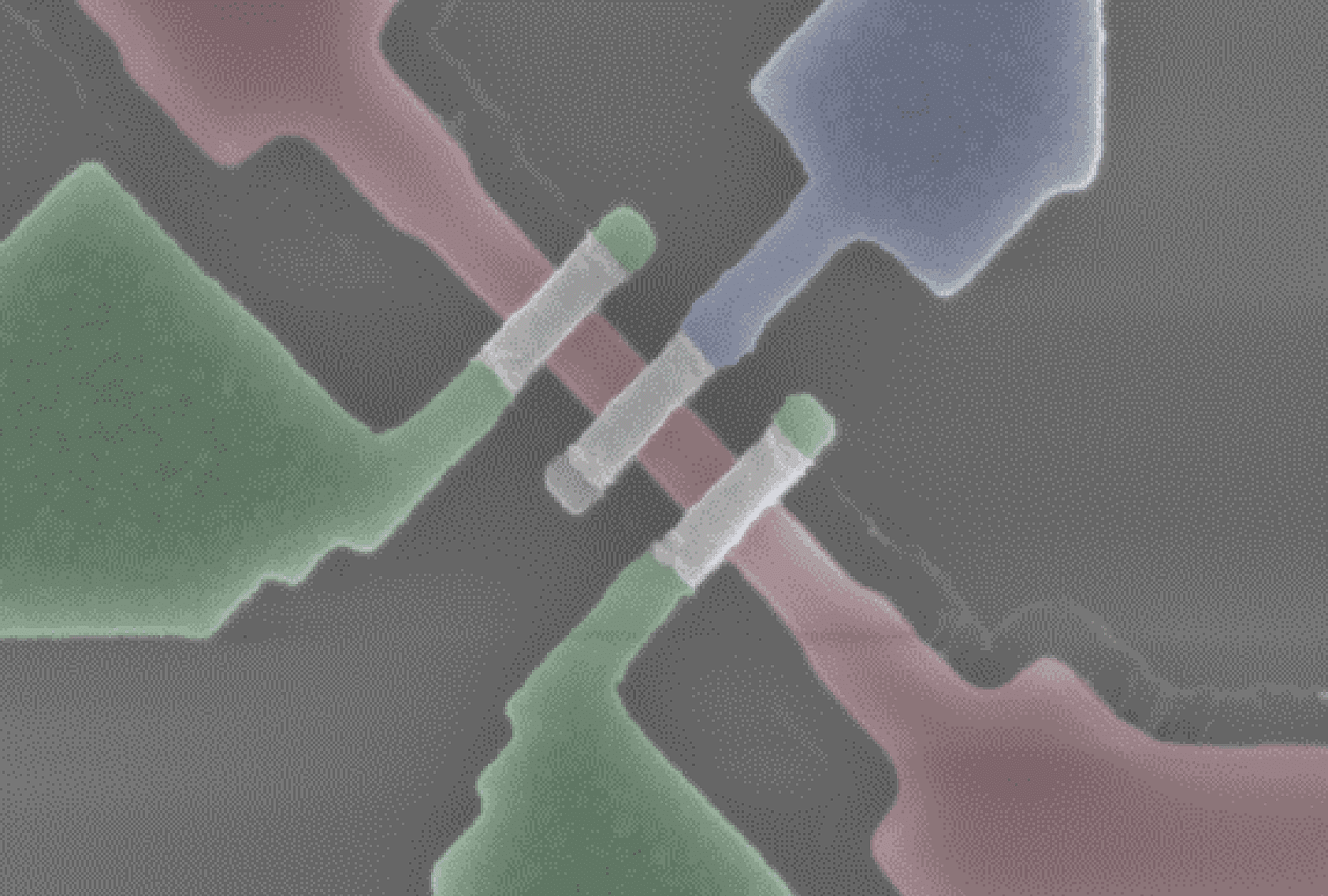}
  \caption{Example of a single-electron transport device (SET) used in the NIST for the definition of the ampere. The interaction between the blue and green electrodes (gates) controls the movement of individual electrons in and out of an ``island'' in the center.
Explanations in the text. The colors are illustrative only.  (Credit NIST).}
  \label{fig:SET}
\end{figure}

\noindent {\bf Integer Quantum Hall Effect}: 
a two-dimensional sample contains electrons constrained to move in that plane subject to a longitudinally aligned coplanar electric field and a very intense constant magnetic field $B$ applied perpendicularly to the sample (see Fig.\ref{fig:Hall_bar}). In addition, the electronic sample is cooled down to temperatures nearby the absolute zero. Then, the system departs from the classical Ohm's law and enters into a quantum regime. As in the classic case, a transverse electric current appears that induces a transverse voltage bias called Hall voltage $V_{\text H}$. 
The electron system enters a new quantum behavior characterized by the appearance of jumps and plateaus in the relationship between the transverse current and the magnetic field \cite{Hallcuantico}. In particular, the Hall resistance $R_{\text H}$ associated to that Hall potential is quantized
\begin{equation}\label{HallEntero}
R_{\text H} = \frac{1}{n^\prime} \frac{h}{e^2},
\end{equation}
where  $n^\prime$ is an integer, giving rise to those plateaus appearing in the curves of the Hall resistivity. 
The von Klitzing constant $R_{\text K}$ is defined as
\begin{equation}\label{HallEnteroconstante}
R_{\text K} :=  \frac{h}{e^2},
\end{equation}
which has dimensions of resistance and is the elementary resistance.
The integer quantum Hall effect allows resistance to be measured with an uncertainty of a few parts in $10^{-11}$ ohms. For this reason it is used to perform the resistance standard \cite{press_kit,cem,SI:BIPM}.

\qed

\begin{figure}[t]
  \includegraphics[width=0.35\textwidth]{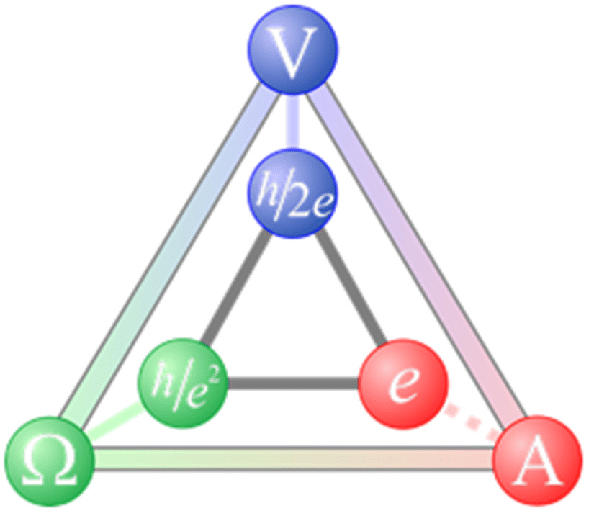}
  \caption{Quantum metrological triangle: schematic relationship between the universal constants of the Hall effect \eqref{HallEnteroconstante}, Josephson effect \eqref{constanteJosephson} and the charge of the electron in a SET device. The triangle establishes a constraint between them and the Planck  constant $h$ and the  elementary charge $e$.  (Credit: Piquemal et al. \cite{triangulocuantico}).}
  \label{fig:TMC}
\end{figure}

\noindent {\bf Josephson Effect}: 
when a superconducting wire is interrupted at a point with a contact made of insulating material that joins two superconducting portions, the superconducting current can be maintained due to a tunnel effect of the superconducting Cooper pairs. This is known as a Josephson junction (see Fig.\ref{fig:Josephson_union}). Under these circumstances, if a radiofrequency radiation $\nu$ is applied, a potential $V$ is induced through the junction that is proportional to the frequency and is quantized \cite{josephson1,josephson2}: 
\begin{equation}\label{Josephson}
V_{\text J} = n \frac{h}{2e} \nu,
\end{equation}
where $n$ is an integer, $2e$ is the Cooper pair charge and the  Josephson constant is defined as 
\begin{equation}\label{constanteJosephson}
K_{\text J} := \frac{2e}{h}.
\end{equation}

This is the so-called Josephson DC effect and Josephson junctions can be made with metallic points or with constrictions, in addition to insulators. It allows measuring voltages with an uncertainty of $10^{-9,-10}$ volts, that is, of the order of nano volts or less. For this reason it is used for the realization of the voltage standard \cite{press_kit,cem,SI:BIPM}.

\qed

Now we can relate the test mass  \eqref{masaKibble} that is used in the weighing mode with the Planck constant that appears when measuring the resistance, also in the weighing mode, and the voltage in the moving mode. Using the integer quantum Hall effect \eqref{HallEntero} to measure the resistance and the Josepshon effect \eqref{Josephson} to measure the voltages $V_{\text R}$ and $V$, we obtain the desired relationship
\begin{equation}\label{masaKibble-h}
m  = \left( n^\prime n_1 n_2\frac{\nu_1 \nu_2}{4 gv} \right) h,
\end{equation}
where the integer numbers that appear come from the concrete measurements of the corresponding quantum effects \eqref{HallEntero}, \eqref{Josephson}. To measure $g$ a high precision absolute gravimeter is used and $v$  with interferometric methods.
With all these high precision measures, the expression \eqref{masaKibble-h} has a dual utility: on the one hand, given a standard mass $m$ like that of the old IPK, we can determine $h$ with great precision. On the other hand, since $h$ can be measured with this method with great precision, we can set this value of $h$ as exact and define the unit of mass based on $h$: this way we can carry out the `quantum kilo' route and detach the mass unit from the kilo IPK artifact.

%\'{\i}

\subsubsection{Quantum Metrology Triangle}
\label{Triangulo}

The new definition of the ampere linked to the value of the electron elementary charge $e$ stands out for its clarity and simplicity compared to the old definition based on the Ampere's law and an unrealizable construction using infinite and null-thick conductor wires \cite{SI:BIPMrealization}. However, it also entails the need to materialize it in some way, and it is not easy for the number of electrons in an ordinary system is immensely large. The BIPM has approved three methods for the practical realization of the ampere
\cite{cem,SI:BIPM,SI:BIPMrealization}. 
One of them uses the direct definition of ampere $A=C/s$ and a single-electron transport device (SET), which has to be cooled to temperatures close to absolute zero (see Fig.\ref{fig:SET}). Through a SET, electrons pass from a source to a drain. A SET consists of a region made of silicon, called an island, between two gates that serve to electrically manipulate the current. The island temporarily stores the electrons coming from the source using another voltage gate. By controlling the voltages at the two gates, you can get a single electron to remain on the island before moving to the drain. Repeating this process many times and very quickly, it is possible to establish a current from which its electrons can be counted.

The electrical sector is the most quantum of them all within the SI system of units. 
Now that the ampere has been redefined in the new SI by linking it to a fixed value of the electron charge, it is possible to relate the three magnitudes that appear in Ohm's law,  $V=IR$, in terms of only two universal constants, $ h $ and $ e $. This is visualized by the so-called quantum metrological triangle
(see Fig.\ref{fig:TMC}). 
This triangle represents an experimental constraint that the voltage, resistance and intensity standards must fulfil, so that the three are not independent. Thus, if we measure Josephson's constant \eqref{constanteJosephson} on the one hand and von Klitzing's on the other  \eqref{HallEnteroconstante}, they allow us to obtain values for the unit of charge $e$ of the electron that must be compatible, within experimental uncertainties, with the value of $e$ obtained with a single-electron transport. And the same goes for any pair of magnitudes that we take in the triangle. Therefore, the quantum metrological triangle allows us to test experimentally, as better accuracy and precision are achieved, whether the constants $h$ and $e$ are really constants as assumed in the new SI. The uncertainties in these constants must be compatible using these three experimental realizations.
If at some point these uncertainties do not overlap, then this is an indication of new physics as it would affect the very foundations of quantum mechanics or quantum electrodynamics as explained in section \ref{sec:constantesfundamentales}. Again, this is an example of how the Metrology not only serves to maintain unit standards, but to open up new paths to new fundamental laws of nature.

%%%%%%%%%%%%%%%%%%%%%%%%%%%%%%%%%%%%%%
%%%%%%%%%%%%%%%%%%%%%%%%%%%%%%%%%%%%%%

%%%%%%%%%%%%%%%%%%%%%%%%%%%%%%%%%%%%%%
%%%%%%%%%%%%%%%%%%%%%%%%%%%%%%%%%%%%%%
\section{A `Gravitational Anomaly' in the  SI}
\label{sec:anomaliaG}
%%%%%%%%%%%%%%%%%%%%%%%%%%%%%%%%%%%%%%
%%%%%%%%%%%%%%%%%%%%%%%%%%%%%%%%%%%%%%

Despite the fact that the new system of SI units amounts to a complete linking
from the base units to fundamental constants of nature, it is still
striking the absence of one of the oldest universal constants of Physics:
Newton's universal gravitation constant $G$ (see Fig.\ref{fig:Ganomalia}), colloquially called big $G$, as opposed to the small $g$ representing local acceleration
of gravity at a point on Earth.

The main reason to exclude $G$ from the new SI system of units is the lack of precision
enough to define a unit of mass. As explained in \ref{sec:nuevasdefiniciones}, this is the origin of the `quantum way' for the definition of the kilo when you want to detach it from a material artifact such as the cylinder of the kilo IPK.
This fact is related to the so-called `Newton's Big G Problem'
\cite{nistbigG,quinn2000,quinn2013,sanfernando2017}.
This problem is the lack of compatibility of the $G$ measures in the last thirty years. Various metrology laboratories around the world have tried to measure $G$ with experimental devices designed to reduce uncertainty in the value of $G$.
The result is surprising: $G$ values do not converge to a consistent single value and their uncertainties do not overlap in a compatible way. This can be seen in Fig.\ref{fig:G}, which shows the results of multiple experiments and a vertical zone where a $G$ commitment value is chosen. The situation has become desperate and the NSF (National Science Foundation) has launched a global initiative trying to clarify the problem \cite{nsfG}.

A fundamental question then arises: what is the origin of Newton's big $G$ problem?
The most natural solution is that it is due to possible systematic errors in the experiments. Favoring
this interpretation is the fact that, despite the increasing sophistication trying to measure $G$ more accurately,
all the experimental methods used are variants of the famous Cavendish balance
\cite{amoreno,springer}.
However, in Fig.\ref{fig:G} there appears a value \cite{AFG} whose experimental method is completely different from the methods based on the Cavendish balance. It is a quantum method of measuring $G$. Their device uses a technique based on atomic interferometry with ultra-cold atoms. With it, the quantum nature of atoms is used at temperatures close to absolute zero, to obtain an accurate measure of the acceleration of gravity.

The Cold Atoms in Gravity (CAG) method consists of 2 steps:
Step 1: measure of the constant small $g$: value of the local terrestrial gravity.
Step 2: measure of the constant big $G$.

The technique consists of turning cold atoms vertically, up and down, repeatedly. This serves to probe Earth's gravity with a cloud of rubidium atoms Rb in free fall. With this procedure it is possible to measure the force of gravity between an atom of Rb and a reference mass of 516 kg. The result is a measure of $G$ with a relative uncertainty of $0.015\%$. Remarkably, it is the first time that a quantum method is admitted to be part of the set of values used to determine $G$.

This gravitational anomaly is still a reflection of the big problem that affects
modern physics: the lack of compatibility between the two great theories of our
time, quantum mechanics and general relativity.
An important observation (see Conclusions) is that the CAG method is an indirect measurement method for Newton's big $G$: it is done by first measuring small $g$. This contrasts with the classic methods based on the Cavendish balance where $G$ can be measured directly.
A direct quantum measure of $G$ would be a first experimental indication of quantum effects on gravity and a first step for a quantum gravity theory. As seen in Fig.\ref{fig:G}, the CAG value is still outside the shaded vertical zone of the most recent recommended value for $G$. This may be indicative that the CAG method does not suffer from possible systematic errors as in the classical methods of measurement of $G$ and could be the beginning for the solution of Newton's big $G$ problem. The way to confirm this hypothesis is to encourage more CAG experiments in more independent laboratories and use quantum metrology techniques to reduce their uncertainties.
If the result of all these new classical and quantum experiments were that there are no systematic errors, then the conclusion would be even more exciting as it would be again a door opened by metrology to new physics.

\begin{figure}
  \includegraphics[width=0.35\textwidth]{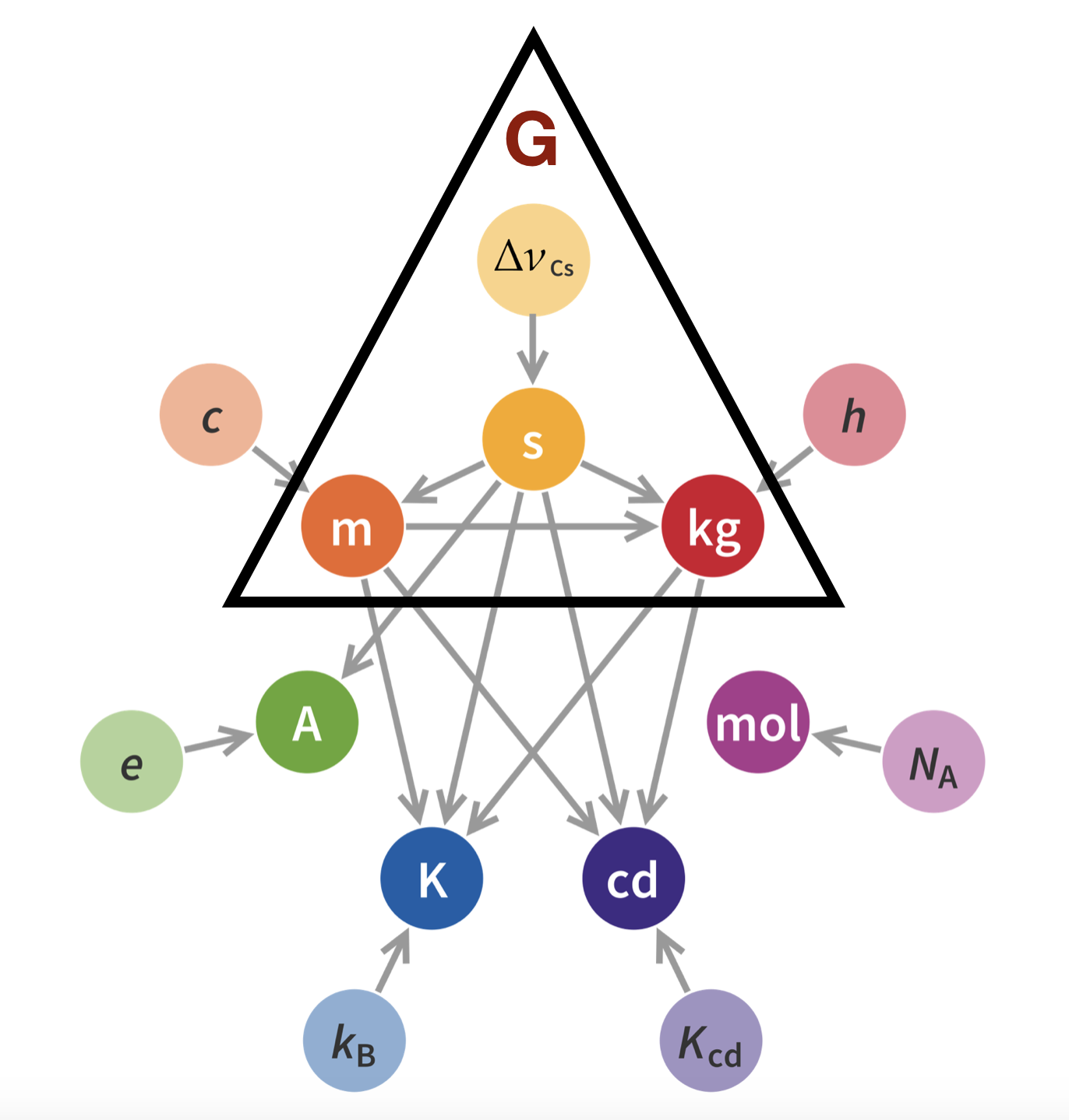}
  \caption{The scheme with the dependencies of the natural constants and the base units of the new SI in Fig.\ref{fig:dependencias} presents a notable absence: the constant $G$ of Newton's universal gravitation. The triangle encompasses the excluded $ G $ with the units on which it depends \eqref{Gunidades} that are the same as Planck's constant $ h $ \eqref{hunidades}, but in another proportion. Credit: Emilio Pisanty/Wikipedia (adapted).}
  \label{fig:Ganomalia}
\end{figure}

Direct methods to measure $G$ are not known. In fact, there are few physics equations where $G$ and $h$ appear together. One of them can help us to see the difficulties of getting a direct quantum method to measure 
$G$. This is the equation of the Chandrasekhar limit for the radius of a white dwarf star. If we thought naively of producing a gravitational condensate of nucleons (fermions) that were the result of compensating the gravitational pressure of $N$ nucleons of mass $M$ with the degeneration pressure due to the Pauli exclusion principle, using non-relativistic quantum mechanics and Newtonian gravitation to simplify, 
we get \cite{enanablanca} a value of the equilibrium radius given by
\begin{equation}\label{Chandrashekar}
R_0 = \left(\frac{9\pi}{4} \right)^{2/3} \frac{\hbar q^{5/3}}{G m M^2 N^{1/3}},
\end{equation}
where $N$ is the number of nucleons and $q$ the number of electrons per nucleon with mass $m$. It has been assumed that the density of the spherical condensate is uniform. To make a simple estimate, consider a system of only neutrons ($q=1$, $m=M_n$ mass of the neutron), and substituting the known experimental values, we obtain
\begin{equation}\label{Chandrashekar2}
R_0 =  5.51 \times 10^{24}N^{-1/3} \text{m}.
\end{equation}
If we want to have a fermion system condensed in a sphere with a radius of the order of one meter to be manageable  in a terrestrial laboratory, we can estimate the number of necessary neutrons obtained with
\eqref{Chandrashekar2} as something of the order of $N\sim 10^{74}$, an intractable amount if we take into account that the number of atoms in the observable universe is of the order of $10^{80}$. This difficulty is a reflection of the disparity of scales where gravity acts against quantum effects.

\begin{figure}
 \includegraphics[width=0.55\textwidth]{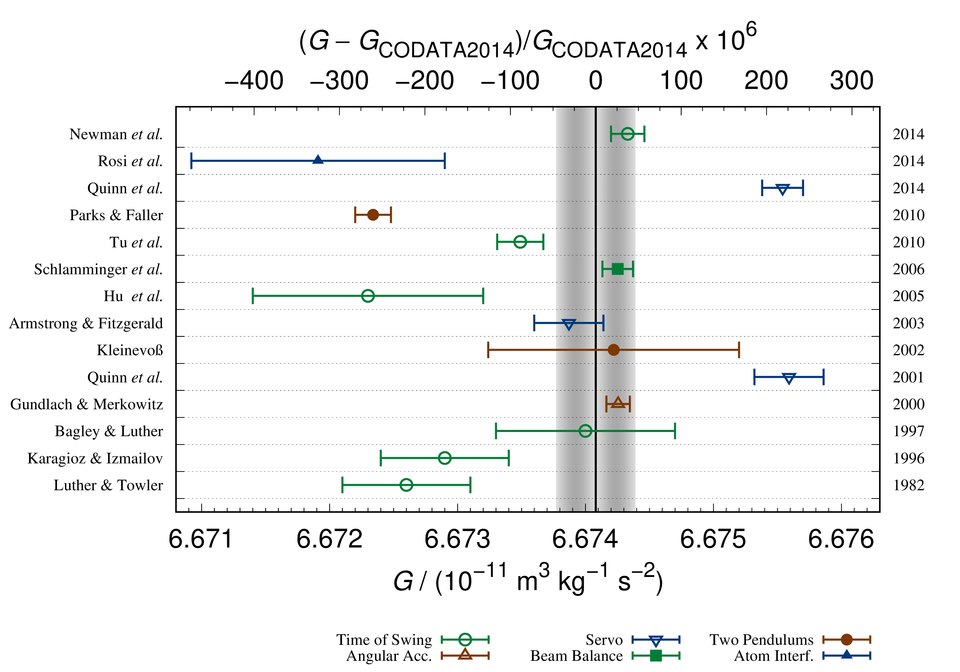}
  \caption{Comparative diagram of the measurements of the constant $G$ over time using classical methods based on variants of the torsion balance, except that of Rossi et al. \cite{AFG} using the Cold Atoms in Gravity (CAG) method.  (Credit: NIST \cite{nistbigG}).}
  \label{fig:G}
\end{figure}

%%%%%%%%%%%%%%%%%%%%%%%%%%%%%%%%%%%%%%
%%%%%%%%%%%%%%%%%%%%%%%%%%%%%%%%%%%%%%
\section{Conclusions}
\label{sec:conclusions}
%%%%%%%%%%%%%%%%%%%%%%%%%%%%%%%%%%%%%%
%%%%%%%%%%%%%%%%%%%%%%%%%%%%%%%%%%%%%%

The adoption of the new SI system of units brings several concrete advantages over the
previous system: it solves the ampere problem and the electrical units that had been
left out of the SI, it eliminates the dependence of the kilo with the artifact of the kilo IPK, conceptually it is more satisfactory to define them in terms of natural constants, future technological improvements will no longer affect the definitions etc.

The new SI of units has no direct impact on our daily life, but it does in research laboratories and in national metrology centers where they need measures of great accuracy and precision to conduct these investigations and to guard and disseminate the primary unit standards.
As usual, in the long run these new discoveries result in applications
that do modify our daily life for the better.

Therefore, it is a great conceptual challenge to explain and convey what it entails and is behind the new system of units. In section  \ref{sec:nuevasdefiniciones}  a common view of all the new definitions has been presented using as an unifying principle the discrete nature of energy, matter and information in the fundamental laws of Physics and Chemistry to which each base unit is linked.  Interestingly, the only thing that remains non-discrete is spacetime.
An advantage of the new SI is that it facilitates the explanation of the new definitions
as it does not need to explain the measuring devices necessary to perform these units.

Metrology has a double mission: 1) To maintain the unit standards and their definitions compatible with the current laws of physics. 2) To measure with increasing accuracy and precision in order to open new doors to discover new laws of physics.

As for its more traditional mission 1), the adoption of the new SI allows to get rid of a material device to define the kilo, which was a long sought-after goal. With this, it is possible to materialize primary standards of the base units in different national metrology centers for the first time. In particular, quantum metrology will drastically change the dissemination and traceability of units  by ensuring that they can be materialized autonomously without the need for a single stored standard.

It is interesting to note that in the course of the construction of the new SI the three most famous balances of physics have appeared:
Cavendish \cite{amoreno,springer}, E{\"o}tv{\"o}s \cite{equivalence} and Kibble \cite{kibblebalance}. 
As well as the seminal papers of Einstein in his  {\it annus mirabilis}  1905
\cite{einstein1905d, einstein1905c,einstein1905a,einstein1905b}.

As for the second mission, we have seen how the new SI uses five universal constants of nature.
Of these, three have a special status, $c,h$ and $e$, as they are associated with symmetry principles of the universe, such as the principle of relativity, unitarity and gauge symmetry. The other two are the Botzmann 
 constant $k$ and the Avogadro constant $N_{\text A}$, neither of which has an associated  symmetry.

Now that the physical units are defined by the fundamental physics of the universe,
and not by a human construct using artifacts, then as for the fundamental constants of the universe:
Are they a machination of something? Why do they take those values? And until when?
We have thus reached the most fundamental questions of physics.
That is why metrology really goes beyond maintaining measurement standards.

%\'{\i}

\begin{acknowledgments}
These notes are the result of several lectures given during 2017, 2018 and 2019.
I would like to thank the organizers Jos\'e Manuel Bernab\'e and Jos\'e \'Angel Robles from Centro Espa\~{n}ol
de Metrolog\'{\i}a (CEM) for their kind invitation to the $6^{o}$ Congreso Espa\~{n}ol de Metrolog\'{\i}a (2017), to  $8^{o}$ Seminario Intercongresos de Metrolog\'{\i}a (2018) and  the Congreso del 30 Aniversario del CEM (2019); to  Alberto Galindo and Arturo Romero from Real Academia de Ciencias Exactas, F\'{\i}sicas y Naturales de Espa\~{n}a for their kind invitation to Ciclo Ciencia para Todos (2018) and Jornada sobre ``La revisi\'on del Sistema Internacional de Unidades, (SI). Un gran paso para la ciencia" (2019); 
to Federico Finkel and Piergiulio Tempesta for their kind invitation to the homage of Artemio Gonz\'alez L\'opez on the occasion of his 60th anniversary.
M.A.M.-D. acknowledges financial support from the Spanish MINECO, FIS 2017-91460-EXP, PGC2018-099169-B-I00 FIS-2018 and the CAM research consortium QUITEMAD+, Grant S2018-TCS-4243. The research of M.A.M.-D. has been supported in part by the U.S. Army Research Office through Grant No. W911N F- 14-1-0103.

%\'{\i}

\end{acknowledgments}

%%%%%%%%%%%%%%%%%%%%%%%%%%%%%%%%%%%%%%%%%%%%%%%%%%%%%%%%%%%%%%%%%%%%%%%%%%%%%%
%\begin{references}

\newpage

\title{El Nuevo SI y las Constantes Fundamentales de la Naturaleza}
\author{Miguel A. Martin-Delgado}
\affiliation{Departamento de F\'{\i}sica Te\'orica, Universidad Complutense, 28040 Madrid, Spain.\\
CCS-Center for Computational Simulation, Campus de Montegancedo UPM, 28660 Boadilla del Monte, Madrid, Spain.}

%\begin{abstract} 
%La puesta en marcha en 2019 del nuevo sistema internacional de unidades es una oportunidad
%para resaltar el papel fundamental que las leyes fundamentales de la F\'{\i}sica y la Qu\'{\i}mica juegan
%en nuestra vida y en todos los procesos de la investigaci\'on fundamental, la industria y el comercio.
%El principal objetivo de estas notas es presentar el nuevo SI de forma accesible para una audiencia amplia. Tras repasar las constantes fundamentales de la naturaleza y sus leyes universales, se presentan las nuevas definiciones de las unidades SI utilizando como principio unificador la naturaleza discreta de la energ\'{\i}a, la materia y la informaci\'on en esas leyes universales.
%El nuevo sistema SI tiene vocaci\'on de futuro: aunque las realizaciones experimentales cambien por mejoras tecnol\'gicas, las definiciones permanecer\'an inalteradas. La Metrolog\'{\i}a cu\'antica est\'a llamada a ser uno de las fuerzas motrices para conseguir
%nuevas tecnolog\'{\i}as cu\'anticas de segunda generaci\'on.
%\end{abstract}

\maketitle

%\tableofcontents

%%%%%%%%%%%%%%%%%%%%%%%%%%%%%%%%%%%%%%
%%%%%%%%%%%%%%%%%%%%%%%%%%%%%%%%%%%%%%
\section{Introducci\'on}
\label{sec:intro}
%%%%%%%%%%%%%%%%%%%%%%%%%%%%%%%%%%%%%%
%%%%%%%%%%%%%%%%%%%%%%%%%%%%%%%%%%%%%%

El d\'{\i}a 20 de mayo de 2019 coincidiendo con el D\'{\i}a Mundial de la Metrolog\'{\i}a
ha entrado en vigor el nuevo sistema internacional (SI) de unidades que fue aprobado
en la asamblea de la 26 Conferencia General de Pesos y Medidas (CGPM) reunida
en Versalles del 13 al 16 de noviembre de 2018 \cite{press_kit}. 
Esto es un logro hist\'orico. Supone la culminaci\'on de muchos esfuerzos durante muchos
a\~{n}os de trabajo conjunto entre los centros nacionales de metrolog\'{\i}a de los estados miembros
y el BIPM (Bureau International des Poids et Mesures), dando un magn\'{\i}fico ejemplo de colaboraci\'on internacional.

La CGPM aprob\'o revisar en 2018 cuatro de las unidades b\'asicas (el kilogramo, el amperio, el kelvin y el  mol).
De esta forma, la totalidad de las unidades b\'asicas de medida quedan vinculadas a constantes f\'{\i}sicas en vez de a referencias arbitrarias. Esto supone la jubilaci\'on del famoso patr\'on de masa, el kilo IPK \cite{press_kit,cem,SI:BIPM}, que era el \'unico patr\'on ligado a un artefacto material que quedaba. Ahora todas las unidades b\'asicas quedan asociadas a reglas de la naturaleza para crear nuestras reglas de medida \cite{press_kit}. Lo que subyace a todas estas redefiniciones el la posibilidad de realizar medidas a escalas at\'omicas y cu\'anticas para realizar las unidades a escala macrosc\'opica.

Aunque la eliminaci\'on de artefactos para definir las unidades b\'asicas sirve para garantizar mejor su estabilidad y universalidad, sin embargo el nuevo sistema trae consigo el enorme desaf\'{\i}o de explicar su funcionamiento a la sociedad en general, y a los centros educativos. Los artefactos son tangibles
(ver Fig.\ref{fig:IPK}), mientras que el conocimiento de las leyes fundamentales de la naturaleza f\'{\i}sica y qu\'{\i}mica dista mucho de ser de amplio dominio p\'ublico. En este sentido, en  \ref{sec:nuevasdefiniciones} se presenta un tratamiento unificado de todas las definiciones de las unidades SI usando como marco com\'un la discretizaci\'on de la energ\'{\i}a, la materia y la informaci\'on que es el ingrediente fundamental de las leyes de la F\'{\i}sica y la Qu\'{\i}mica
a las que aparecen vinculadas las nuevas unidades SI.
Estas notas surgen de varias conferencias de divulgaci\'on con el fin explicar la relaci\'on de las constantes fundamentales  con las nuevas definiciones de la unidades. 

Al introducir las constantes de la naturaleza en la secci\'on \ref{sec:constantesfundamentales}
se ha resaltado un rasgo diferenciador entre las cinco constantes universales asociadas a leyes
fundamentales de la naturaleza. Mientras que $h,c$ y $e$ est\'an asociadas a principios de simetr\'{\i}a, no es el caso de las constantes de Botzmann $k$ y de Avogadro $N_{\text A}$.

Se presentan todas las nuevas definiciones de las unidades b\'asicas, sin embargo no se hace una presentaci\'on exhaustiva de todas sus realizaciones experimentales por ser demasiado t\'ecnico para la intenci\'on de estas notas. Existe documentaci\'on m\'as detallada \cite{cem,SI:BIPM,SI:BIPMrealization}. Excepci\'on aparte merece el caso del nuevo `kilo cu\'antico', que por ser tan novedoso y la masa algo de uso diario tan habitual, se ha hecho una descripci\'on sencilla del funcionamiento de la balanza de Kibble con la que se lleva a la pr\'actica su materializaci\'on. 

El resto del art\'{\i}culo se organiza como sigue: en la secci\'on \ref{sec:nuevoSI} se explica la nueva metodolog\'{\i}a de separar las definiciones de las unidades de sus realizaciones experimentales; en la secci\'on \ref{sec:constantesfundamentales} se describen las constantes fundamentales de la naturaleza que aparecen en la nueva definici\'on de las unidades SI como preparaci\'on a la definici\'on expl\'{\i}cita en la secci\'on \ref{sec:nuevasdefiniciones} de las siete unidades b\'asicas, as\'{\i} como la descripci\'on en \ref{sec:MetrologiaSI} del papel de la metrolog\'{\i}a cu\'antica en el nuevo SI a trav\'es de tres ejemplos: los relojes cu\'anticos, la balanza de Kibble y el `kilo cu\'antico', y el tri\'angulo metrol\'ogico cu\'antico. En la secci\'on \ref{sec:anomaliaG} se reflexiona sobre la ausencia de la constante de la gravitaci\'on universal $G$ en el nuevo sistema de unidades y sus implicaciones.
La secci\'on \ref{sec:conclusions} se dedica a conclusiones.

%\'{\i}

\begin{figure}[t]
  \includegraphics[width=0.3\textwidth]{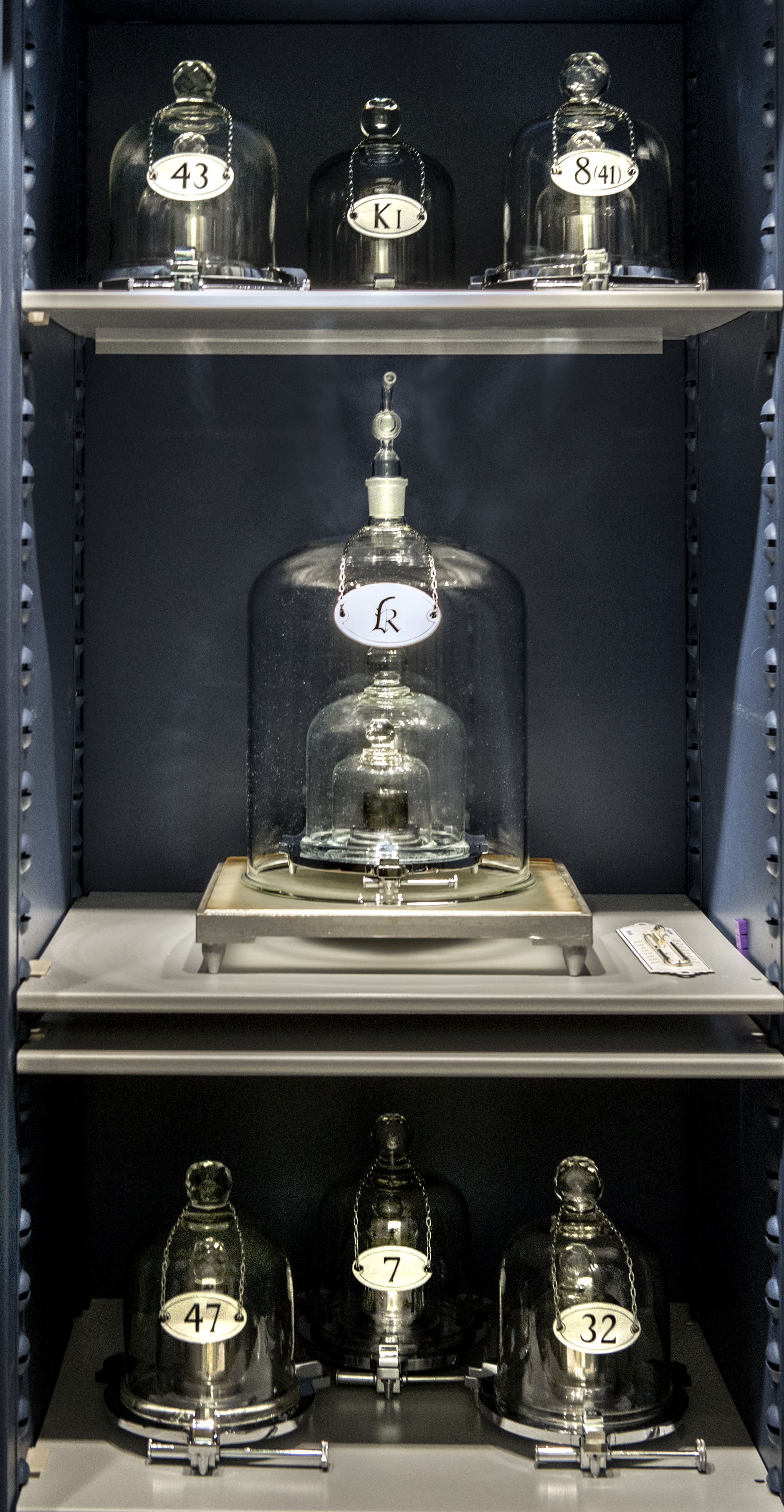}
  \caption{El prototipo internacional del kilogramo IPK, guardado en el BIPM cerca de Paris, y sus seis copias oficiales, {\it t\'emoins}. (Cr\'edito: BIPM).}
  \label{fig:IPK}
\end{figure}

\begin{table}
\begin{center}
    \begin{tabular}{ | c | c |}
    \hline \hline
    Constante & Valor  \\ \hline \hline
    $h$ & $6.62607015 \times 10^{-34} \text{J s}$  \\ \hline
    $e$ & $1.602176634 \times 10^{-19} \text{C}$  \\  \hline
    $k$ & $1.380649  \times 10^{-23} \text{J} \text{K}^{-1}$  \\  \hline
    $N_{\text{A}}$ &  $6.02214076  \times 10^{23} \text{mol}^{-1}$     \\ 
    \hline\hline
    \end{tabular} 
    \label{CODATA2018}
\end{center}
\caption{Los valores CODATA 2018  \cite{CODATA2018,newell2018} para las constantes universales cuyo valor ha sido fijado 
para definir el kilo, el amperio, el kelvin y el mol en el nuevo SI de unidades puesto en vigor por
el BIPM desde el 20 de mayo de 2019.} 
\end{table}

%%%%%%%%%%%%%%%%%%%%%%%%%%%%%%%%%%%%%%
%%%%%%%%%%%%%%%%%%%%%%%%%%%%%%%%%%%%%%
\section{ El Nuevo SI de Unidades}
\label{sec:nuevoSI}
%%%%%%%%%%%%%%%%%%%%%%%%%%%%%%%%%%%%%%
%%%%%%%%%%%%%%%%%%%%%%%%%%%%%%%%%%%%%%

El nuevo Sistema Internacional (SI) de unidades que ha entrado en vigor el 20
de mayo de 2019 representa una gran revoluci\'on a todos los niveles pues 
por primera vez todas las unidades se vinculan a constantes de la naturaleza,
muchas de ellas universales, y constituye un sue\~{n}o de la F\'{\i}sica y la Qu\'{\i}mica hecho realidad.
\begin{figure}[h]
  \includegraphics[width=0.32\textwidth]{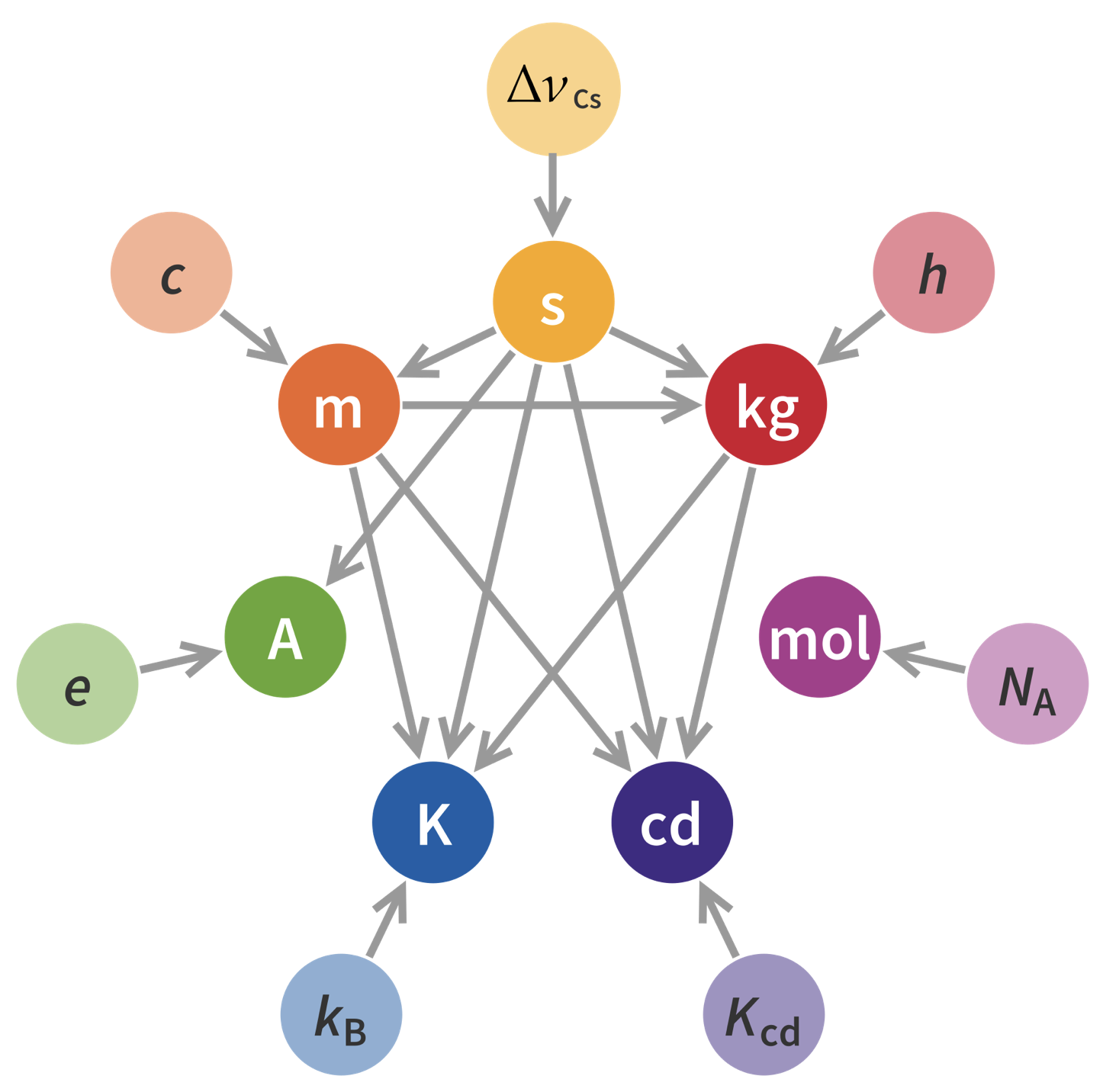}
  \caption{Relaci\'on esquem\'atica entre las unidades b\'asicas del nuevo SI y sus constantes naturales asociadas.
  En la parte central aparecen las unidades y sus dependencias entre s\'{\i}: el segundo influye en la definici\'on de cinco unidades, mientras que el mol aparece desacoplado. En la parte exterior aparecen los s\'{\i}mbolos de las constantes que sirven para su definici\'on. Ver subsecci\'on \ref{sec:nuevasdefiniciones}. (Cr\'edito: Emilio Pisanty/Wikipedia).}
  \label{fig:dependencias}
\end{figure}

El fundamento del nuevo SI se basa en las siguientes premisas \cite{cem,SI:BIPM}:
\begin{enumerate}
\item La separaci\'on de las definiciones de las unidades de sus realizaciones experimentales particulares.

\item Las vinculaci\'on de las definiciones de las unidades  a constantes de la naturaleza.

\item El nuevo sistema de unidades est\'a dise\~{n}ado para que perdure en el tiempo y no est\'e sujeto
a cambios debidos a los continuos avances en los m\'etodos de medida experimental.
\end{enumerate}

El gran avance conceptual del nuevo SI consiste en la separar la realizaci\'on pr\'actica de las unidades
de sus definiciones. Esto permite materializar las unidades de forma independiente en cualquier lugar y 
en cualquier tiempo propmulgada por el Comit\'e de Sabios del Sistema M\'etrico Decimal en 1789.
Esto permite que nuevas realizaciones sean a\~{n}adidas en el futuro seg\'un se vayan desarrollando nuevas
tecnolog\'{\i}as, sin por ello tener que modificar la definici\'on de la propia unidad. Un ejemplo de esto proviene de las
nuevas tecnolog\'{\i}as cu\'anticas y el desarrollo del reloj cu\'antico que cambiar\'a la realizaci\'on experimental del
segundo (ver subsecci\'on \ref{sec:nuevasdefiniciones}).

En el nuevo SI, las unidades de masa (kg), corriente el\'ectrica (A), temperatura (K) y candidad de sustancia (mol)
se redefinen vincul\'andolas a las cuantro constantes universales que aparecen en la tabla \ref{CODATA2018}, mientras
que las unidades de tiempo (s), longitud (m) y eficacia luminosa (cd) 
permanecen asociadas a constantes de la naturaleza como anteriormente (ver Fig.\ref{fig:dependencias}).

Las nuevas dependencias entre las unidades en el nuevo SI son ahora mucho m\'as sim\'etricas que en el anterior sistema como se ve a simple vista en la   Fig.\ref{fig:dependencias}. El segundo sigue siendo la unidad b\'asica de la
cual dependen todas las dem\'as, excepto el mol que aparece desacoplado del resto de unidades. Las constantes fundamentales que se fijan a un valor exacto aparecen en el exterior del esquema y sus unidades vinculadas aparecen en el interior. Estas redefiniciones tienen consecuencias fundamentales en ciertas magnitudes, como las
constantes el\'ectrica $\epsilon_0$ y magn\'etica $\mu_0$ en el vacio que dejan de ser exactas y pasan a tener una indeterminaci\'on experimental en el nuevo SI. As\'{\i}, la constante magn\'etica se determina por la f\'ormula \cite{SI:BIPM}:
\begin{equation}\label{cmagnetica}
\mu_0 = \frac{2 \alpha}{e^2}  \frac{h}{c},
\end{equation}
de modo que todas las constantes en esta ecuaci\'on tienen un valor fijo (ver tabla \ref{CODATA2018}) salvo la constante de estructura fina electromagn\'etica $\alpha$ que se mide experimentalmente, resultando un valor de (CODATA2018)
\begin{align}\label{cmagnetica2}
\mu_0 &=  4\pi[1 + 0.0(6.8) \times 10^{-10}] \times 10^{-7}  \text{N} \text{A}^{-2} \\
 & =1.256 637 062 12 (19) \times 10^{-6} \text{N} \text{A}^{-2}. 
\end{align}
Entonces, la constante el\'ectrica en el vacio c\'asico se obtiene de la relaci\'on,
\begin{equation}\label{celectrica}
\epsilon_0 = \frac{1}{\mu_0 c^2},
\end{equation}
resultando un valor actual de (CODATA2018)
\begin{equation}\label{celectrica2}
\epsilon_0 = 8.854 187 8128(13)  \times 10^{-12} \text{F} \text{m}^{-1}.
\end{equation}
Sin embargo, en nuestra vida cotidiana los cambios no suponen ning\'un trastorno pues t\'{\i}picamente corresponden a  cambios de una parte en $10^8$, o a veces menos. Sus efectos s\'{\i} son muy importantes en las medidas de alta precisi\'on que se necesitan en los laboratorios de investigaci\'on e institutos de metrolog\'{\i}a donde es imprescindible poder hacer medidas exactas y precisas para saber si se ha encontrado un nuevo descubrimiento.

%\'{\i}

%%%%%%%%%%%%%%%%%%%%%%%%%%%%%%%%%%%%%%
%%%%%%%%%%%%%%%%%%%%%%%%%%%%%%%%%%%%%%
\section{Las Constantes Fundamentales de la Naturaleza}
%\section{Revisi\'on del SI basada en las Constantes Fundamentales}
\label{sec:constantesfundamentales}
%%%%%%%%%%%%%%%%%%%%%%%%%%%%%%%%%%%%%%
%%%%%%%%%%%%%%%%%%%%%%%%%%%%%%%%%%%%%%

Detr\'as de cada constante universal de la naturaleza hay una de las leyes fundamentales
de la F\'{\i}sica y la Qu\'{\i}mica. De las siete unidades del nuevo SI, cinco est\'an asociadas 
a constantes universales de la naturaleza como se muestra en la tabla \ref{Constantes-Leyes}.

Las constantes fundamentales son como el ADN de nuestro universo. Otros universos, si existen, pueden tener otro conjunto distinto de constantes universales. En nuestro universo, dependiendo del fen\'omeno f\'{\i}sico de que se trate y de su escala, necesitaremos algunas de constantes fundamentales para poder explicarlo. Con este pu\~{n}ado de constantes podemos describir nuestro mundo 
f\'{\i}sico desde escalas at\'omicas, mesosc\'opicas, microsc\'opicas, macrosc\'opicas, astron\'omicas hasta cosmol\'ogicas.

Adem\'as de estas 5 constantes universales, en el nuevo SI hay otras 2 constantes que se utilizan para determinar la unidad de tiempo y la de eficacia luminosa. La primera de ellas es la frecuencia de transici\'on hiperfina del estado fundamental no perturbado del \'atomo de cesio 133, denotada por $\Delta \nu_{\text{Cs}}$. Si bien esta es una constante de la naturaleza, no la podemos considerar como fundamental en pie de igualdad con las otras cinco. Si lo hici\'eramos, tambi\'en ser\'{\i}an fundamentales cualquier salto de energ\'{\i}a en el espectro de cualquier \'atomo y resultar\'{\i}an infinitas constantes fundamentales. Adem\'as, esta frecuencia en principio es calculable usando las leyes de la electrodin\'amica cu\'antica, mientras que constantes como las de la tabla \ref{CODATA2018} no son calculables desde primeros principios conocidos actualmente.
En cuanto a la eficiencia luminosa $K_{\text{cd}}$, la constante asociada ni siquiera es universal y es puramente convencional. En resumen, solo 5 de las 7 constantes empleadas en el nuevo SI son realmente fundamentales en este sentido aqu\'{\i} expresado.

\begin{table}
\begin{center}
    \begin{tabular}{ | c | l | l |  p{3cm} |}
    \hline \hline
    S\'{\i}mbolo & Constante  & Ley \\ \hline \hline
    $c$ & velocidad de la luz & Teor\'{\i}a de la Relatividad \\ \hline
    $h$ & Planck  & F\'{\i}sica Cu\'antica \\ \hline
    $k$ & Botzmann  & Termodin\'amica \\  \hline
    $e$ & Carga del electr\'on & Electrodin\'amica Cu\'antica \\  \hline
    $N_{\text{A}}$ &  Avogadro   & Teor\'{\i}a At\'omica \\ 
    \hline\hline
    \end{tabular} 
\end{center}
\caption{Las cinco constantes universales de la naturaleza y sus leyes correspondientes a las que est\'an asociadas.
Las leyes de la F\'{\i}sica y la Qu\'{\i}mica permiten describir fen\'omenos naturales una vez que los valores de las constantes son conocidos.} 
    \label{Constantes-Leyes}
\end{table}

Hay otro aspecto esencial que merece ser resaltado: tres de estas cinco constantes universales est\'an asociadas a principios de simetr\'{\i}a de la naturaleza. 
La constante $c$ de la velocidad de la luz es la responsable de la unificaci\'on del espacio y el tiempo en la teor\'{\i}a de la Relatividad \cite{einstein1905d}, uno de los pilares de la f\'{\i}sica moderna. 
 Lo que subyace a esta ley fundamental es el Principio de Relativiad, que declara f\'{\i}sicamente equivalentes todos los sistemas de referencia inerciales en movimiento relativo. Es esta simetr\'{\i}a la responsable de la constancia de la velocidad de la luz. Si $c$ no es constante, las transformaciones de Lorentz y por ende el Principio de Relativiad, se rompen.

La constante de Planck $h$ es la responsable de que las magnitudes f\'{\i}sicas como la energ\'{\i}a, momento angular, etc. puedan tomar valores discretos, llamados cuantos. Es la constante fundamental de la Mec\'anica Cu\'antica, otro de los pilares de la f\'{\i}sica moderna.  Lo que subyace a esta ley fundamental es el Principio de Unitariedad, que fuerza a que la probabilidad de encontrar a las part\'{\i}culas en su estado cu\'antico se preserve a lo largo de su evoluci\'on temporal. M\'as b\'asico a\'un es la linealidad de la Mec\'anica Cu\'antica representada por el Principio de Superposici\'on de estados y que es necesario para garantizar la unitariedad. Si $h$ no fuese constante, el Principio de Unitariedad se quebrar\'{\i}a. Es el Principio de Superposici\'on (linealidad) de la Mec\'anica Cu\'antica el que est\'a en la ra\'{\i}z de todas las sorpresas contra-intuitivas que nos brinda la f\'{\i}sica cu\'antica como nos ense\~{n}a R. Feynman \cite{feynmannSIX,doblerendija}. Se pueden realizar test de  precisi\'on sobre la posible falta de no-linealidad de la 
Mec\'anica Cu\'antica usando modelos no-lineales y poni\'endolos a prueba obteni\'endose cotas de validez de la linealidad de solo un error de  $10^{-21}$ con estimaciones te\'oricas usando resultados experimentales \cite{weinberg1989}, y de hasta $4 \times 10^{-27}$ con medidas directas \cite{wineland1989}. Precisamente, para obtener estas cotas se usan las medidas  de las transiciones de radiofrecuencia tan exactas empleadas en los est\'andares de frecuencia. Resulta que una posible no-linealidad producir\'{\i}a una desintonizaci\'on de esas transiciones resonantes en los est\'andares.

La carga del electr\'on $e$ es el valor de la fuente elemental  (no confinada) de campo el\'ectrico en la Electrodin\'amica Cu\'antica, la primera de las teor\'{\i}as de part\'{\i}culas elementales conocidas y que sirve de referencia para el resto de interacciones fundamentales. Lo que subyace a esta ley fundamental es el Principio de Invariancia Gauge que describe las
interacciones elementales conocidas. En el caso del electromagnetismo, el grupo de invariancia es el m\'as simple $U(1)$.
Es esta simetr\'{\i}a la responsable de la constancia del valor de la carga del electr\'on: si $e$ no es constante, la simetr\'{\i}a gauge se rompe.

En estos tres ejemplos, los valores de las constantes fundamentales $c, h$ y $e$ est\'an protegidos por sendas simetr\'{\i}as de la naturaleza. Una medici\'on cada vez m\'as precisa de ellas puede resultar en una falta de constancia, y por tanto, la violaci\'on de un de las leyes fundamentales de la F\'{\i}sica. Luego la metrolog\'{\i}a es tambi\'en una fuente de descubrimiento de nueva f\'{\i}sica a trav\'es de la mejora con el tiempo de sus m\'etodos de medida. Un ejemplo muy importante de este hecho dentro del nuevo SI es el llamado Tri\'angulo Metrol\'ogico Cu\'antico que veremos en la subsecci\'on \ref{sec:MetrologiaSI}, y tambi\'en las posibles variaciones en la constante de estructura electromagn\'etica $\alpha$ o la raz\'on de la masa del prot\'on al electr\'on (ver \ref{sec:MetrologiaSI}).

La constante de Boltzmann $k$ es el factor de conversi\'on que permite relacionar la temperatura termodin\'amica $T$ macrosc\'opica de un cuerpo  con la energ\'{\i}a t\'ermica de sus grados de libertad (constituyentes) microsc\'opicos. Esta es la constante fundamental en la  F\'{\i}sica Estad\'{\i}stica que estudia la relaci\'on entre la f\'{\i}sica macrosc\'opica y sus constituyentes microsc\'opicos, y es otro de los pilares de la f\'{\i}sica. La constante $k$ aparece en la descripci\'on del mundo macrosc\'opico en la probabilidad $P_i$, o factor de Boltzmann, de encontrar un sistema  en un estado microsc\'opico $i$ cuando est\'a en equilibrio termodin\'amico a la temperatura $T$:
\begin{equation}\label{probabilidad}
P_i = \frac{e^{-\frac{E}{kT}}}{Z},
\end{equation}
donde $Z$ es la funci\'on de partici\'on caracter\'{\i}tica del sistema.
Boltzmann estableci\'o la relaci\'on entre los mundos macrosc\'opicos y microsc\'opicos en su f\'ormula para la entrop\'{\i}a $S$:
\begin{equation}\label{entropia}
S = k \ln W,
\end{equation}
donde $W$ es el n\'umero de estados microsc\'opicos distintos correspondientes a un estado macrosc\'opico del sistema con energ\'{\i}a $E$. Esta es la famosa ecuaci\'on que aparece en el frontispicio de la tumba de Boltzman en Viena. Sin embargo, Boltzmann estableci\'o la relaci\'on \eqref{entropia} como una ley de proporcionalidad, sin introducir expl\'{\i}citamente su constante. Hist\'oricamente, fue Planck el primero en escribirla en su art\'{\i}culo donde enunci\'o la ley de radiaci\'on del cuerpo negro, junto con su constante $h$ de los cuantos de energ\'{\i}a \cite{planck1901}. Planck fue el primero en dar valores num\'ericos a estas dos constantes usando los valores experimentales de las constantes universales que aparecen en la ley de Wien del desplazamiento y la ley de Stefan-Boltzman, que describen propiedades esenciales de la radiaci\'on en equilibrio t\'ermico. Estos primeros valores resultaron estar muy cercanos a los actuales  \cite{planck1901} (ver Tabla \ref{CODATA2018}):
\begin{equation}\label{probabilidad}
h = 6.55 \times 10^{-27} \ \text{erg s}; \quad k =  1.346 \times 10^{-16}  \text{erg} \ \text{grad}^{-1}.
\end{equation}
La constante de Boltzmann no tiene asociado ning\'un principio de simetr\'{\i}a, a diferencia de las otras tres constantes mencionadas m\'as arriba.

La constante de Avogadro $N_{\text A}$ es un factor de conversi\'on que relaciona la cantidad macrosc\'opica de
una sustancia con el n\'umero de sus constituyentes elementales, ya sean estos \'atomos, iones, mol\'eculas etc.
Es una constante fundamental en la Teor\'{\i}a At\'omica de la materia en la F\'{\i}sica y la Qu\'{\i}mica. El mol se 
introduce para poder manejar cantidades macrosc\'opicas de una sustancia que est\'a hecha de un n\'umero enorme
de entidades elementales. La constante de Avogadro $N_{\text A}$  es el factor de proporcionalidad 
entre la masa de un mol de sustancia (masa molar) y la masa promedio de una de sus mol\'eculas, o cualesquiera
que sean sus constituyentes elementales.  $N_{\text A}$  tambi\'en es aproxim\'adamente igual al n\'umero de nucleones en un gramo de materia. Para definir el mol, inicialmente se tom\'o como referencia el \'atomo de oxigeno
y posteriormente al carbono. En el nuevo SI, la masa de un mol de cualquier sustancia, ya sea hidrogeno, oxigeno o carbono, es $N_{\text A}$ veces la masa promedio de cada una de sus part\'{\i}culas constituyentes, la cual es una 
cantidad f\'{\i}sica cuyo valor debe determinarse experimentalmente para cada sustancia.

El origen de la palabra mol proviene del lat\'{\i}n {\it moles} que significa masa, y {\it mol\'ecula} que significa porci\'on peque\~{n}a de masa. La constante de Avogadro tampoco tiene asociado ning\'un principio de simetr\'{\i}a.

Al ser tanto la constante de Boltzmann $k$ como la de Avogadro $N_{\text A}$ factores de conversi\'on de propiedades macrosc\'opicas y microsc\'opicas, tambi\'en est\'an relacionados entre s\'{\i}:
\begin{equation}\label{gases}
R = N_{\text A} k,
\end{equation}
donde $R$ es la constante de los gases ideales que relaciona presi\'on, volumen y temperatura: $P V=nRT$, con
$n$ el n\'umero de moles del gas.

La Hip\'otesis At\'omica juega un papel fundamental en la descripci\'on de la naturaleza. Establece que la materia no es un continuo, sino que es discreta y est\'a hecha de entidades elementales llamadas \'atomos. Feynman la consider\'o la idea m\'as importante de toda la ciencia porque contiene una gran cantidad de informaci\'on en muy pocas palabras, y a partir de la cual, se puede reconstruir muchas de las propiedades de la materia que nos rodea, como que hay distintos estados de la materia dependiendo de la temperatura y sus cambios de fase \cite{feynmannSIX}. En \'epoca de Boltzmann en la segunda mitad del siglo XIX la existencia de \'atomos y mol\'eculas esta todav\'{\i}a objeto de debate y es una de las razones por las que la constante de Boltzmann se introdujo tard\'{\i}amente pues entonces se empleaban energ\'{\i}as macrosc\'opicas, no por mol\'ecula, y se empleaba la constante de los gases \eqref{gases} \cite{planck1920}. Los trabajos sobre el movimiento browniano,
te\'oricos de Einstein \cite{einstein1905c} y experimentales de Perrin \cite{perrin1909}, fueron esenciales para establecer la validez de la Hip\'otesis At\'omica a principios del siglo XX.

%\'{\i}

\section{Revisi\'on del SI basado en las Constantes Fundamentales}
\label{sec:revisionSI}
%%%%%%%%%%%%%%%%%%%%%%%%%%%%%%%%%%%%%%
%%%%%%%%%%%%%%%%%%%%%%%%%%%%%%%%%%%%%%

\subsection{Las Nuevas Definiciones}
\label{sec:nuevasdefiniciones}

Las explicaciones del nuevo SI se ven enormemente facilitadas por la nueva
visi\'on adoptada por el SI de separar las definiciones de unidades, que van vinculadas
a las constantes de la naturaleza, de sus realizaciones experimentales concretas, que son
cambiantes con la tecnolog\'{\i}a y los desarrollos de nuevos m\'etodos de medici\'on en el laboratorio
(secci\'on \ref{sec:nuevoSI}).

El universo visible est\'a hecho de materia y radiaci\'on. La F\'{\i}sica es la ciencia que estudia la materia y la radiaci\'on, y sus interacciones.
El nuevo SI utiliza la naturaleza discreta de la materia y la la radiaci\'on para definir sus unidades
en funci\'on de constantes de la naturaleza. El car\'acter discreto de la materia se denomina hist\'oricamente la hip\'otesis at\'omica y el car\'acter discreto de la radiaci\'on, la hip\'otesis cu\'antica.

Empecemos por la radiaci\'on electromagn\'etica, una de cuyas formas es la luz. Su velocidad $c$ tiene una 
propiedad que la hace especial para medir tiempos y distancias: es una constante universal y tiene el mismo
valor para todos los observadores inerciales, es decir, los que miden las magnitudes f\'{\i}sicas u observables.

Como el tiempo es lo m\'as dif\'{\i}cil de definir, se define lo primero como lo m\'as b\'asico y para ello usamos
un oscilador que sea muy estable: los ciclos de los \'atomos de cesio en un reloj at\'omico. Galileo utiliz\'o p\'endulos, o incluso su propio pulso, para medir tiempos.  Es famosa la definici\'on de tiempo dada por Einstein cuando le preguntaron:
``?`Qu\'e es el tiempo? Lo que miden los relojes". Es una definici\'on muy simple y a la vez muy profunda. De hecho, es una  definici\'on metrol\'ogica de tiempo que encaja muy bien en el nuevo SI: una vez definido el tiempo de forma gen\'erica en t\'erminos de un oscilador o reloj, queda para la materializaci\'on de la unidad segundo (s) la elecci\'on de un reloj adecuado. 
Seg\'un la normas del nuevo SI, la definici\'on y realizaci\'on del segundo queda as\'{\i}:

\noindent {\bf segundo}:
``El segundo, de s\'{\i}mbolo s, se define estableciendo el valor num\'erico fijo de la frecuencia $\Delta \nu_{\text{Cs}}$ del cesio, la frecuencia de  transici\'on hiperfina del estado fundamental no perturbado del \'atomo de cesio 133, en 9 192 631 770, cuando se expresa en la unidad de Hz (herzio), igual a 1/s." 
\qed

La realizaci\'on del segundo por medio de la frecuencia de transici\'on del cesio
\begin{equation}\label{cesio}
\Delta \nu_{\text{Cs}} = 9 192 631 770 \ \text{Hz},
\end{equation}
implica que el segundo es igual a la duraci\'on de 9 192 631 770 per\'{\i}odos de la radiaci\'on correspondiente a la transici\'on entre los dos niveles hiperfinos del estado fundamental no perturbado del \'atomo de ${}^{133}\text{Cs}$.

Esta materializaci\'on del patr\'on de tiempo es un ejemplo de la provisionalidad de las realizaciones experimentales de las unidades del SI. Los patrones basados en relojes de cesio llevan us\'andose desde los a\~{n}os 60 del siglo XX. Actualmente tenemos realizaciones m\'as precisas usando relojes cu\'anticos y ya hay prevista por el BIPM una renovaci\'on de la realizaci\'on del segundo usando esta tecnolog\'{\i}a cu\'antica antes de 2030 (ver \ref{sec:MetrologiaSI}). Sin embargo, la definici\'on de tiempo permanecer\'a inalterada.

El metro entonces se define con el  tiempo y de la velocidad de la luz $c$:

\noindent {\bf metro}:
``El metro, de s\'{\i}mbolo $m$, se define tomando el valor num\'erico fijo de la velocidad de la luz en el vac\'{\i}o $c$ como 299 792 458 cuando se expresa en la unidad  $\text{m} \text{s}^{-1}$, donde el segundo se define en t\'erminos de la frecuencia de cesio $\Delta \nu_{\text{Cs}}$".
\qed

Con esta definici\'on, un metro es la longitud del camino recorrido por la luz en el vac\'{\i}o durante un intervalo de tiempo con una duraci\'on de 1/299 792 458 de segundo. Esta definici\'on se basa en fijar la velocidad de la luz en el vac\'{\i}o exactamente en
\begin{equation}\label{luz}
c = 299 792 458 \ \text{m} \text{s}^{-1}.
\end{equation}
Los m\'etodos para medir la velocidad de la luz han ido cambiado a trav\'es de los tiempos, desde el inicial de Ole R\"{o}mer en 1676, basado en el tr\'ansito de la luna \'Io de  J\'upiter mediante un telescopio, hasta las modernas t\'ecnicas usando interferometr\'{\i}a con l\'aseres.

A continuaci\'on lo natural es definir la unidad de masa, el kilo.
Resulta que la luz tambi\'en tiene otra propiedad que la hace muy \'util para la definici\'on del kilo:
la luz de una frecuencia fija (monocrom\'atica) tiene un m\'{\i}nimo de energ\'{\i}a discreto llamado fot\'on
cuya energ\'{\i}a es proporcional a su frecuencia seg\'un descubrieron Planck \cite{planck1901}, y luego  Einstein
\cite{einstein1905a}. Esa constante de proporcionalidad es
la constante $h$ de Planck. Las unidades de esta constante son las tres b\'asicas de lo que una vez se llam\'o 
el Sistema MKS, precursor del SI actual: metro, kilo y segundo en las siguiente proporci\'on,
\begin{equation}\label{hunidades}
[h] = \text{kg}\ \text{m}^2 \ \text{s}^{-1}.
\end{equation}
Es importante hacer notar que la constante de la gravitaci\'on universal $G$ de Newton tambi\'en tiene unidades del sistema MKS, aunque en otra proporci\'on:
\begin{equation}\label{Gunidades}
[G] = \text{kg}^{-1}\ \text{m}^3 \ \text{s}^{-2}.
\end{equation}
Resulta que $h$ y $G$ son las \'unicas constantes fundamentales con unidades MKS. Hay otras constantes
asociadas a interacciones fundamentales, pero no contienen  la masa sino otras cargas elementales.
Pero con $G$ esto no es suficiente para definir la unidad de masa con la exactitud que se necesita en metrolog\'{\i}a. El problema radica en que la precisi\'on con la que se mide $G$ es mucho peor que la de $h$.
El `kilo gravitatorio' no es una buena unidad metrol\'ogica pr\'actica. Este hecho es el origen de la `via cu\'antica' del kilo como vamos a ver. En definitiva, podemos usar $h$ para definir el kilo a partir del segundo y el metro que ya est\'an definidos una vez fijado el valor de $c$.

Si no fuera por esta falta de precisi\'on en la medida de $G$, la constante $h$ se podr\'{\i}a desacoplar del kilo y fijarla de forma independiente a trav\'es de los efectos puramente cu\'anticos de Hall y de Josephson (ver \ref{sec:MetrologiaSI}):
\begin{equation}
h = \frac{4}{K_J^2 R_K} = 6.626068854 \ldots \times 10^{-34} \text{Js}.
\end{equation}
Pero si se tomara esta otra via cu\'antica, tan natural en teor\'{\i}a, entonces desacoplamos el kilo de $h$ y se volver\'{\i}a a quedar vinculado a un artefacto: nos vemos pues abocados al `kilo cu\'antico'.

Una vez tomada la via del `kilo cu\'antico', la siguiente cuesti\'on es c\'omo usar la constante de Planck $h$ para definirlo. Para ello, usamos  la prescripci\'on del nuevo SI de usar las unidades de $h$ y las definiciones de segundo y metro ya introducidas anteriormente. Entonces, la definici\'on de kilo queda as\'{\i}:

\noindent {\bf kilogramo}:
``El kilogramo, s\'{\i}mbolo kg, se define tomando el valor num\'erico fijo de la constante de Planck, h como $6. 626 070 15 \times 10^{-34}$, cuando se expresa en la unidad J s, igual a $\text{kg} \ \text{m}^2 \ \text{s}^{-1}$, donde el metro y el segundo se definen en t\'erminos de $c$ y $\Delta \nu_{\text{Cs}}$".
\qed

O en ecuaciones,
\begin{equation}\label{kilo}
\begin{matrix}
1 \text{kg} &= \left(\frac{h}{6,626 070 15 \times 10^{-34}}\right)   \text{m}^{-2} \ \text{s} \\
&= 1,475 521 \ldots
\times 10^{40} \frac{h \Delta \nu_{\text{Cs}}}{c^2}
\end{matrix}
\end{equation}
Esta definici\'on es equivalente a la relaci\'on exacta
\begin{equation}\label{h}
h = 6. 626 070 15 \times 10^{-34} \ \text{Js}.
\end{equation}

Tras la definici\'on del `kilo cu\'antico' surge el problema de como realizarlo experimentalmente.
Lo m\'as sencillo a primera vista ser\'{\i}a usar las relaciones fundamentales de energ\'{\i}a de Einstein \cite{einstein1905b} y Planck \cite{planck1901}, respectivamente:
\begin{equation}\label{energia}
E = mc^2 \quad \text{y} \quad E=h \nu.
\end{equation}
El fundamento del `kilo cu\'antico' consiste en tener un m\'etodo muy preciso para medir $h$ y luego usarlo para definir el kilo. Pero para esto, las relaciones fundamentales anteriores presentan un problema. El fot\'on como cuanto de energ\'{\i}a luminosa no tiene masa. Si queremos que el cuanto tenga una masa, lo que est\'a mejor definido es su longitud de onda de de Broglie \cite{debroglie1924}:
\begin{equation}
E = \frac{h c}{\lambda}.
\end{equation}
Sin embargo, medir una longitud de onda es f\'acil para un movimiento tipo onda plana, que de nuevo es m\'as t\'{\i}pico de una radiaci\'on monocrom\'atica. Para tener una masa real $m$, necesitamos una part\'{\i}cula con una longitud de onda asociada, es decir, una masa localizada en el espacio la cual se describe m\'as naturalmente con un paquete de ondas, el cual no tiene una sola longitud de onda. As\'{\i} que usar las relaciones m\'as b\'asicas de de la energ\'{\i}a \eqref{energia} no es lo metrol\'ogicamente m\'as \'util.

De ah\'{\i} que la via cu\'antica del kilo se realize a trav\'es de la Balanza de Kibble (ver \ref{sec:MetrologiaSI}). Esto nos lleva a una pregunta importante: ?`Qu\'e tipo de masa, inercial o gravitatoria, aparece en las unidades de $h$, y por ende en la nueva definici\'on de kilo? Para el caso del fot\'on que 
no tiene masa, tal distinci\'on no existe.  Cuando tenemos una part\'{\i}cula con masa, entonces depender\'a de la relaci\'on
mec\'anica que usemos para relacionarla con $h$, y as\'{\i} decidir si el kilo que definimos es inercial o gravitatorio.
Por ejemplo, si usamos la relaci\'on de Einstein, el kilo ser\'a inercial, si usamos una balanza, entonces el kilo ser\'a gravitatorio. Por tanto, la Balanza de Kibble nos proporciona una definici\'on de kilo gravitatorio cu\'antico. Ahora bien,
el Principio de Equivalencia nos dice que ambos tipos de masa son iguales y est\'a comprobado experimentalmente con una
precisi\'on superior a la medici\'on de las constantes fundamentales que intervienen en el SI: una incertidumbre 
de $(0.3 \pm 1.8)\times 10^{-13}$  \cite{equivalence}. Con lo cual, podemos omitir el adjetivo gravitatorio mientras que la precisi\'on del Principio de Equivalencia sea superior a la de las constantes fundamentales.

\begin{figure}[t]
  \includegraphics[width=0.25\textwidth]{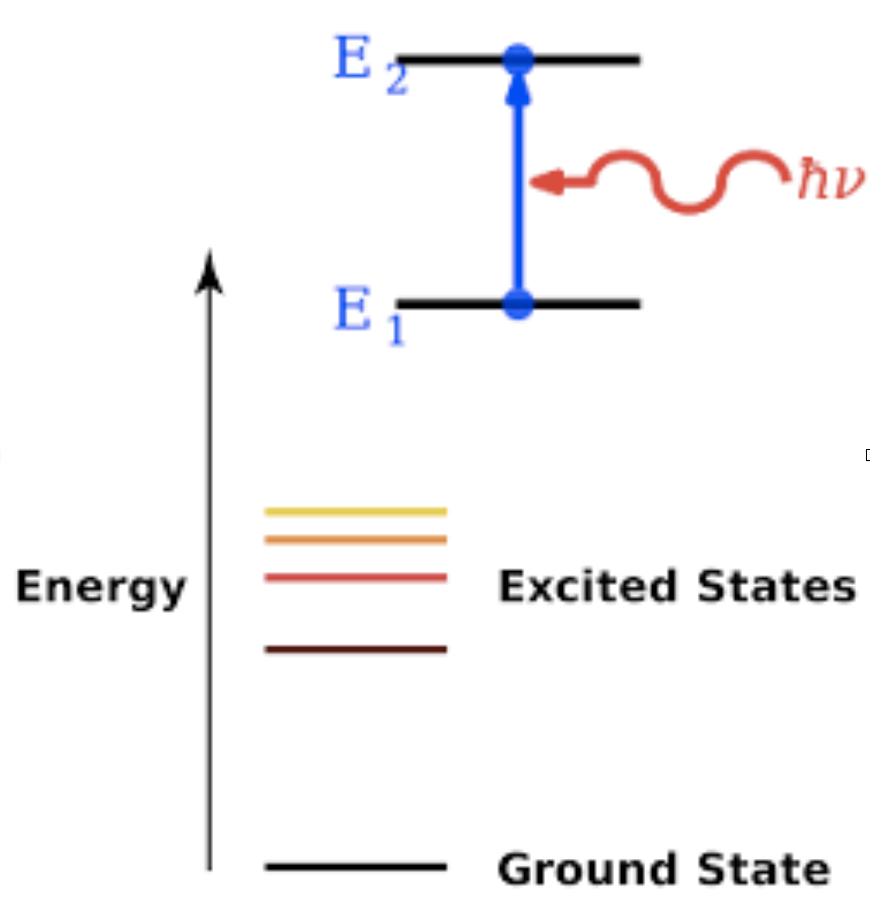}
  \caption{La primera revoluci\'on cu\'antica de las tecnolog\'{\i}as se basa en el car\'acter discreto de las  magnitudes f\'{\i}sicas como los estados de energ\'{\i}a en los \'atomos. Los fotones de la radiaci\'on  electromagn\'etica permiten manipular los estados de energ\'{\i}a bien definida \eqref{energia}. Cr\'edito: Wikipedia (adapted).}
  \label{fig:primerarevolucion}
\end{figure}

La descripci\'on del `kilo cu\'antico' mediante la Balanza de Kibble pertenece a la parte del nuevo sistema SI correspondiente a la realizaci\'on pr\'actica de la unidad kilo, no a su definici\'on (ver \ref{sec:MetrologiaSI}). 

Para continuar definiendo el resto de unidades SI y derivarlas de las anteriores ya definidas, nos volvemos a fijar en el car\'acter discreto de la materia. As\'{\i}, existen \'atomos (neutros) y electrones (cargados). La materia m\'as elemental cargada
(sin confinar) es el electr\'on y con el se define el amperio usando el segundo ya definido:

\noindent {\bf amperio}:
``El amperio, s\'{\i}mbolo A, se define tomando el valor num\'erico fijo de
carga elemental, $e$, como:
\begin{equation}
e = 1.602176634 \times 10^{-19} \ \text{C},
\end{equation}
cuando se expresa en la unidad culombio, C, igual a A s, y el segundo se define en t\'erminos de  $\Delta \nu_{\text{Cs}}$."
\qed

En consecuencia, un amperio es la corriente el\'ectrica correspondiente al flujo de $1/(1.602 176 634 \times 10^{-19})$ cargas elementales por segundo.
La ventaja del nuevo amperio es que se puede medir, no como el antiguo que ten\'{\i}a una definici\'on impracticable que le dejaba de hecho fuera del sistema SI. Adem\'as,  es independiente del kilogramo y se reduce la incertidumbre de las magnitudes el\'ectricas.

Para la realizaci\'on experimental del amperio se han propuesto varias t\'ecnicas: a) la m\'as directa es usar la definici\'on actual del SI mediante el transporte mono-electr\'onico (ver \ref{sec:MetrologiaSI}) \cite{SI:BIPMrealization,cem} aunque
todav\'{\i}a est\'a en vias de desarrollo para ser competitiva; b) usando la ley de Ohm y los efectos Hall y Josephson para definir el voltio y el ohmio (ver \ref{sec:MetrologiaSI}) \cite{SI:BIPMrealization,cem}; c) la relaci\'on entre la corriente el\'ectrica y la variaci\'on temporal del potencial en un condensador  \cite{SI:BIPMrealization,cem}.

%\'{\i}

\begin{figure}[t]
  \includegraphics[width=0.45\textwidth]{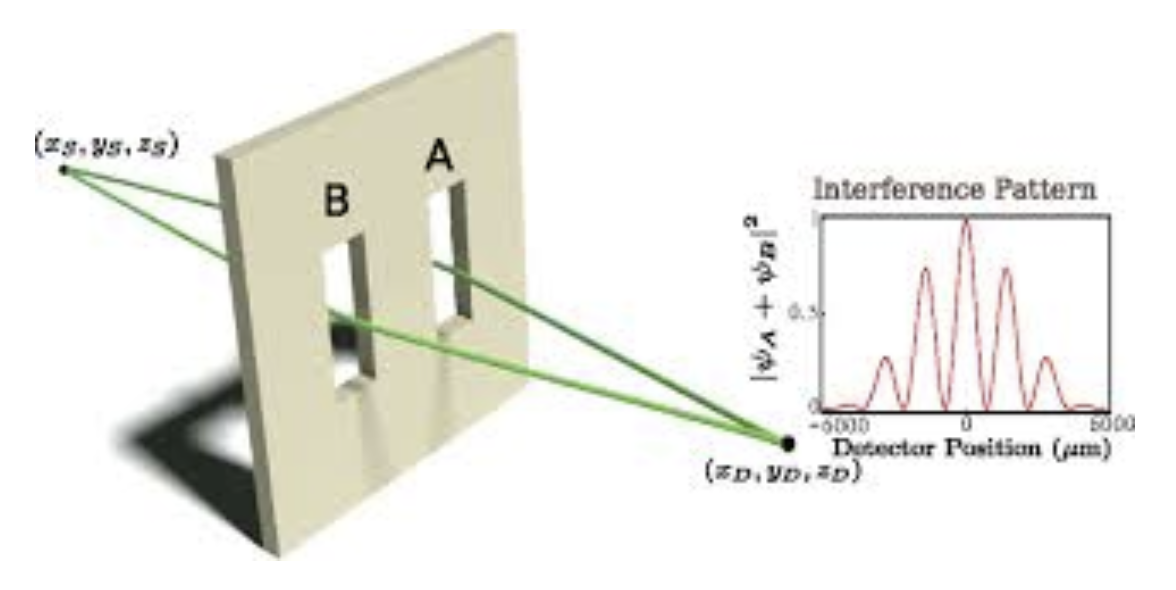}
  \caption{La segunda revoluci\'on cu\'antica de las tecnolog\'{\i}as se basa en el principio de superposici\'on de la
  mec\'anica cu\'antica. El caso m\'as sencillo se ejemplifica con el experimento de la doble rendija \cite{feynmannSIX,doblerendija} donde se   superponen las propiedades de una sola part\'{\i}cula. Cuando la superposici\'on cu\'antica involucra varias part\'{\i}culas da lugar al entrelazamiento cu\'antico que es el recurso fundamental en la informaci\'on cu\'antica \cite{rmp}.
   Cr\'edito: R. Sawant et al. \cite{segundarevolucion}.}
  \label{fig:segundarevolucion}
\end{figure}

En cuanto a los \'atomos, hist\'oricamente, son las unidades elementales de sustancia y con ellos se puede definir el mol como la unidad de cantidad de una cierta sustancia. Detr\'as de esta definici\'on est\'a la naturaleza discreta de la materia a trav\'es de la hip\'otesis at\'omica y la constante de la naturaleza asociada es la de Avogadro \eqref{gases}:

\noindent {\bf mol}:
``El mol, s\'{\i}mbolo mol, es la unidad de la cantidad de sustancia. Un mol contiene exactamente $6.02214076 \times 10^{23}$ entidades elementales."
\qed

Este valor sale de fijar el valor num\'erico de la constante de Avogadro a
\begin{equation}
N_{\text A} = 6.02214076 \times 10^{23} \ \text{mol}^{-1},
\end{equation}
en unidades de $\text{mol}^{-1}$.  
Como consecuencia, el mol es la cantidad de sustancia de un sistema que contiene $6.022 140 76 \times 10^{23}$ entidades elementales especificadas.
La cantidad de sustancia de un sistema es una medida del n\'umero de cantidades elementales. Una cantidad elemental puede ser un \'atomo, una mol\'ecula,
un ion, un electr\'on, cualquier otra part\'{\i}cula o un grupo espec\'{\i}fico de part\'{\i}culas.

Para la realizaci\'on experimental del amperio se han propuesto varias t\'ecnicas \cite{SI:BIPMrealization,cem}: a) el proyecto Avogadro (International Avogadro Coordination), b) m\'etodos grav\'{\i}m\'etricos, c) ecuaci\'on de los gases, y d) m\'etodos electrol\'{\i}ticos.

%\'{\i}

Nos falta todav\'{\i}a la unidad fundamental para medir la temperatura, el kelvin. Siguiendo el nuevo SI, la debemos vincular
a una constante fundamental de la naturaleza, en este caso la constante de Boltzmann. La constante de Boltzmann sirve como un factor de conversi\'on entre energ\'{\i}a y temperatura:
\begin{equation}\label{kT}
E = k T.
\end{equation}
Tambi\'en esta constante la podemos ver como resultado del car\'acter discreto de la naturaleza.  Por ejemplo, la hip\'otesis
at\'omica para gases nobles ideales permite calcular la energ\'{\i}a cin\'etica de estos como:
\begin{equation}\label{nobles}
E_{\text c} = \frac{3}{2} k T.
\end{equation}
La nueva definici\'on resultante del kelvin como la unidad de temperatura termodin\'amica es:

\begin{figure}[t]
  \includegraphics[width=0.45\textwidth]{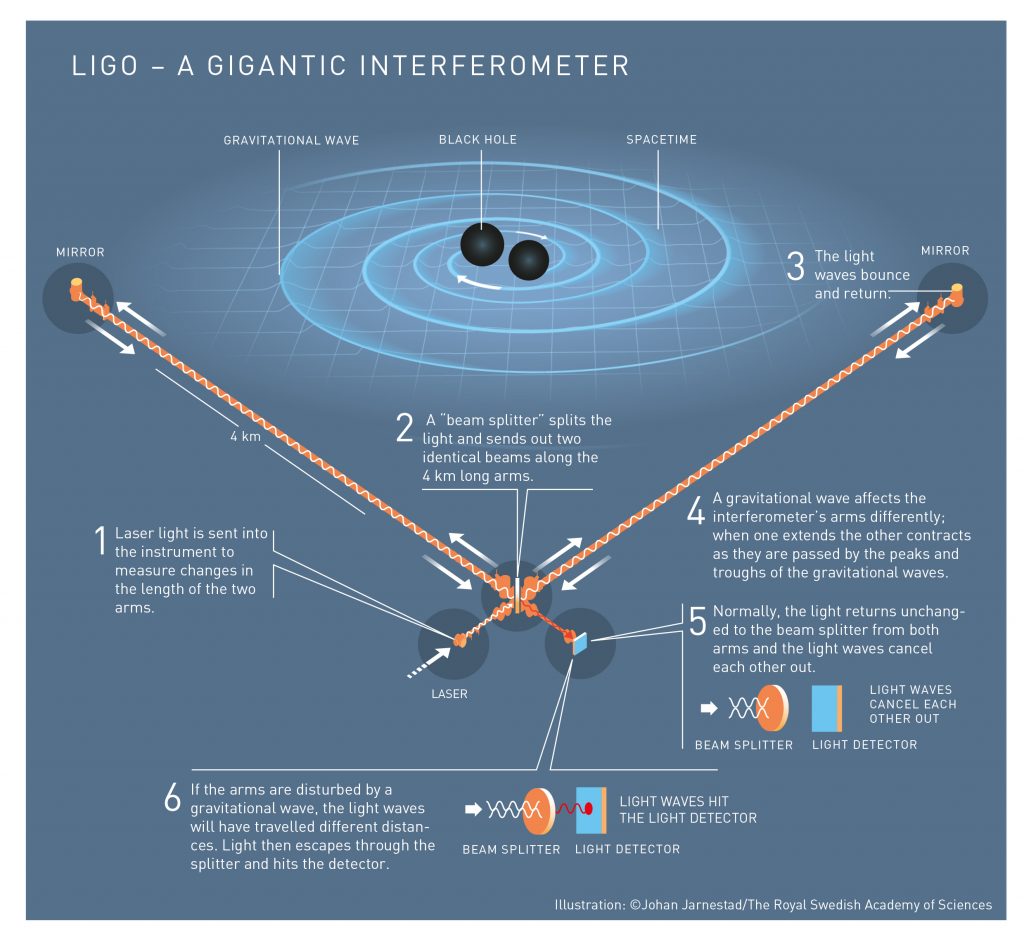}
  \caption{Esquema de un detector de ondas gravitacionales tipo LIGO. Se muestran los dos brazos perpendiculares de cuatro km cada uno. Las ondas gravitacionales causadas por la colisi\'on de dos agujeros negros se pueden detectar en el patr\'on de interferencias gracias a la extremada sensibilidad del aparato que permite resolver distancias miles de veces m\'as peque\~{n}as  que el n\'ucleo at\'omico. (Cr\'edito: Johan Jarnestad/The Royal Swedish Academy of Sciences \cite{ligonobel})}
  \label{fig:LIGO}
\end{figure}

%\'{\i}

\noindent {\bf kelvin}:
``El kelvin, s\'{\i}mbolo K, se define tomando el valor num\'erico fijo de la constante de Boltzman, k, como
\begin{equation}\label{Boltzmann}
k = 1.380649  \times 10^{-23} \ \text{J} \text{K}^{-1},
\end{equation}
cuando se expresa en las unidades kg $\text{m}^2$  $\text{s}^{-1}$ $\text{K}^{-1}$,
donde el kilogramo, el metro y el segundo se definen de acuerdo con $h, c$ y $\Delta \nu_{\text{Cs}}$".
\qed

Esta definici\'on implica que el kelvin es igual a un cambio termodin\'amico de temperatura que resulta en un cambio en la energ\'{\i}a t\'ermica kT de $1.380649  \times 10^{-23} \ \text{J}$. Como consecuencia de la nueva definici\'on,
el punto triple del agua deja de tener un valor exacto y ahora tiene una incertidumbre dada por:
\begin{equation}\label{triple}
T_{\text{TPW}} = 273.160 00 \text{K} \pm 0.000 10 \text{K},
\end{equation}
como consecuencia de heredar la incertidumbre que ten\'{\i}a la constante de Boltzmann $k$ antes de la nueva definici\'on.

\begin{figure}[t]
  \includegraphics[width=0.45\textwidth]{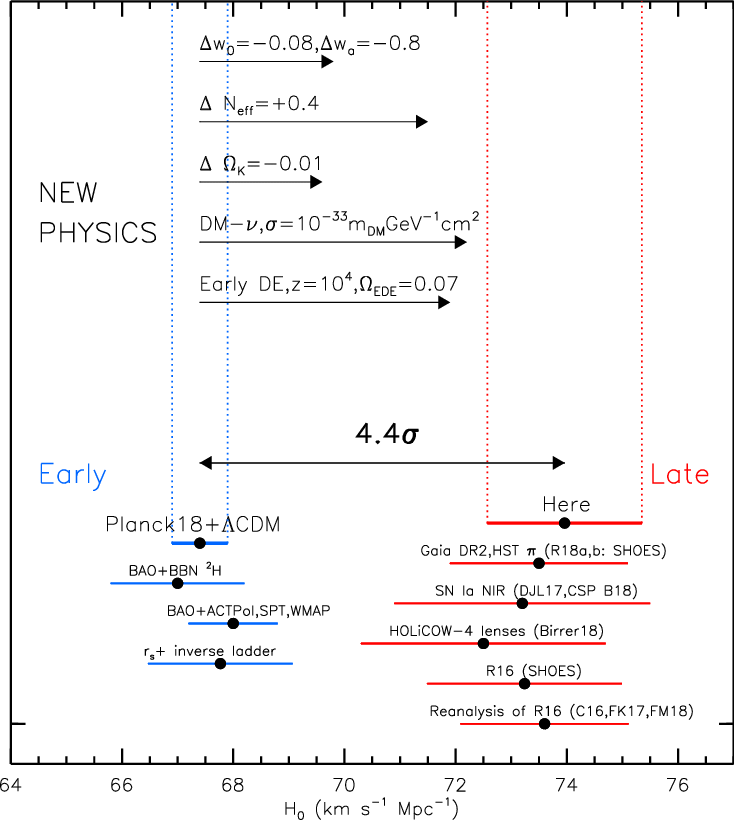}
  \caption{Diferencias de valores para la constante de Hubble $H_0$ medidas con el sat\'elite Planck 2018 en el universo temprano  y extrapoladas con el modelo cosmol\'ogico es\'andar (azul) \cite{planck2018}, y las medidas locales realizadas con las escaleras de distancias de cefeidas y supernovas \cite{shoes2019,H0tension}.
(Cr\'edito: Riess et al. \cite{H0tension}).}
  \label{fig:Tension}
\end{figure}

Siguiendo el principio gu\'{\i}a usado hasta ahora para introducir las nuevas definiciones de las unidades SI utilizando 
el car\'acter discreto de la energ\'{\i}a \eqref{energia} y la materia \eqref{gases}, podemos ir m\'as all\'a y usar el car\'acter discreto de la informaci\'on para introducir la constante de Boltzmann.  As\'{\i},
otra manera de fundamentar el origen discreto de la constante de Boltzmann es a trav\'es del Principio de Landauer 
\cite{landauer1961} que establece que el m\'{\i}nimo de energ\'{\i}a disipada en forma de calor a una temperatura $T$ en el sistema elemental m\'as 
simple, ya sea cl\'asico (bit) o cu\'antico, c\'ubit, es:
\begin{equation}\label{kT}
E = \ln 2 \ k T.
\end{equation}
El origen del Principio de Landauer reside en el car\'acter irreversible del borrado de informaci\'on \cite{bennett1982,bennett2003} en un sistema
que conlleva una disipaci\'on m\'{\i}nima de energ\'{\i}a \cite{informationengine2014}. Su verificaci\'on experimental ha sido posible directamente en varios trabajos recientes \cite{expland1,expland2,expland3,expland4}.
De esta manera, tenemos una visi\'on unificada de todas las unidades del SI bas\'andonos en la propiedad fundamental que poseen tanto la energ\'{\i}a, la materia y la informaci\'on: su naturaleza discreta en nuestro universo.

Para la realizaci\'on experimental del kelvin se han propuesto varias t\'ecnicas \cite{SI:BIPMrealization,cem}:
a) mediante termometr\'{\i}a ac\'ustica con gases, b) termometr\'{\i}a radiom\'etrica de banda espectral, c) termometr\'{\i}a de gas polarizante y d) termometr\'{\i}a de ruido de Johnson.

%\'{\i}

\begin{figure}[t]
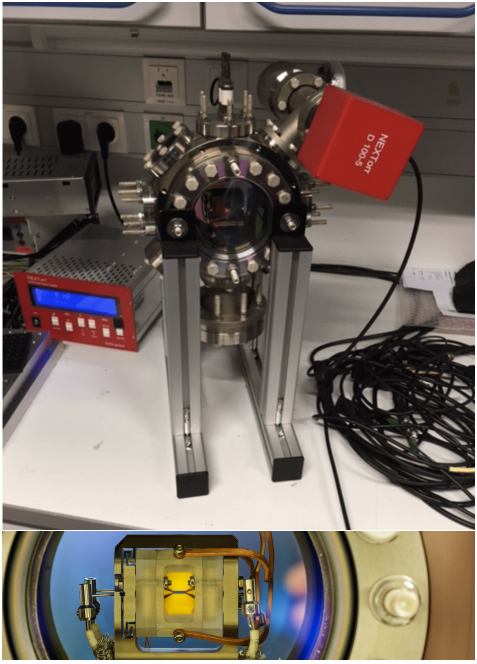

  \includegraphics[width=0.35\textwidth]{quantumclock1.png}
    \includegraphics[width=0.35\textwidth]{BR3A0278-3.jpg}
  \caption{Arriba: Visi\'on general de un reloj cu\'antico de iones atrapados desarrollado por el grupo de Rainer Blatt
  en el laboratorio de la Universidad de Innsbruck. Es un prototipo compacto que ocupa un espacio funcional reducido y es comercializado por la empresa Alpine Quantum Technologies \cite{monz2019}. Abajo: detalle de la parte central del reloj donde se muestra la trampa donde se confinan los iones usando campos electromagn\'eticos. (Cr\'edito: Rainer Blatt Lab).}
  \label{fig:QuantumClock}
\end{figure}

Menci\'on aparte merece la s\'eptima unidad b\'asica del SI empleada para medir la eficacia luminosa de una fuente.
Es una medida de la bondad de una fuente de luz cuando su luz visible se percibe por el ojo humano.  Es claramente 
convencional y subjetiva. Se utilizan valores promedio del comportamiento del ojo humano.
Se cuantifica mediante la relaci\'on del flujo luminoso a la potencia, medida en l\'umenes (lm) por vatio en el SI.
No hay una ley universal de la f\'{\i}sica asociada a esta unidad, porque la fuente de luz no tiene por qu\'e estar
en equilibrio termodin\'amico. El concepto b\'asico para la candela se mantiene en el nuevo SI:

\noindent {\bf candela}:
``La candela, s\'{\i}mbolo cd, es la unidad SI de intensidad luminosa en una direcci\'on dada. Se define tomando el 
valor num\'erico fijo de la eficacia luminosa de la radiaci\'on monocrom\'atica de frecuencia 
$540 \times 10^{12} \text{Hz}$, 
$K_{\text cd}$, como 683 cuando se expresa en la unidad lm $\text{W}^{-1}$, que es igual a cd sr  
$\text{W}^{-1}$, o 
cd sr  $\text{kg}^{-1}$ $\text{m}^{-2}$ $\text{s}^3$, donde el kilogramo, el metro y el segundo se definen en t\'rminos de $h, c$ y $\Delta \nu_{\text{Cs}}$".
\qed

Esta definici\'on se basa en tomar el valor exacto para la constante
\begin{equation}\label{luminosidad}
K_{\text {cd}} = 683 \ \text{cd}  \ \text{sr} \  \text{kg}^{-1} \ \text{m}^{-2} \ \text{s}^3,
\end{equation}
para una raciaci\'on monocrom\'atica de frecuencia $\nu = 540 \times 10^{12 }\text{Hz}$.
Como consecuencia, una candela es la intensidad luminosa, en una direcci\'on dada, de una fuente que emite radiaci\'on monocrom\'atica de frecuencia $\nu = 540 \times 10^{12 }\text{Hz}$ y tiene una intensidad radiante en esa direcci\'on de (1/683) W/sr.

Para la realizaci\'on experimental de la candela se han propuesto varias t\'ecnicas \cite{SI:BIPMrealization,cem}
como la realizaci\'on pr\'actica de unidades radiom\'etricas, empleando dos tipos de m\'etodos primarios: los basados
en detectores patr\'on como el radi\'ometro el\'ectrico de sustituci\'on y fotodiodos de eficiencia cu\'antica predecible,
y los basados en fuentes patr\'on como el radiador de Planck y la radiaci\'on de sincotr\'on. En la pr\'actica se utiliza
m\'as habitualmente una l\'ampara patr\'on con dise\~{n}o optimizado para emitir en una direcci\'on definida y a larga distancia desde el detector.

%\'{\i}

\subsection{Metrolog\'{\i}a Cu\'antica y el Nuevo SI}
\label{sec:MetrologiaSI}

El establecimiento de las leyes de la mec\'anica cu\'antica en el primer cuarto del siglo XX permit\i'o comprender
la naturaleza a escala at\'omica. Como consecuencia de esa mejor comprensi\'on del mundo at\'omico surgieron 
aplicaciones en forma de nuevas tecnolog\'{\i}as cu\'anticas. A este primer per\'{\i}odo se le conoce como la primera revoluci\'on cu\'antica y ha producido tecnolog\'{\i}as tan innovadoras como el transistor y el l\'aser. Incluso los ordenadores cl\'asicos actuales son consecuencia de estos avances cu\'anticos de primera generaci\'on. Con el comienzo del siglo XXI estamos viendo emerger nuevas tecnolog\'{\i}as cu\'anticas de segunda generaci\'on \cite{manifesto} que constituyen lo que se conoce como la segunda revoluci\'on cu\'antica. Ambas revoluciones se basan en explotar aspectos concretos de las leyes de la mec\'anica cu\'antica. As\'{\i}, podemos clasificar estas revoluciones tecnol\'ogicas en dos grupos:

\noindent {\bf Primera Revoluci\'on Cu\'antica}: se basa en el car\'acter discreto de las propiedades del mundo cu\'antico: los cuantos de energ\'{\i}a (como los fotones), los cuantos de momento angular, etc. Esta naturaleza discreta de las magnitudes f\'{\i}sicas es lo primero que sorprende en la f\'{\i}sica cu\'antica (ver Fig.\ref{fig:primerarevolucion}).
\qed

\noindent {\bf Segunda Revoluci\'on Cu\'antica}:  se basa en el principio de superposici\'on de estados cu\'anticos, en los cuales se puede guardar y procesar informaci\'on como consecuencia de sus propiedades de entrelazamiento cu\'antico (ver Fig.\ref{fig:segundarevolucion}). 
\qed

En la segunda revoluci\'on cu\'antica se han identificado cinco \'areas de trabajo que dar\'an lugar a nuevos desarrollos tecnol\'ogicos. Enumeradas de menor a mayor complejidad son: metrolog\'{\i}a cu\'antica, sensores cu\'anticos, criptograf\'{\i}a cu\'antica, simulaci\'on cu\'antica y ordenadores cu\'anticos.

La metrolog\'{\i}a cu\'antica est\'a considerada una de las tecnolog\'{\i}as cu\'anticas de desarrollo m\'as inmediato. Es la parte de la metrolog\'{\i}a que se ocupa de c\'omo realizar medidas de par\'ametros f\'{\i}sicos con gran resoluci\'on y sensibilidad usando la mec\'anica cu\'antica, especialmente explotando sus propiedades de entrelazamiento. Una pregunta fundamental en metrolog\'{\i}a cu\'antica es c\'omo escala la precisi\'on, medida en t\'erminos de la variancia $\Delta \textfrak{m}$, con que se estima un par\'ametro f\'{\i}sico con el n\'umero $N$ de part\'{\i}culas empleadas o repeticiones del experimento. Resulta que con los interfer\'ometros cl\'asicos no se puede superar el llamado l\'{\i}mite bal\'{\i}stico est\'andar \cite{qmetrology,quantummetrology,quantummetrology2} dado por 
\begin{equation}\label{limiteclasico}
\Delta \textfrak{m} \geq \frac{1}{\sqrt{N}},
\end{equation}
mientras que con metrolog\'{\i}a cu\'antica de segunda generaci\'on se puede el l\'{\i}mite de Heisenberg dado por
\begin{equation}\label{limitecuantico}
\Delta \textfrak{m} \geq \frac{1}{N}.
\end{equation}
Un ejmplo de aplicaci\'on muy importante de la metrolog\'{\i}a cu\'antica a la ciencia b\'asica lo constituye la detecci\'on de ondas gravitacionales por experimentos como el LIGO (Laser Interferometer Gravitational-Wave Observatory),
  \cite{ligo2016} (ver Fig.\ref{fig:LIGO}),  donde se necesita medir distancias entre masas separadas del orden de unos kil\'ometros con una alt\'{\i}sima precisi\'on (mil\'esima parte del di\'ametro de un prot\'on). Estas variaciones en distancia se producen en las longitudes de los brazos de los interfer\'ometros cuando pasa una onda gravitacional por ellos. Utilizando `luz comprimida', una forma de \'optica cu\'antica \cite{orzag2016}, se puede conseguir mejorar la sensibilidad de los interfer\'ometros de acuerdo con el l\'{\i}mite cu\'antico \eqref{limitecuantico} \cite{squeezed,squeezed_LIGO,squeezed_LIGO2}.

\begin{figure}[t]
  \includegraphics[width=0.5\textwidth]{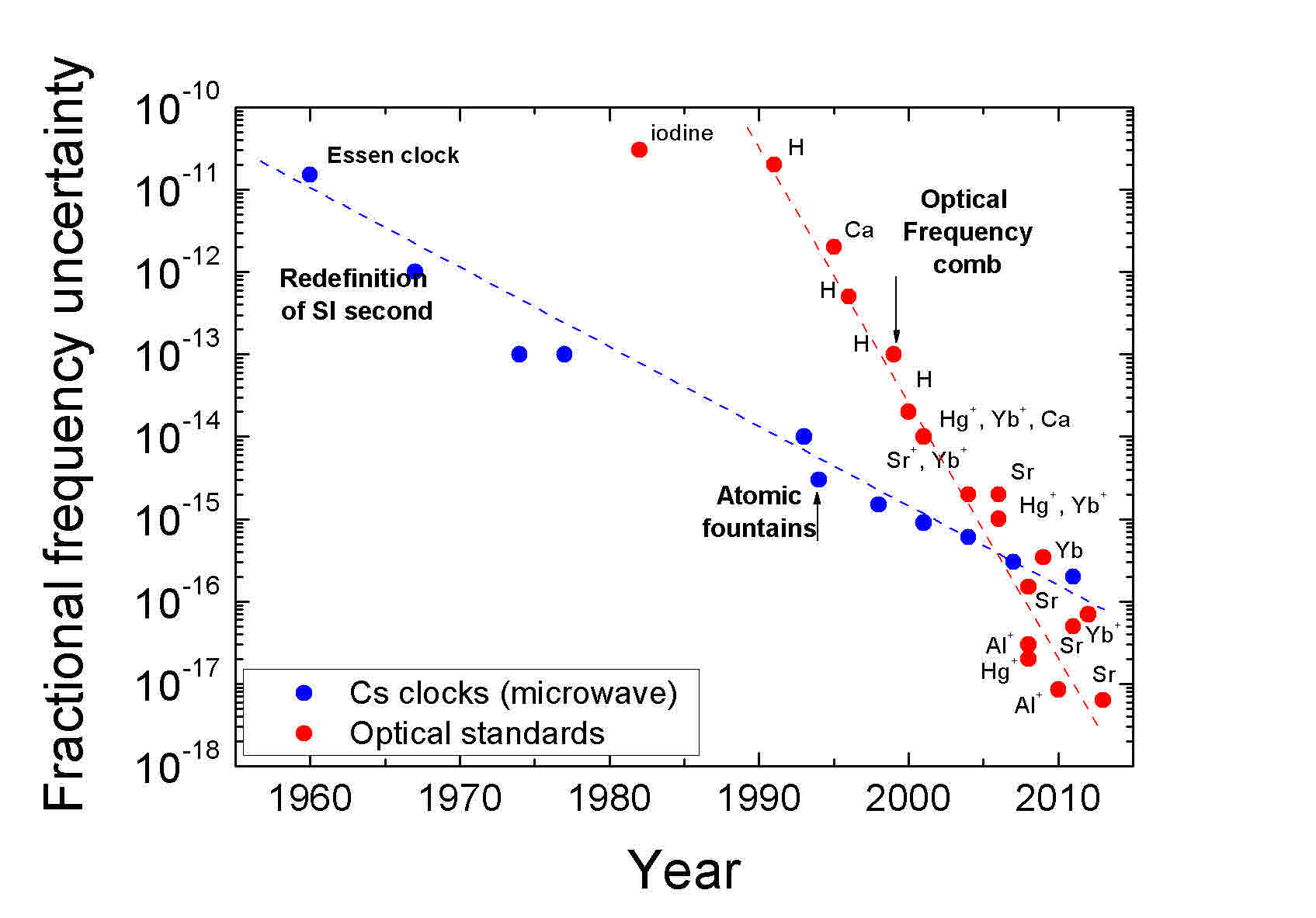}
  \caption{Comparativa de la evoluci\'on temporal reciente de las incertidumbres en
relojes at\'omicos de cesio (microondas) y relojes cu\'anticos (\'opticos),
de iones y de \'atomos neutros en redes \'opticas. Claramente se observa un cambio de tendencia
con una ganancia de exactitud en los relojes cu\'anticos. (Cr\'edito: Reproducido por cortes\'{\i}a  de la  Societ\`a Italiana di Fisica de N. Poli et al. \cite{comparativeclocks}).}
  \label{fig:Comparativa}
\end{figure}

Las medidas de experimentos tipo LIGO pueden tener otra aplicaci\'on b\'asica para tratar de elucidar la controversia que ha surgido recientemente con la llamada constante de Hubble $H_0$. En el modelo cosmol\'ogico est\'andard denotado por $\Lambda \text{CDM}$ ($\Lambda$=energ\'{\i}a oscura, CDM=materia oscura fr\'{\i}a) $H_0$ mide la velocidad con que se expande actualmente el universo de acuerdo con la ley de Hubble. Es un par\'ametro de capital importancia en cosmolog\'{\i}a. Hay dos fuentes de medidas que dan valores discrepantes. Por una lado, el valor 
proporcionado por el sat\'elite Planck en 2018 analizando el fondo de radiaci\'on de microondas del universo temprano y extrapolando el par\'ametro de Hubble a su valor actual con el modelo est\'andar  $\Lambda \text{CDM}$ da un valor de $H_0=67.4 \pm 0.5$ kil\'ometros por segundo y por megaparsec de distancia \cite{planck2018}. Por otro lado, medidas realizadas en el universo actual utilizando escaleras de distancias basadas en cefeidas y supernovas de tipo 1a da como resultado $H_0=74.03 \pm 1.42 \ \text{km} \text{s}^{-1}\text{Mpc}^{-1}$ \cite{shoes2019}. Esta discrepancia supone una confidencia estad\'{\i}stica de 4.4 sigmas (desviaciones est\'andar), muy cercana a la barrera de las 5 sigmas que se considera como clara evidencia de que son resultados distintos y supondr\'{\i}a ser que hay nueva f\'{\i}sica que el modelo est\'andar no ha tenido en cuenta (ver Fig.\ref{fig:Tension}).
Adem\'as, para aumentar la controversia, tambi\'en existe una medida del universo actual con otra escalera de distancias que arroja un valor intermedio entre los dos discrepantes, $H_0=69.8 \ \text{km} \text{s}^{-1}\text{Mpc}^{-1}$ \cite{friedman2019}.
Todas las medidas realizadas con el universo actual dan valores por encima de los valores obtenidos con el universo temprano y el modelo est\'andar. Luego, o se trata de errores sistem\'aticos, o puede ser el indicio de nueva f\'{\i}sica m\'as all\'a del modelo cosmol\'ogico vigente, como por ejemplo la existencia de energ\'{\i}a oscura din\'amica, por citar solo un ejemplo \cite{H0tension}.

\begin{figure}[t]
  \includegraphics[width=0.45\textwidth]{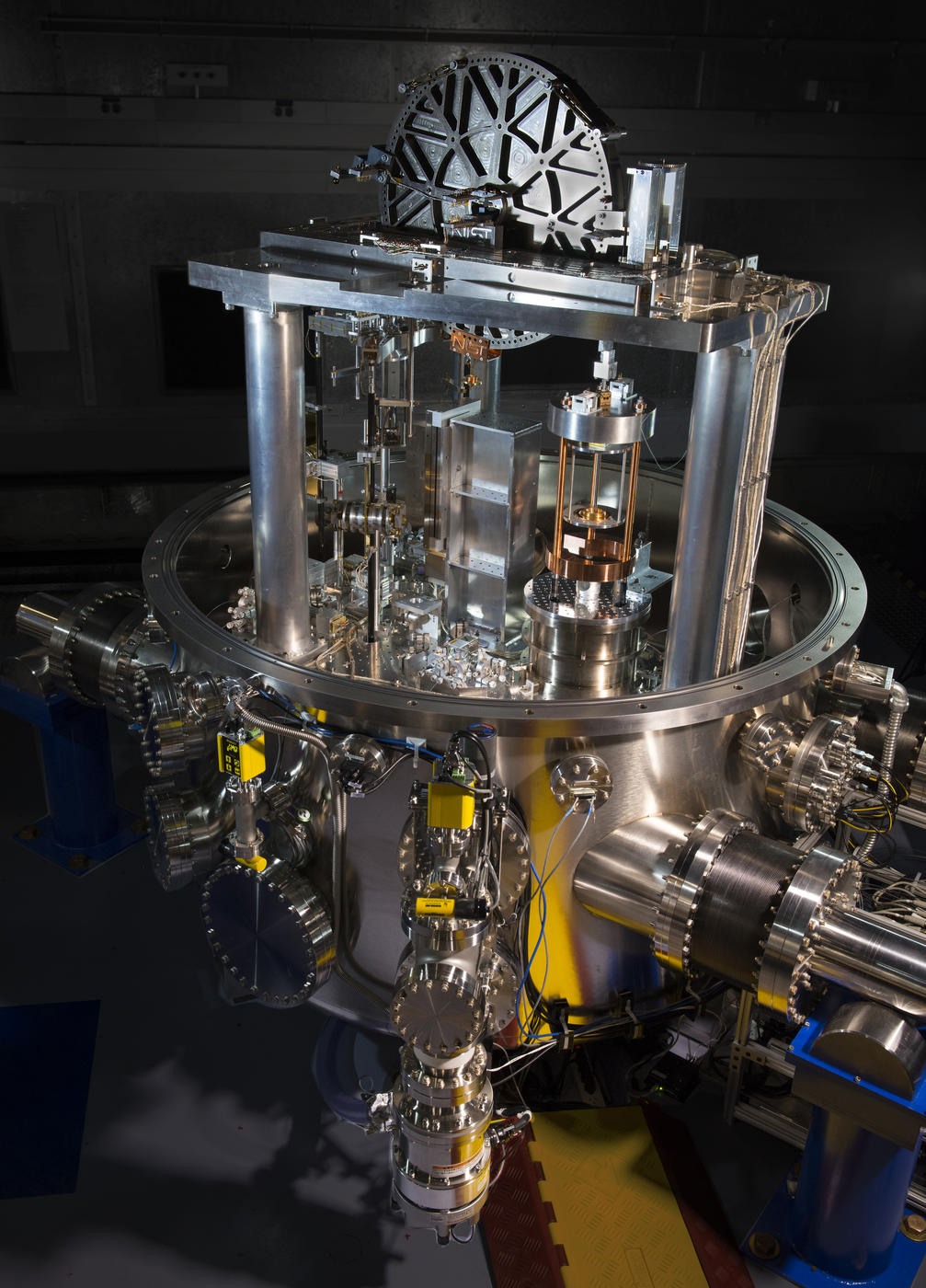}
  \caption{ Balanza de Kibble NIST-4 empleada para medir la constate de Planck $h$ con una incertidumbre de 13 partes en mil millones en 2017, contribuyendo a la redefinici\'on del kilogramo como unidad de masa den el nuevo SI
  en 2019.(Cr\'edito: NIST).}
  \label{fig:KibbleBalance}
\end{figure}

Los experimentos tipo LIGO pueden ser muy \'utiles en el futuro para elucidar ensta controversia, una m\'as, sobre la constante de Hubble. No se trata de analizar colisiones de agujeros negros como en el descubrimiento original de las ondas gravitacionales, sino de las colisiones de estrellas binarias de neutrones. Resulta que cuando dos estrellas de neutrones se fusionan, las ondas gravitacionales resultantes pueden usarse para obtener informaci\'on sobre la posici\'on de las estrellas, y por ende de las galaxias donde se encuentran. Haciendo estad\'{\i}stica de con al menos 50 sucesos de estas colisiones se podr\'{\i}a tener una medida directa de la constante de Hubble con una exactitud no lograda hasta la fecha y resolver la controversia \cite{neutronstars1}. Y resultados recientes mejoran estas expectativas. Ya se ha podido detectar un suceso de fusi\'on de estrellas de neutrones con el que estimar la constante de Hubble, cuyo valor resultante es precisamente $H_0=70 \ \text{km} \text{s}^{-1}\text{Mpc}^{-1}$ con una indeterminaci\'on del orden de $\sim 7\%$ \cite{neutronstars2}. Se estima que con 15 sucesos se pod\'ra reducir el error al 1\% y empezar a resolver la discrepancia.

%\'{\i}

 Veamos a continuaci\'on tres ejemplos importantes de aplicaciones de la metrolog\'{\i}a cu\'antica al nuevo SI, tanto de tecnolog\'{\i}as cu\'anticas de primera y segunda revoluci\'on.

\subsubsection{Relojes Cu\'anticos}

El mecanismo b\'asico de un reloj consiste en un sistema de oscilaciones per\'{\i}odicas muy estables donde cada per\'{\i}odo define la unidad de tiempo y el reloj cuenta esos per\'{\i}odos para medir el tiempo. En el pasado se han usado movimientos per\'{\i}odicos naturales como el de la Tierra alrededor de su eje o del sol, osciladores mec\'anicos como p\'endulos o resonadores de cristal de cuarzo. En los a\~{n}os 60 del siglo pasado se empez\'o a usar oscilaciones at\'omicas para medir el tiempo usando \'atomos de cesio por su mayor exactitud y estabilidad que los sistemas mec\'anicos. La referencia actual para definir el est\'andar de tiempo es  el cesio 133 donde se usa
la frecuencia de resonancia correspondiente a la diferencia de energ\'{\i}a entre los dos niveles hiperfinos de su estado fundamental  (ver \ref{sec:nuevasdefiniciones}). Los relojes at\'omicos de cesio pueden medir  el tiempo t\'{\i}picamente con una exactitud de un segundo en 30 millones de a\~{n}os. 
En general, la exactitud y estabilidad de un reloj at\'omico es mayor cuanto m\'as alta es la frecuencia de la transici\'on at\'omica y menor es la anchura de la l\'{\i}nea de transici\'on electr\'onica. 
Si hacemos la frecuencia del oscilador m\'as grande podemos aumentar la resoluci\'on del reloj al hacerse m\'as peque\~{n}o el per\'{\i}odo que usamos como referencia.

Los relojes at\'omicos han permitido mejorar m\'ultiples desarrollos tecnol\'ogicos a los que estamos acostumbrados en nuestra vida diaria: controlar la frecuencia de las ondas de las emisiones de televisi\'on, los sistemas mundiales de navegaci\'on por sat\'elite como el GPS, transacciones financieras, internet, tel\'efonos m\'oviles etc.

Los nuevos relojes cu\'anticos son un tipo de relojes at\'omicos donde el  aumento de su exactitud  se debe a utilizar frecuencias de transici\'on at\'omicas en el rango \'optico en vez del rango de microondas usado por los relojes de cesio. Las frecuencias \'opticas de la luz visible son como cinco \'ordenes de magnitud mayores que las microondas. Para conseguir dar este salto se necesit\'o utilzar trampas de iones (ver Fig.\ref{fig:QuantumClock}) con t\'ecnicas de l\'ogica cu\'antica empleadas en la computaci\'on cu\'antica \cite{cz1,cz2} que forma parte de las tecnolog\'{\i}as cu\'anticas de la segunda revoluci\'on 
\cite{quantumclock1,quantumclock2,quantumclock3}. Los relojes cu\'anticos consiguen una incertidumbre de solo una parte en $10^{17}$s, lo que equivale a un segundo de error en un reloj cu\'antico que hubiese empezado a medir la edad del universo actual de 13.700 millones de a\~{n}os. Un reloj de iones basado en l\'ogica cu\'antica es un ejemplo de cooperaci\'on entre dos iones donde cada uno proporciona cualidades complementarias. Por ejemplo, el ion del aluminio $\text{Al}^+$ tiene una frecuencia de transici\'on en el rango \'optico que es \'util para la frecuencia de referencia del reloj. Sin embargo, su estructura de niveles le hace un mal candidato para enfriarlo a las temperaturas necesarias para estabilizarlo. En cambio, esto s\'{\i} es posible con el ion de berilio $\text{Be}^+$. Usando protocolos de computaci\'on cu\'antica, la informaci\'on sobre el estado interno del ion espectrosc\'opico $\text{Al}^+$ despues de sondear su transici\'on con un l\'aser, puede ser fielmente transferido al ion l\'ogico $\text{Be}^+$, en donde esta informaci\'on puede ser detectada con eficiencia casi del 100\% \cite{blattwineland2003}. Cada especie de ion aporta una funcionalidad distinta, la frecuencia de referencia o el m\'etodo de enfriamiento, y el entrelazamiento cu\'antico entre los estados de ambos iones permite al conjunto funcionar como un reloj cu\'antico.

%\'{\i}

Los relojes cu\'anticos actuales utilizan, o bien a) uno o dos iones atrapados, o bien b) \'atomos ultrafr\'{\i}os confinados en campos electromagn\'eticos en forma de redes \'opticas. 

Cada una de estas realizaciones tiene sus ventajas y su inconvenientes. Los relojes de iones tienen una exactitud muy alta pues pueden confinar al ion enfri\'andolo en una trampa de forma que nos acercamos mucho al ideal de un sistema aislado de las perturbaciones exteriores. Sin embargo, al utilizar solo un ion para la se\~{n}al de absorci\'on  se consigue menos estabilidad pues se reduce la raz\'on de la se\~{n}al frente al ruido externo. Por el contrario, los relojes de \'atomos en redes \'opticas pueden trabajar con un gran n\'umero de \'atomos consiguiendo una mayor estabilidad y mejor se\~{n}al frente al ruido externo.

Diferentes equipos trabajando con ambas opciones  est\'an desarrollando t\'ecnicas para conseguir versiones cada vez mejores.  El record m\'as reciente lo tienen los iones atrapados con una incertidumbre de $9.4 \times 10^{-19}$s
 \cite{quantumclockrecord2019}.
Los relojes de \'atomos en redes \'opticas tambi\'en consiguen los $10^{-18} \text{s}$ de exactitud. Todav\'{\i}a queda tiempo para decidir cu\'al de las dos realizaciones se escoger\'a para realizar un futuro nuevo segundo, o si las dos son complementarias (ver Fig.\ref{fig:Comparativa}).

Las mejoras en la medici\'on del tiempo proporcionadas por los relojes cu\'anticos tambi\'en tienen importantes aplicaciones. Las aplicaciones tecnol\'ogicas son similares a las anteriormente citadas y es de gran inter\'es poder enviar uno de estos relojes cu\'anticos en misiones espaciales y para mejorar los sistemas de navegaci\'on. 
Otra aplicaci\'on es la medida con gran precisi\'on  del campo gravitatorio pues de acuerdo con la relatividad general de Einstein, existe una dilataci\'on temporal por efectos gravitatorios adem\'as de la debida a la velocidad. Con un reloj cu\'antico se pueden distinguir campos gravitatorios de solo 30 cm de altura \cite{dilataciongravitacional}, e incluso menos. 
Estas medidas van a permitir definir mejor las alturas por encima del nivel del mar, ya que \'este no se mide de igual manera en distintas partes del mundo y es crucial para conocer la actividad de los oc\'eanos. De forma similar se pueden aplicar estos dispositivos cu\'anticos a geodesia, hidrolog\'{\i}a, sincronizaci\'on de redes de telescopios.

La ciencia b\'asica es una de las primeras aplicaciones fundamentales de ellos. Comparando el funcionamiento de varios relojes cu\'anticos a lo largo del tiempo podemos descubrir si alguna de las constantes fundamentales de la f\'{\i}sica cambian con el tiempo, lo cual es fundamental para encontrar nueva f\'{\i}sica y para la definici\'on de las unidades b\'asicas de acuerdo con el nuevo SI (ver \ref{sec:nuevasdefiniciones}). Ejemplos de constantes fundamentales que se pueden sondear su dependencia temporal son la constante de estructura fina electromagn\'etica $\alpha$ \eqref{cmagnetica} y la raz\'on de las masas del prot\'on y el electr\'on $\mu:=m_p/m_e$. En el pasado ha habido controversia sobre posibles variaciones temporales de $\alpha$ y $\mu$ detectadas mediante medidas de transiciones at\'omicas en cu\'asares distantes comparadas con las medidas actuales en el laboratorio \cite{alphavariacion1,alphavariacion2}. Como sucede que todas las transiciones at\'omicas dependen funcionalmente de $\alpha$  y tambi\'en las transiciones hiperfinas dependen sensiblemente de la raz\'on $\mu$, resulta que los relojes cu\'anticos permiten mejorar las cotas de variaci\'on de estas constantes fundamentales. Con estos experimentos, se ha podido medir la raz\'on de frecuencias \'opticas entre iones de $\text{Al}^+$ y $\text{Hg}^+$ proporcionando una cota a la variaci\'on temporal en $\alpha$ de $-1.6\pm 2.3 \times 10^{-17}$ por a\~{n}o, y con el ion yterbio $\text{Yb}^+$ se obtiene una cota para $\mu$ de 
$0.2\pm 1.1 \times 10^{-16}$ por a\~{n}o, las cuales son mejores en un factor diez que las medidas astrof\'{\i}sicas
\cite{alphavariacion3,alphavariacion4}.
Estos resultados negativos para la variaci\'on temporal de las constantes fundamentales sirven de justificaci\'on para el nuevo sistema de unidades SI y su universalidad independientemente del espacio y del tiempo, al menos mientras los experimentos sigan confirmando tales resultados.

%\'{\i}

\begin{figure}[t]
  \includegraphics[width=0.35\textwidth]{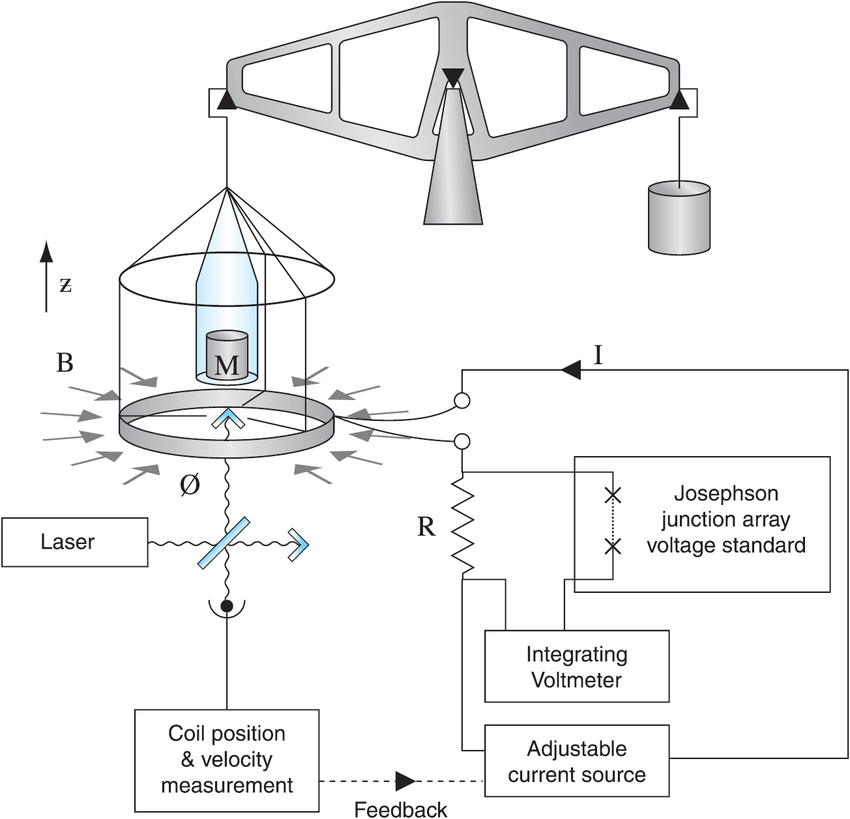}
  \caption{Balanza de Kibble funcionando en modo pesada. Explicaciones en el texto. (Cr\'edito: Robinson-Schlamminger \cite{kibblebalance}).}
  \label{fig:KibbleModoPesada}
\end{figure}

\subsubsection{ Balanza de Kibble}

La balanza de Kibble es la realizaci\'on experimental actual de la unidad de masa  a trav\'es de la via cu\'antica en el nuevo SI \cite{kibble1,kibble2,kibble3,kibblebalance}. De este modo se sigue la nueva metodolog\'{\i}a de separar definiciones de unidades
de su materializaci\'on pr\'actica (ver \ref{sec:nuevoSI}). Mientras que la definici\'on del nuevo kilo vinculado a la constante de Planck ha sido ya explicada en \ref{sec:nuevasdefiniciones}, ahora veremos c\'omo realizarlo en el laboratorio con la tecnolog\'{\i}a actual.

La realizaci\'on del `kilo cu\'antico' consiste en dos partes bien diferenciadas: a) la balanza de Kibble y b) la determinaci\'on cu\'antica de la potencia el\'ectrica. La balanza de Kibble  (ver Fig.\ref{fig:KibbleBalance}) tiene por objetivo establecer la equivalencia
de una potencia mec\'anica en potencia el\'ectrica. Despues \'esta se relaciona con la constante de Planck mediante procedimientos de metrolog\'{\i}a de la primera revoluci\'on cu\'antica: efectos Hall cu\'antico entero y efecto Josephson.

\begin{figure}[t]
  \includegraphics[width=0.35\textwidth]{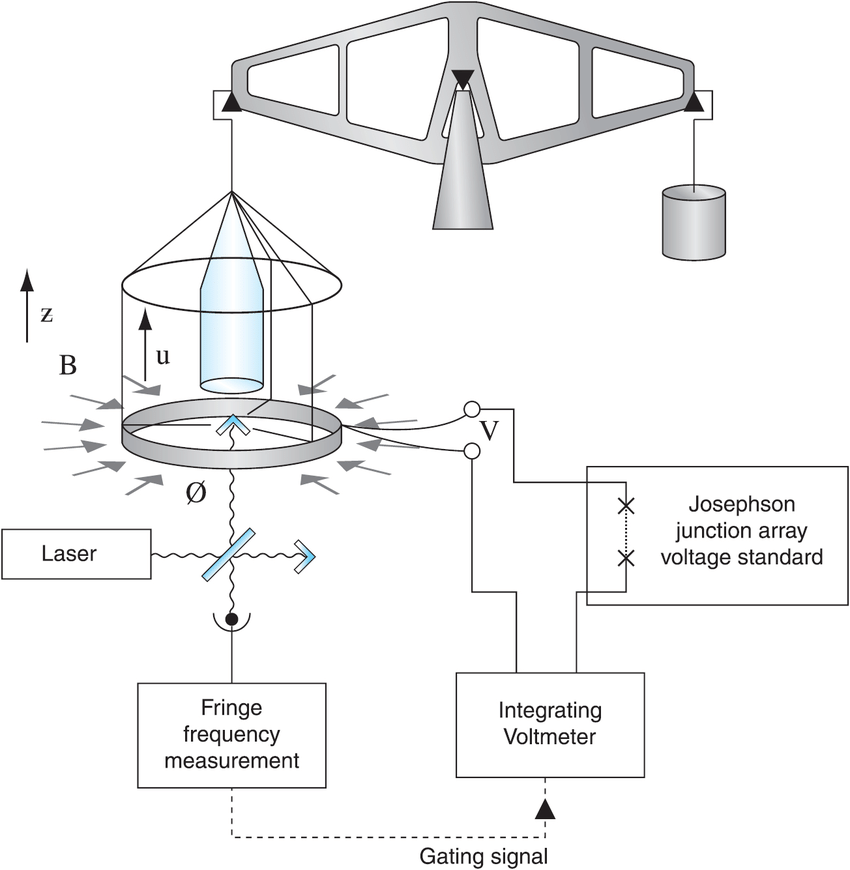}
  \caption{Balanza de Kibble funcionando en modo velocidad. Explicaciones en el texto. (Cr\'edito Robinson-Schlamminger \cite{kibblebalance}).}
  \label{fig:KibbleModoVelocidad}
\end{figure}

Empecemos por la balanza de Kibble. Aunque se presenta una discusi\'on simplificada de ella, es suficiente para entender sus fundamentos. Se parece a una balanza ordinaria en que tambi\'en tiene dos brazos, pero
mientras que en la balanza ordinaria se comparan dos masas, una patr\'on y otra inc\'ognita, en la de Kibble se 
comparan fuerzas mec\'anicas gravitatorias con fuerzas electromagn\'eticas.
Su funcionamiento consta de dos modos de funcionamiento: i/ Modo Pesada y ii/ Modo Velocidad.

\noindent {\bf Modo Pesada}: Se dispone en uno de los brazos una masa $m$ de prueba, que  podr\'{\i}a ser por ejemplo el patr\'on IPK. En el otro plato se monta un circuito de cables el\'ectricos por donde se hace pasar una corriente $I$ (ver Fig.\ref{fig:KibbleModoPesada}).
El circuito  est\'a suspendido en un campo magn\'etico muy fuerte creado por imanes con un campo estacionario y permanente $B$. La longitud del circuito es $L$. Entonces, la corriente induce un campo electromagn\'etico que interacciona con el campo magn\'etico constante del im\'an.  La fuerza electromagn\'etica vertical resultante se iguala al peso de la masa prueba,
\begin{equation}\label{}
B L I = m g. 
\end{equation}
Durante este modo de funcionamiento, se mide la intensidad de corriente el\'ectrica continua de forma muy precisa mediante instrumentos apropiados (efecto Hall cu\'antico entero), y que es proporcional a la fuerza vertical. La corriente se ajusta para que la fuerza resultante iguale al peso de la masa prueba.

\qed

\noindent {\bf Modo Velocidad}: Este es un modo de calibraci\'on que es necesario pues la cantidad $B L$ es muy dif\'{\i}cil de medir con precisi\'on. Si no fuera por esto, bastar\'{\i}a con el modo pesada. 
Se usa un motor el\'ectrico para mover el circuito de cables verticalmente a trav\'es del campo magn\'etico externo a una velocidad constante $v$ (ver Fig.\ref{fig:KibbleModoVelocidad}). Este movimiento induce un voltaje $V$ en el circuito cuyo origen es tambi\'en una fuerza de Lorentz y est\'a dado por,
\begin{equation}\label{}
B L v= V. 
\end{equation}
Durante este modo de funcionamiento, se mide el voltaje de forma muy precisa mediante instrumentos apropiados (efecto Josephson), y por ende el campo magn\'etico que es proporcional. Tambi\'en se usan sensores l\'aser para monitorizar el movimiento vertical del circuito el\'ectrico mediante interferometr\'{\i}a. Con esto se pueden detectar variaciones en posici\'on del orden de la semi-longitud de onda del l\'aser empleado. Con todo esto se asegura que el movimiento vertical sucede a velocidad constante y se puede medir el campo magn\'etico constante.

%\'{\i}
\qed 

\begin{figure}
  \includegraphics[width=0.45\textwidth]{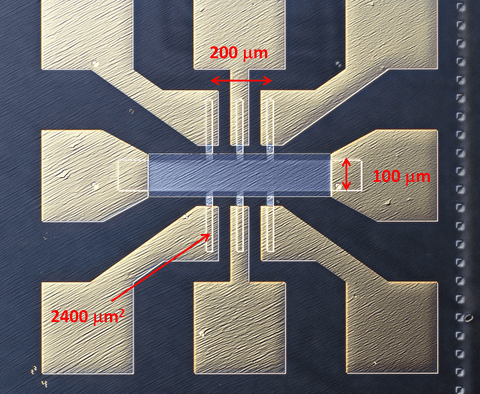}
  \caption{Un ejemplo de dispositivo de efecto Hall cu\'antico entero empleado por el NIST para medir resistencias.
  Esta barra Hall utiliza componentes con grafeno que aparecen delineadas por l\'{\i}neas blancas.
   La fuente y el drenaje de electrones est\'an en los extremos izquierdo y derecho de la barra. No se muestran los contactos el\'ectricos que hay encima y debajo de la barra. (Cr\'edito: NIST).}
  \label{fig:Hall_bar}
\end{figure}
El efecto Hall cu\'antico entero permite medir resistencias  con una indeterminaci\'on de unas pocas partes en $10^{-11}$ de ohmio. Por esta raz\'on se utiliza para realizar el patr\'on est\'andar de resistencia \cite{press_kit,cem,SI:BIPM}.

El resultado de comparar el modo de pesada con el modo de velocidad, eliminando la cantidad $BL$, es la equivalencia entre potencias mec\'anica y el\'ectrica:
\begin{equation}\label{potencias}
m g v = I V.
\end{equation}
Aunque es habitual llamar a la balanza de Kibble como balanza de potencia o de watt, sin embargo, n\'otese que la balanza de Kibble no mide potencias reales, sino virtuales. Este punto es de crucial importancia en metrolog\'{\i}a:
si la potencia mec\'anica se midiera realmente, entonces el dispositivo estar\'{\i}a sujeto a p\'erdidas por fricci\'on incontrolables; otros\'{\i}, si se midiera la potencia el\'ectrica directamente, entonces estar\'{\i}a sujeta a disipaci\'on por calor. Vemos que el modo velocidad es esencial y proporciona la calibraci\'on adecuada.

Resulta que experimentalmente es m\'as preciso medir resistencias que intensidades de corriente. Usando la ley de Ohm, podemos despejar la masa en la balanza de Kibble en funci\'on de medidas de resistencia y de voltaje:
\begin{equation}\label{masaKibble}
m  = \frac{V_{\text R} V}{ g v R},
\end{equation}
donde $V_{\text R}$ y $V$ son las dos mediciones del voltaje necesarias.

En la segunda parte de la realizaci\'on experimental del `kilo cu\'antico' necesitamos relacionar la potencia el\'ectrica en \eqref{potencias} con la constante de Planck $h$. Esto se hace a trav\'es de la medida de la intensidad el\'ectrica $I$ en el modo pesada y el voltaje $V$ en el modo velocidad de la balanza de Kibble. Para ello se utilizan los siguientes efectos cu\'anticos.

\begin{figure}
  \includegraphics[width=0.45\textwidth]{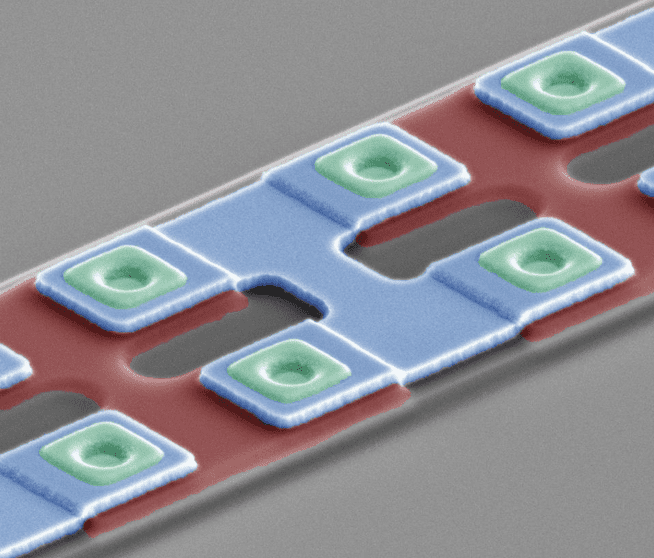}
  \caption{
 Vista de cerca de una uni\'on Josephson moderna empleada en el NIST. Las uniones de Josephson se forman en los pozos circulares de color verde donde las dos capas superconductoras se superponen. Ver explicaci\'on en texto. (Cr\'edito: M. Malnou/NIST/JILA).}
  \label{fig:Josephson_union}
\end{figure}

\noindent {\bf Efecto Hall Cu\'antico Entero}: cuando una muestra bidimensional contiene electrones constre\~{n}idos a moverse en dicho plano sujetos a un campo el\'ectrico coplanar alineado longitudinalmente y un campo magn\'etico constante $B$ muy intenso aplicado perpendicularmente a la muestra (ver Fig.\ref{fig:Hall_bar}), y adem\'as la muestra electr\'onica se somete a temperaturas pr\'oximas al cero absoluto, el sistema se aparta de la ley de Ohm cl\'asica y entra en un r\'egimen cu\'antico.  Al igual que en el caso cl\'asico, aparece una corriente electr\'onica transversal que induce un voltaje transversal llamado voltaje de Hall $V_{\text H}$. 
El sistema de electrones entra en un  nuevo comportamiento cu\'antico caracterizado por la aparici\'on de saltos y mesetas en la relaci\'on entre la corriente transversal y el campo magn\'etico \cite{Hallcuantico}. En particular, la resistencia Hall $R_{\text H}$ asociada a dicho potencial Hall est\'a cuantizada 
\begin{equation}\label{HallEntero}
R_{\text H} = \frac{1}{n^\prime} \frac{h}{e^2},
\end{equation}
donde $n^\prime$ es un n\'umero entero, dando lugar a dichas mesetas aparecen en las curvas de la resistividad Hall. Se define la constante de von Klitzing $R_{\text K}$ como
\begin{equation}\label{HallEnteroconstante}
R_{\text K} :=  \frac{h}{e^2},
\end{equation}
que tiene dimensiones de resistencia y es la resistencia elemental. 
\qed

\begin{figure}[t]
  \includegraphics[width=0.45\textwidth]{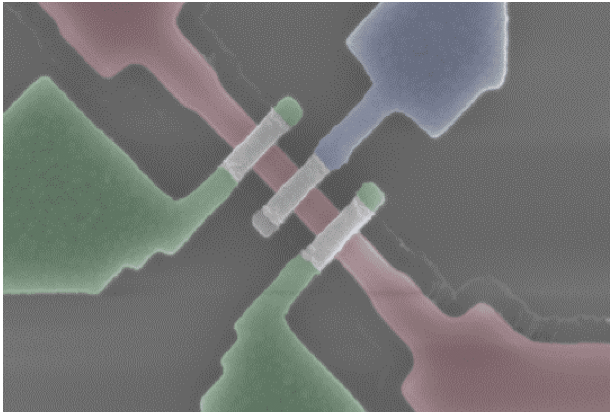}
  \caption{Ejemplo de un dispositivo de transporte mono-eletr\'onico (TME) empleado en el NIST para la definici\'on
  del amperio. La interacci\'on entre los electrodos (puertas) azul y verde controla el movimiento de electrones individuales dentro y fuera de una ``isla" en el centro. Explicaciones en el texto. Los colores son solo ilustrativos. (Cr\'edito: NIST).}
  \label{fig:SET}
\end{figure}

\noindent {\bf Efecto Josephson}: cuando un cable superconductor se interrumpe en un punto con un contacto hecho de material aislante que une dos porciones superconductoras, la corriente superconductora puede mantenerse debido a un efecto t\'unel de los pares de Cooper del superconductor. Esto se conoce como una uni\'on Josephson
(ver Fig.\ref{fig:Josephson_union}). En estas circunstancias, si se aplica una radiaci\'on de radiofrecuencia $\nu$ se induce un potencial $V$ a trav\'es de la uni\'on que es proporcional a la frecuencia y est\'a cuantizado \cite{josephson1,josephson2}: 
\begin{equation}\label{Josephson}
V_{\text J} = n \frac{h}{2e} \nu,
\end{equation}
donde $n$ es un entero, $2e$ es la carga del par de Cooper y se define la constante de Josephson como
\begin{equation}\label{constanteJosephson}
K_{\text J} := \frac{2e}{h}.
\end{equation}
Este es el llamado efecto Josephson de corriente continua (CC) y las uniones Josephson se pueden realizar con puntos met\'alicos o con constricciones, adem\'as de aislantes. Permite medir voltajes con una indeterminaci\'on de $10^{-9,-10}$ voltios, es decir, del orden de nano voltios o menos. Por esta raz\'on se utiliza para la realizaci\'on el patr\'on est\'andar de voltaje \cite{press_kit,cem,SI:BIPM}.

\qed

Ahora podemos relacionar la masa prueba \eqref{masaKibble} que se emplea en el modo pesada con la constante de Planck que aparece al medir la resistencia tambi\'en en el modo pesada y el voltaje en el modo velocidad. Usando el efecto Hall cu\'antico entero \eqref{HallEntero} para medir la resistencia y el efecto Josepshon \eqref{Josephson} para medir los voltajes $V_{\text R}$ y $V$, obtenemos la relaci\'on deseada
\begin{equation}\label{masaKibble-h}
m  = \left( n^\prime n_1 n_2\frac{\nu_1 \nu_2}{4 gv} \right) h,
\end{equation}
donde los n\'umeros enteros que aparecen provienen de las mediciones concretas de los correspondientes efectos cu\'anticos. Para medir $g$ se utiliza un grav\'{\i}metro absoluto de alta precisi\'on y $v$ con m\'etodos interferom\'etricos.
Con todas estas medidas de alta precisi\'on, la expresi\'on \eqref{masaKibble-h} tiene una una utilidad ambivalente: por un lado,
dada una masa patr\'on $m$ como la del antiguo IPK podemos determinar $h$ con gran precisi\'on. Por otro lado,
dado que $h$ se puede medir  con este m\'etodo con mucha precisi\'on, podemos fijar este valor de $h$ como exacto y definir la unidad de masa en funci\'on de $h$: conseguimos as\'{\i} realizar la v\'{\i}a del `kilo cu\'antico' y desvincular a la unidad de masa del artefacto kilo IPK.

%\'{\i}

\begin{figure}[t]
  \includegraphics[width=0.35\textwidth]{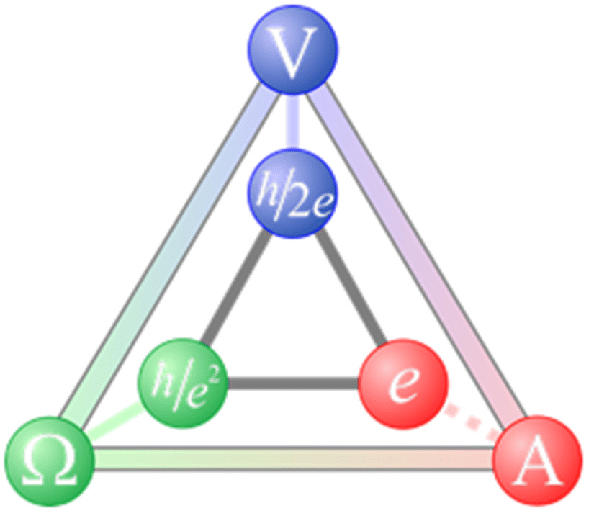}
  \caption{Tri\'angulo metrol\'ogico cu\'antico: relaci\'on esquem\'atica entre las constantes universales del efecto Hall \eqref{HallEnteroconstante},  efecto Josephson \eqref{constanteJosephson} y la carga del electr\'on en un dispositivo TME. El tri\'angulo establece una ligadura entre ellas y las constantes de Planck $h$ y la carga elemental $e$. (Cr\'edito: Piquemal et al. \cite{triangulocuantico}).}
  \label{fig:TMC}
\end{figure}

\subsubsection{Tri\'angulo Metrol\'ogico Cu\'antico}
\label{Triangulo}

La nueva definici\'on del amperio vincul\'andose al valor de la carga elemental del electr\'on $e$ resalta por su claridad y simplicidad comparada con la vetusta definici\'on basada en la ley de Ampere y una construcci\'on irrealizable utilizando hilos conductores infinitos y de grosor nulo \cite{SI:BIPMrealization}. Pero tambi\'en conlleva la necesidad de materializarlo de alguna manera, y no es nada f\'acil pues el n\'umero de electrones en un sistema ordinario es inmensamente grande. El BIPM ha aprobado tres m\'etodos para la realizaci\'on pr\'actica del amperio \cite{cem,SI:BIPM,SI:BIPMrealization}. Uno de ellos hace uso de la definici\'on direacta de amperio $A=C/s$ y de  un dispositivo de transporte mono-electr\'onico (TME),  que ha de enfriarse hasta temperaturas cercanas al cero absoluto (ver Fig.\ref{fig:SET}). Por un TME, los electrones pasan desde una fuente hasta un drenaje. El TME consta de una regi\'on hecha de silicio, llamada isla, entre dos puertas que sirven para manipular el\'ectricamente la corriente. La isla almacena temporalmente los electrones provenientes de la fuente usando otra puerta de voltaje. Mediante el control de los voltajes en las dos puertas, se puede conseguir que un solo electr\'on permanezca en la isla antes de pasar al drenaje. Repitiendo este proceso muchas veces y muy r\'apidamente, se consigue establecer una corriente de la que se pueden contar sus electrones.

El sector el\'ectrico es el m\'as cu\'antico de todas las unidades en el SI.
Ahora que el amperio ha sido redefinido en el nuevo SI vincul\'andolo a un valor fijo de la carga del electr\'on, es posible relacionar las tres magnitudes que aparecen en la ley de Ohm, $V=IR$, en t\'erminos de solo dos constantes universales, $h$ y $e$. Esto se visualiza mediante el llamado tri\'angulo metrol\'ogico cu\'antico (ver Fig.\ref{fig:TMC}). Este tri\'angulo representa una ligadura experimental que deben cumplir los patrones de voltaje, resistencia e intensidad, de modo que los tres no son independientes. As\'{\i}, si medimos la constante de Josephson \eqref{constanteJosephson} por un lado y la de von Klitzing \eqref{HallEnteroconstante} por otro, nos permiten obtener valores para la unidad de carga $e$ del electr\'on que debe ser compatible, dentro de las indeterminaciones experimentales, con el valor de $e$ obtenido con un transporte mono-electr\'onico. Y lo mismo para cualquier pareja de magnitudes que tomemos en el tri\'angulo. Por tanto, el tri\'angulo metrol\'ogico cu\'antico nos permite comprobar experimentalmente a medida que se consiguen mejores exactitudes y precisiones, si las constantes $h$ y $e$ son realmente constantes como se ha supuesto en el nuevo SI. Las indeterminaciones en estas constantes deben ser compatibles usando estas tres realizaciones experimentales. Si en alg\'un momento estas indeterminaciones no se solapan, entonces esto es un indicio de nueva f\'{\i}sica pues afectar\'{\i}a a los fundamentos de la mec\'anica cu\'antica o de la electrodin\'amica cu\'antica como se explic\'o en la secci\'on \ref{sec:constantesfundamentales}. De nuevo, esto es un ejemplo de c\'omo la metrolog\'{\i}a no solo sirve para mantener patrones de unidades, sino para abrir nuevos caminos hacia nuevas leyes fundamentales de la naturaleza.

%%%%%%%%%%%%%%%%%%%%%%%%%%%%%%%%%%%%%%
%%%%%%%%%%%%%%%%%%%%%%%%%%%%%%%%%%%%%%

%%%%%%%%%%%%%%%%%%%%%%%%%%%%%%%%%%%%%%
%%%%%%%%%%%%%%%%%%%%%%%%%%%%%%%%%%%%%%
\section{Una `Anomal\'{\i}a Gravitacional' en el SI}
\label{sec:anomaliaG}
%%%%%%%%%%%%%%%%%%%%%%%%%%%%%%%%%%%%%%
%%%%%%%%%%%%%%%%%%%%%%%%%%%%%%%%%%%%%%

A pesar de que el nuevo sistema de unidades SI supone la vinculaci\'on completa
de las unidades base a constantes fundamentales de la naturaleza, no deja de ser
llamativo la ausencia de una de las constantes universales m\'as antiguas de la F\'{\i}sica:
la constante de la gravitaci\'on universal de Newton $G$ (ver Fig.\ref{fig:Ganomalia}), llamada coloquialmente la
$G$ grande, en contraposici\'on a la $g$ peque\~{n}a que representa la aceleraci\'on local
de la gravedad en un punto terrestre.

La principal raz\'on para excluir a $G$ del nuevo sistema SI de unidades es la falta de precisi\'on
suficiente para poder definir una unidad de masa. Como se explic\'o en \ref{sec:nuevasdefiniciones}, este es el origen de la `v\'{\i}a cu\'antica' para la definici\'on del kilo cuando se quiere desvincularlo de un artefacto material
como el cilindro del kilo IPK.
Este hecho est\'a relacionado con el llamado `Problema de la G grande de Newton' \cite{nistbigG,quinn2000,quinn2013,sanfernando2017}. Este problema consiste
en la fata de compatibilidad de las medidas de $G$ en los \'ultimos treinta a\~{n}os. Diversos laboratorios
de metrolog\'{\i}a de todo el mundo han tratado de medir $G$ con dispositivos experimentales dise\~{n}ados
para conseguir reducir la incertidumbre en el valor de $G$.
El resultado es sorprendente:  los valores de $G$ no convergen a un valor \'unico consistente y sus incertidumbres
no se solapan de forma compatible. Esto se puede apreciar en la Fig.\ref{fig:G}, donde se muestran los resultados de m\'ultiples experimentos y una zona vertical donde se elige un valor de $G$ de compromiso.  La situaci\'on se ha vuelto desesperante y la NSF (National Science Foundation) ha lanzado una iniciativa mundial para tratar de esclarecer el problema \cite{nsfG}.

Surge entonces una pregunta fundamental: ?`Cu\'al es el origen del problema de la $G$ grande de Newton?
La soluci\'on m\'as natural es que sea debido a posibles errores sistem\'aticos en los experimentos. A favor de
esta interpretaci\'on est\'a el hecho de que,
a pesar de la cada vez mayor sofistificaci\'on para tratar de medir $G$ con mayor precisi\'on,
todos los m\'etodos experimentales empleados son variantes de la c\'elebre  balanza de Cavendish \cite{amoreno,springer}.
Sin embargo, en la Fig.\ref{fig:G} aparece un valor \cite{AFG} cuyo m\'etodo experimental es completamente diferente de los m\'etodos basados en la balanza de Cavendish. Se trata de un m\'etodo cu\'antico de medir $G$. Su dispositivo utiliza una t\'ecnica basada en interferometr\'{\i}a at\'omica con \'atomos ultrafr\'{\i}os. Con ella, se hace uso de la naturaleza cu\'antica de los \'atomos a temperaturas cercanas al cero absoluto, para obtener una medida precisa de la aceleraci\'on de la gravedad.

El m\'etodo de los \'Atomos Fr\'{\i}os en Gravedad (AFG) consta de 2 pasos:
Paso 1: medida de la constante $g$ peque\~{n}a: valor de la gravedad local terrestre.
Paso 2: medida de la constante $G$ grande.

La t\'ecnica consiste en voltear \'atomos fr\'{\i}os verticalmente, arriba y abajo, de forma repetida. Esto sirve para sondear la gravedad terrestre con una nube de \'atomos de rubidio Rb en ca\'{\i}da libre. Con esto se consigue medir la fuerza de la gravedad entre un \'atomo de Rb y una masa de referencia de 516 kg. El resultado es una medida de $G$ con una incertidumbre relativa del $0.015\%$. Es la primera vez que un m\'etodo cu\'antico es admitido a formar parte del conjunto de valores que sirven para determinar $G$.

\begin{figure}
  \includegraphics[width=0.35\textwidth]{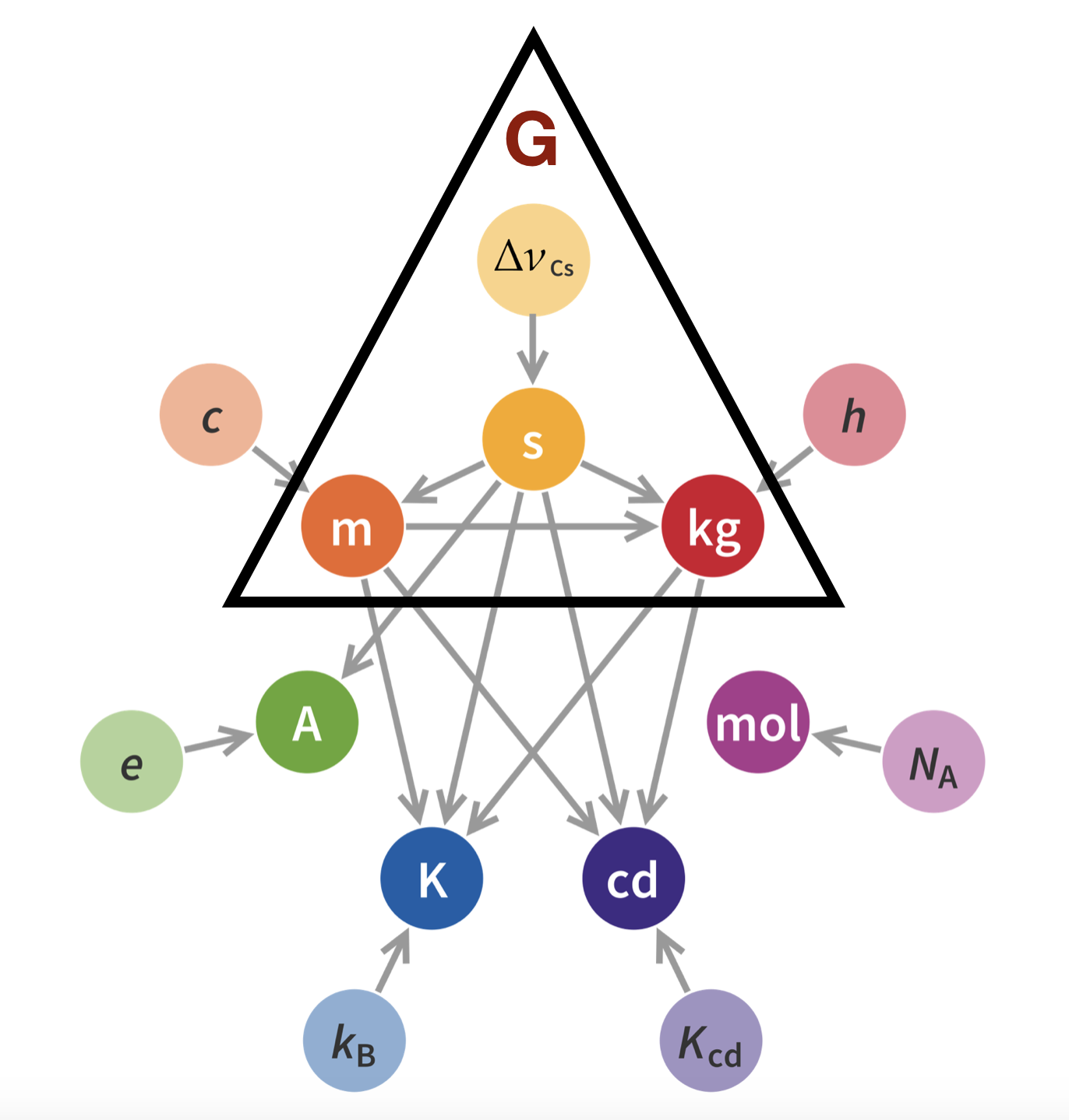}
  \caption{El esquema de las dependencias de las constantes naturales y las unidades b\'asicas del nuevo SI en la Fig.\ref{fig:dependencias} presenta una ausencia notable: la constante $G$ de la graviaci\'on universal de Newton. El tri\'angulo engloba a la excluida $G$ con las unidades de las que depende  \eqref{Gunidades}
  que son las mismas que la constante $h$ de Planck \eqref{hunidades}, pero en otra proporci\'on. (Cr\'edito: Emilio Pisanty/Wikipedia (adapted)).}
  \label{fig:Ganomalia}
\end{figure}

Esta anomal\'{\i}a gravitacional no deja de ser un reflejo del gran problema que afecta
a la f\'{\i}sica moderna: la falta de compatibilidad entre las dos grandes teor\'{\i}as de nuestro
tiempo, la mec\'anica cu\'antica y la relatividad general.
Una observaci\'on importante (ver Conclusiones) es que
el m\'etodo AFG es un m\'etodo indirecto de medici\'on para la $G$ grande de Newton: se hace a trav\'es de medir primero $g$ peque\~{n}a. Esto contrasta con los m\'etodos c\'asicos basados en la balanza de Cavendish donde se puede medir $G$ directamente.
Una medida cu\'antica directa de $G$ ser\'{\i}a un primer indicio experimental de efectos cu\'anticos en la gravedad y un primer paso para una teor\'{\i}a de gravedad cu\'antica. Como se ve en la Fig.\ref{fig:G}, el valor de AFG todav\'{\i}a queda fuera de la zona vertical sombreada del valor recomendado m\'as reciente para $G$. Esto puede ser indicativo de que el m\'etodo AFG no sufre de los posibles errores sistem\'aticos en los m\'etodos cl\'asicos de medida de $G$ y podr\'{\i}a ser el comienzo para la soluci\'on del problema de la $G$ grande de Newton. La manera de confirmar esta hip\'otesis es fomentar la realizaci\'on de m\'as experimentos tipo AFG en m\'as laboratorios independientes y usar t\'ecnicas de metrolog\'{\i}a cu\'antica para disminuir sus incertidumbres.
Si el reultado de todos estos nuevos experimentos cl\'asicos y cu\'anticos fuera que no existen errores sistem\'aticos, entonces la conclusi\'on ser\'{\i}a todav\'{\i}a m\'as emocionante pues ser\'{\i}a de nuevo una puerta abierta por la metrolog\'{\i}a a una nueva f\'{\i}sica.

M\'etodos directos para medir $G$ no se conoce ninguno. De hecho, hay pocas ecuaciones de la f\'{\i}sica donde $G$ y $h$ aparezcan juntas. Una de ellas nos puede servir para ver las dificultades de conseguir un m\'etodo cu\'antico directo para medir $G$. Se trata de la ecuaci\'on del l\'{\i}mite de Chandrasekhar para el radio de una estrella enana blanca. Si pens\'aramos ingenuamente en producir un condensado gravitatorio de nucleones (fermiones) que fuera el resultante de compensar la presi\'on gravitatoria de $N$ nucleones de masa $M$ con la presi\'on de degeneraci\'on debido al principio de exclusi\'on de Pauli, usando mec\'anica cu\'antica no-relativista y gravitaci\'on newtoniana para simplificar, se obtiene \cite{enanablanca} un valor del radio de equilibrio dado por
\begin{equation}\label{Chandrashekar}
R_0 = \left(\frac{9\pi}{4} \right)^{2/3} \frac{\hbar q^{5/3}}{G m M^2 N^{1/3}},
\end{equation}
donde $N$ es el n\'umero de nucleones y $q$ el n\'umero de electrones por nucle\'on con masa $m$. Se ha supuesto que la densidad del condensado esf\'erico es uniforme. Para hacer una estimaci\'on sencilla, consideremos un sistema de solo neutrones ($q=1$, $m=M=m_n$ masa del neutr\'on), y sustityendo los valores experimentales conocidos, obtenemos
\begin{equation}\label{Chandrashekar2}
R_0 =  5.51 \times 10^{24}N^{-1/3} \text{m}.
\end{equation}
Si queremos tener un sistema de fermiones condensado en una esfera con un radio del orden de un metro para ser manipulable en un laboratorio terrestre, podemos estimar el n\'umero de neutrones necesarios que se obtiene con 
\eqref{Chandrashekar2} en algo del orden de $N\sim 10^{74}$, una cantidad intratable si tenemos en cuenta que el n\'umero de \'atomos en el universo observable es del orden de $10^{80}$. Esta dificultad es un reflejo de la disparidad de escalas donde act\'ua la gravedad frente a los efectos cu\'anticos.

\begin{figure}
 \includegraphics[width=0.55\textwidth]{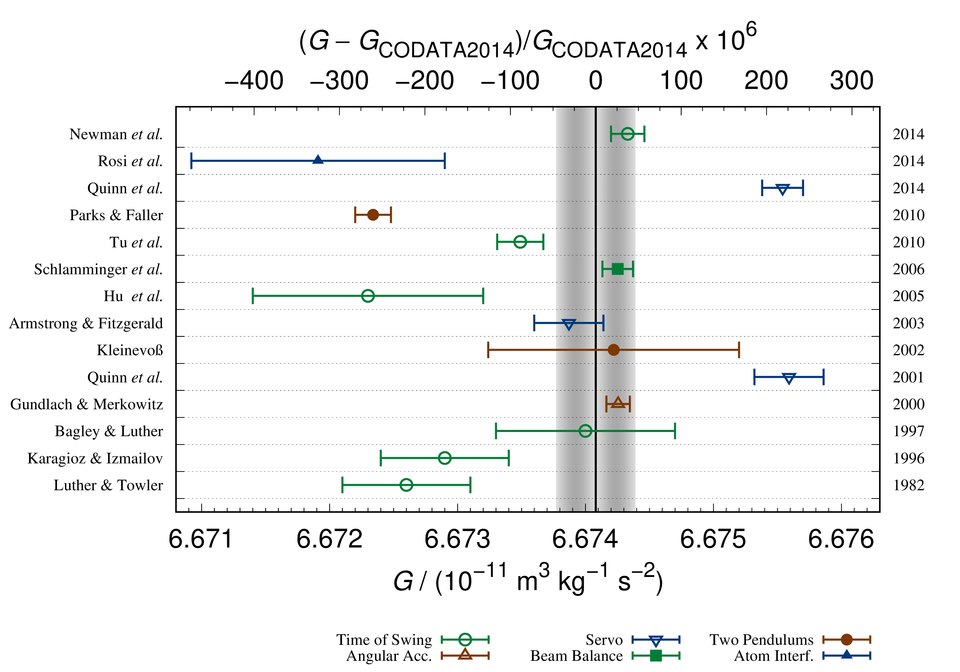}
  \caption{Diagrama comparativo de las mediciones de la constante $G$ a lo largo del tiempo utilizando m\'etodos
  cl\'asicos basados en variantes de la balanza de torsi\'on, salvo el de Rossi et al. \cite{AFG} que emplea el m\'etodo de los \'Atomos Fr\'{\i}os en Gravedad (AFG). (Cr\'edito: NIST \cite{nistbigG}).}
  \label{fig:G}
\end{figure}

%%%%%%%%%%%%%%%%%%%%%%%%%%%%%%%%%%%%%%
%%%%%%%%%%%%%%%%%%%%%%%%%%%%%%%%%%%%%%
\section{Conclusiones}
\label{sec:conclusions}
%%%%%%%%%%%%%%%%%%%%%%%%%%%%%%%%%%%%%%
%%%%%%%%%%%%%%%%%%%%%%%%%%%%%%%%%%%%%%

La adopci\'on del nuevo SI de unidades trae consigo varias ventajas concretas respecto al
anterior sistema: resuelve el problema del amperio y las unidades el\'ectricas que se hab\'{\i}an
quedado fuera del SI, elimina la dependencia del kilo con el artefacto del kilo IPK, conceptualmente es m\'as satisfactorio definirlas en t\'erminos de constantes naturales, las futuras mejoras tecnol\'ogicas no afectar\'an m\'as a las definiciones etc.

El nuevo SI de unidades no tiene repercusiones directas en la vida cotidana, pero s\'{\i} en 
los laboratorios de investigaci\'on y en los centros nacionales de metrolog\'{\i}a donde se
necesitan medidas de gran exactitud y precisi\'on para realizar esas investigaciones y para 
custodiar y diseminar los patrones primarios de las unidades.
Como siempre sucede, a la larga esos nuevos descubrimientos resultan en aplicaciones
que s\'{\i} modifican a mejor nuestra vida diaria.

Por tanto, es un gran reto conceptual explicar y transmitir lo que supone y hay detr\'as del nuevo sistema de unidades. En \ref{sec:nuevasdefiniciones} se ha presentado una visi\'on com\'un de todas las nuevas definiciones utilizando como principio unificador la naturaleza discreta de la energ\'{\i}a, la materia y la informaci\'on en las leyes fundamentales de la F\'{\i}sica y la Qu\'{\i}mica a las que aparecen vinculadas cada una de las unidades. Curiosamente, lo \'unico que queda
sin discretizar es el espacio-tiempo.
Una ventaja del nuevo SI es que facilita la explicaci\'on de las nuevas definiciones 
al no necesitar explicar los aparatos de medida necesarios para realizar dichas unidades.

La metrolog\'{\i}a tiene una doble misi\'on: 1) Mantener los patrones de las unidades y su definiciones compatibles
con las leyes actuales de la f\'{\i}sica. 2) Medir cada vez con mayor exactitud y precisi\'on abriendo nuevas puertas a descubrir nuevas leyes de la f\'{\i}sica.

En cuanto a su misi\'on m\'as tradicional 1), la adopci\'on del nuevo SI permite deshacerse de un artefacto material 
para definir el kilo, que era  un objetivo largamente so\~{na}do. Con ello, es posible materializar patrones primarios de las unidades b\'asicas en centros de metrolog\'{\i}a nacionales distintos por primera vez. En particular, la metrolog\'{\i}a cu\'antica va a cambiar dr\'asticamente la diseminaci\'on y trazabilidad de las unidades de medida al conseguir que las unidades se puedan materializar aut\'onomamente sin necesidad de tener un \'unico patr\'on.

Es interesante hacer notar que en el transcurso de la construcci\'on del nuevo SI han aparecido las tres balanzas m\'as famosas de la f\'{\i}sica: Cavendish \cite{amoreno,springer}, E{\"o}tv{\"o}s \cite{equivalence} y Kibble \cite{kibblebalance}. Como tambien han sido fundamentales los art\'{\i}culos de Einstein en su {\it annus mirabilis} de 1905 \cite{einstein1905d, einstein1905c,einstein1905a,einstein1905b}.

En cuanto a la segunda misi\'on, hemos visto c\'omo el nuevo SI utiliza cinco constantes universales de la naturaleza.
De ellas, tres tienen un estatus especial, $c,h$ y $e$, pues est\'an asociadas a principios de simetr\'{\i}a del universo
como el principio de relatividad, unitariedad y simetr\'{\i}a gauge. Las otras dos son la constante de Botzmann $k$ y la constante de Avogadro $N_{\text A}$, las cuales no tiene ninguna simetr\'{\i}a asociada.

Ahora que las  unidades f\'{\i}sicas  est\'an definidas por la f\'{\i}sica fundamental del universo, 
y no por una maquinaci\'on humana usando artefactos, las constantes fundamentales del universo,
?`son una maquinaci\'on de algo? ?`porqu\'e son las que son? ?`y hasta cu\'ando?
Hemos llegado as\'{\i} a las preguntas m\'as fundamentales de la f\'{\i}sica.
Por eso la metrolog\'{\i}a realmente va m\'as all\'a de mantener los patrones de medida.

%\'{\i}

\begin{acknowledgments}
Estas notas son el resultado de varias conferencias impartidas durante 2017, 2018 y 2019.
Mi agradecimiento a los organizadores Jos\'e Manuel Bernab\'e y Jos\'e \'Angel Robles del Centro Espa\~{n}ol
de Metrolog\'{\i}a (CEM) por su invitaci\'on al $6^{o}$ Congreso Espa\~{n}ol de Metrolog\'{\i}a (2017), al $8^{o}$
Seminario Intercongresos de Metrolog\'{\i}a (2018) y al Congreso del 30 Aniversario del CEM (2019); a Alberto Galindo y Arturo Romero de la Real Academia de Ciencias Exactas, F\'{\i}sicas y Naturales de Espa\~{n}a por su
invitaci\'on al Ciclo Ciencia para Todos (2018) y la Jornada sobre ``La revisi\'on del Sistema Internacional de Unidades, (SI). Un gran paso para la ciencia" (2019); a Federico Finkel y Piergiulio Tempesta por su invitaci\'on al acto homenaje de Artemio Gonz\'alez L\'opez con motivo de su 60 aniversario.
M.A.M.-D. agradece el apoyo financiero del  MINECO grant FIS2015-67411, del consorcio de investigaci\'on de la CAM QUITEMAD+, Grant No. S2013/ICE-2801 y en parte por el  U.S. Army Research Office a trav\'es de Grant No. W911N F- 14-1-0103. 

%\'{\i}

\end{acknowledgments}

%%%%%%%%%%%%%%%%%%%%%%%%%%%%%%%%%%%%%%%%%%%%%%%%%%%%%%%%%%%%%%%%%%%%%%%%%%%%%%
%\begin{references}


\begin{thebibliography}{99}
\bibitem{squeezed}
Squeezed states of light have amplitude quadrature and phase quadrature associated to the electric part of the electromagnetic waves. They are orthogonal to each other but they are not independent since they satisfy an uncertainty constraint: increasing the precision of one comes at the expense of the other. Since for increasing the sensitivity of a position measurement in the arms of a LIGO interferometer what matters is the amplitude quadrature, the fundamental of using squeezed light in these measurements relies in reducing the uncertainty of the amplitude at the expense of that of the phase that plays no role in the position determination.
\end{thebibliography}
\end{document}